\documentclass[a4paper]{article}
\usepackage{amsfonts}
\usepackage{amssymb}
\usepackage{amsmath}
\usepackage{graphicx}
\usepackage{graphics}
\usepackage{geometry}
\usepackage{caption} 
\usepackage{float, times}
\usepackage{media9}
\usepackage[hyperindex=true, colorlinks=true] {hyperref}
\usepackage{mathptmx}
\usepackage{newcent}
\usepackage{mathpazo}
\usepackage{setspace}
\usepackage{enumerate}
\usepackage{epstopdf}
\newcommand{\R}{\mathbb{R}}

\newcommand{\edem}{\hfill $\Box$ }

\newcommand{\dsum}{\displaystyle\sum}
\newcommand{\dlim}{\displaystyle\lim}

\newcommand{\diag}{\mathrm{diag}}
\newcommand{\mb}{\mathbf}

\newcommand{\defn}[1]{\emph{#1}}

\newtheorem{prop}{Proposition}[section] 
\newtheorem{dfn}{Definition}[section] 
\newtheorem{thm}{Theorem}[section] 
\newtheorem{lem}{Lemma}[section] 
\newtheorem{rmq}{Remark}[section] 
 
\newtheorem{cor}{Corollary}[section]

\title{Analysis of Malaria Control Measures' Effectiveness Using Multi-Stage Vector Model}
\author{J. C. Kamgang$^{1, \dag, }$\thanks{UMI 209 IRD/UPMC UMMISCO, Bondy, Projet MASAIE INRIA Grand Est, France}\footnote{LIRIMA -- GRIMCAPE, Cameroun} \;\;\thanks{Corresponding author Jean Claude Kamgang email: jckamgang@gmail.com} ,  C. P. Thron$^{2}$  \\
\\$^1$ Department of Mathematics and Computer Sciences,\\ ENSAI -- University of N'Gaound\'er\'e, P. O. Box 455 N'Gaound\'er\'e (Cameroon)\\$^2$ Department of Sciences and Mathematics,\\  Texas A\& M University -- Central Texas 76549 USA}

\begin{document}

\maketitle
    
\begin{abstract}
We analyze  an epidemiological model  to evaluate the effectiveness of  multiple means of control in malaria-endemic areas. The mathematical model consists of  a system of several ordinary differential equations, and is based on a multicompartment representation of the system. The model takes into account the mutliple resting-questing stages undergone by adult female mosquitos during the period in which they function as disease vectors.  We compute the basic reproduction number $\mathcal R_0$, 
and show  that that if $\mathcal R_0<1$, the disease free equilibrium (DFE)  is globally asymptotically stable (GAS) on the non-negative orthant. If $\mathcal R_0>1$,  the system admits a unique endemic equilibrium (EE) that is GAS. We perform a sensitivity analysis of the dependence of $\mathcal R_0$ and the EE on parameters related to control measures, such as killing effectiveness and bite prevention. Finally, we discuss the implications for a comprehensive, cost-effective strategy for malaria control.  
\end{abstract}


\section{Introduction}\label{sec.Intro}
 Malaria is a vector-borne infectious disease that  is widespread in tropical regions, including parts of America, Asia and much of Africa. Humans contract malaria following effective bites of infectious Anopheles female mosquitoes during blood feeding by the infiltration of parasites contained in the mosquitoes saliva into the host's bloodstream. Among those parasites, {\it Plasmodium falciparum} is the prevalent cause of malaria mortality in Africa. The chain of transmission can be broken through the use of insecticides and anti-malarial drugs, as well as other control strategies. 

Malaria accounts for more than 207 million infections and results in over 627000 deaths globally in 2012 ~\cite{wmr2013}.  About 90\% of these fatalities occur in sub-Saharan Africa~\cite{DCollCZim, wmr2013}. Despite intensive social and medical research and numerous programs to combat malaria,  the incidence of  malaria across the African continent remains high.

Control measures that have been used against malaria include: 
\begin{itemize}
\item
Outdoor application of larvicides (chemical or biological)\cite{floore2006mosquito, walker2007contributions} ; 
\item
Breeding habitat reduction (e.g. draining standing water) \cite{ keiser2005reducing,walker2007contributions,yohannes2005can}; 
\item
Outdoor vector control (mosquito fogging, attractive toxic sugar bait (ATSB)) \cite{zhu2017outdoor};
\item
Indoor residual spraying (IRS)\cite{pluess2010indoor,sharp2007seven};
\item
Bed nets, including insecticide-treated bed nets (ITN), long-lasting insecticidal nets (LLIN), and untreated bed nets \cite{world2015world};
\item
Repellents, including topical repellents, mosquito coils, etc \cite{lawrance2004mosquito, maia2015mosquito};
\item
Rapid diagnosis and treatment (RDT); \cite{awoleye2016improving,shillcutt2008cost}
\item
Preventative drugs: seasonal malaria chemoprevention (SMC), intermittent preventative treatment (IPT). \cite{wilson2011systematic};
\end{itemize}

Numerous empirical studies have been conducted to assess the cost effectiveness of these different methods \cite{akhavan1999cost,utzinger2001efficacy,worrall2011large}.
In \cite{morel2005cost} examined the cost effectiveness of ITN, IRS, IPT, case management with various drugs, and various combinations of these measures as applied to malaria control in sub-Saharan Africa. For  60 alternative strategies involving these measures, costs (in 2000 international dollars) per disability adjusted life years (DALY) averted were estimated.  The most cost-effective intervention found involved the use of artemisinin based combination treatment (ACT) only.  However, this option only acheived a relatively low level of DALY averted.  The study found that the best way to improve DALY averted involved introducing other measures: first ITN and IPT, and subsequently IRS to acheive the maximum DALY averted.
  
A comprehensive (as of 2010) review of studies on cost effectiveness of ITN, IRS, and IPT interventions may be found in \cite{white2011costs}. Costs per death averted and per DALY averted are also given, as are costs per treatment. In general results show that costs are highly situation-dependent.
Estimates for costs of protection per individual  per year are given for numerous studies employing  ITN, IRS, or IPT: results are summarized in Table~\ref{tab.tabvd4}.

\begin{table}[htbp]
\caption{Cost per person per year of protection (2009 US\$) across all studies}\label{tab.tabvd4}

\begin{tabular}{ | l | c | c | c | c |}

\cline{1-3}
\hline
\emph{Control measure} & Mean (Standard Deviation) & Median \\
\hline
Indoor residual spraying (IRS) & 6.3 (3.4 ) & 6.7 \\
Insecticide-treated bed nets (ITN) & 2.9 (2.2) & 2.2 \\
Intermittent preventative treatment (IPT) & 4.3 (5.7) & 2.545  \\
\hline
\end{tabular}
\end{table}
Reference \cite{white2011costs} does  not consider the impacts of different interventions on overall malaria prevalence. For example, there is evidence that use of ITNs decreases the vector population, which may reduce malaria rates even among non-users \cite{bayoh2010anopheles, hawley2003community}.

Besides financial costs,  significant environmental costs may be incurred by control measures, particularly those that involve insecticides\cite{keiser2005reducing}. Some insecticides also pose health risks to humans \cite{ehiri2004mass}.
 Extensive use of insecticides also tends to increase resistance among vectors, which decreases the insecticides' effectiveness. \cite{menze2016multiple, ranson2011pyrethroid}
Similarly, use of preventative drugs tends to produce resistant parasites.  Some sources recommend an integrated approach which incorporates several different control strategies \cite{beier2008integrated, russell2011increased}

In the field of mathematical epidemiology, numerous models have been proposed with the purpose  of understanding various aspects of the disease. The foundational model of Sir Ronald Ross, originally proposed in 1911 \cite{Ross1911} and extended by Macdonald in 1957 \cite{Macdonald78}, serves as the basis for many mathematical investigations of the epidemiology of malaria.  A prominent example is the model of Ngwa and Shu \cite{NgwaMCM00}, which introduces Susceptible ($\mathrm{S}$), Exposed ($\mathrm{E}$), and Infectious ($\mathrm{I}$) classes for both humans and mosquitoes, plus an additional Immune class ($\mathrm{R}$) for humans. This model is extended in the Ph.D. theses of Chitnis \cite{Chit_08} and Zongo \cite{Zongo09},  both of which also provide comprehensive reviews on the state of the art.  Chitnis introduces immigration into the host population, which is a significant effect since hosts migrating from a naive (disease-free)
 region to a region with high endemicity are especially susceptible to infection. Chitnis also performed a sensitivity analysis of model parameters, and identified the mosquito biting rate (and the recovery rate as the two important parameters which should be targeted in controlling malaria \cite{chitnis2008determining}.  Chitnis' conclusion was that  the use of insecticide-treated bed nets, coupled with rapid medical treatment of 
new cases of infection,  is the best strategy to combat malaria transmission.
Zongo further extends the model by dividing the human population into non-immune and semi-immune subpopulations, which are modeled using $(\mathrm{SEIS})$ and $(\mathrm{SEIRS})$ model types, respectively. 
In this paper we include all of the above effects, and  extend  the model by dividing the human population into  groups according to the method(s) they use to protect themselves against the mosquito bites (as in \cite{jckam201411}). This extension improves the applicability of the model because it represents the actual situation in many endemic areas, particularly in poor countries.
 
Malaria is highly seasonal\cite{9129525, 10697865}: the highest endemicity typically occurs during rainy  seasons, when mosquito density is high due to high humidity and the presence of standing water where mosquitoes can breed. During this period, even people with predispositional immunity to malaria infection are at risk of attaining the critical level of malaria parasites in their bloodstream that could make them sick. In our model, we consider conditions characteristic of a rainy season in a region of high malaria endemicity: typically, such conditions last for a period between three to six months. This paper improves on the model in~\cite{jckam201411} by including  the effects of death, birth and migration on each host subpopulation. This inclusion is justified,  
since malaria is a major cause of death in endemic areas. As in~\cite{jckam201411}, we omit Exposed and Removed classes for hosts: the duration of Exposed and Removed states can be assumed to be negligible due to the high density of anopheles mosquitoes on the one hand, and rapid detection and treatment of infectious individuals on the other hand. Results for more sophisticated models that include Exposed and/or Removed state(s) are reserved for forthcoming papers. 

The paper is organized as follows. Section~\ref{sec.Bdnmdel} describes our model and gives the corresponding system of  differential equations.  Section~\ref{dfestabana} establishes the well-posedness of the model by demonstrating invariance of the set of nonnegative states, as well as boundedness properties of the solution. The equilibria of the system, are calculated, and a threshold condition for the stability of the disease free equilibrium (DFE), which is based on the basic reproduction number $\mathcal R_0$ is calculated.  
Section~\ref{sec.analysis} analyzes the stability of equilibria. Section~\ref{subsec.dfestabanan} demonstrates the global asymptotic stability (GAS) of the disease free equilibrium (DFE) when $\mathcal R_0\leq 1$, and  Section~\ref{sec:eeqstana}, establishes the GAS of the endemic equilibrium (EE) when $\mathcal R_0>1$. Sections~\ref{sec.discuss} and~\ref{sec.conclusion} provide discussion and conclusions. Finally, the Appendix contains detailed proofs and computations required by the analysis.  

\section{Model description and mathematical specification}\label{sec.Bdnmdel}

The model assumes an area populated by human hosts and female mosquitoes (disease vectors) under conditions of elevated endemicity of malaria. Mosquitos in the model are assumed to be anthropophilic, and bite only humans: this assumption reflects situations in which malaria poses the biggest danger \cite{besansky2004no}.  Both human and mosquito populations are homogeneously mixed, so no spatial effects are present.  In the following subsections, we provide a detailed description of the population structure and dynamics of hosts and vectors.

\subsection{Host population structure and dynamics}\label{subsec.Assumption2}
The human population is divided into $n+1$ groups, indexed by $0,1, \cdots, n$. Group $0$ consists of humans who use no prevention, while the other $n$ groups correspond to users who take various preventative measures such as bed nets (untreated or treated with insecticides of various degrees of toxicity), repellents, prophylactic drugs, indoor insecticides, and so on. 
At any given time,we let $H_i~(i=0,\,\cdots,\,n)$ denote the size of the $i^{th}$ group; note that $H = {\dsum_{i=0}^n} H_i$.

The dynamics of the $i^{th}$ host group ($i=0,\,\cdots,\,n$) is described by a SIS-based compartment model  as shown in  Figure~\ref{fig:figMulticomAppli1}. The incidence of infection for  humans in the $i^{th}$ group  is given by $am_i{I_q}/{H}$, where $a$ is the average number of bites per mosquito per unit time (the entomological  inoculation rate); $I_q$ is the number of Infectious mosquitoes; $m_i$ is the infectivity of mosquitoes relative to the human of the $i^{th}$ group, which is the probability that a bite by an infectious mosquito on a susceptible human of the $i^{th}$ group will transfer  infection to the human. The transition rate from Infectious to Susceptible state within the $i^{th}$  group is $\gamma_i$. The force of migration into the $i^{th}$  group is $\Lambda_i$. The incoming  $\tilde \nu_i$ and outgoing $\nu_i$  rates in the $i^{th}$  group   include the effects of birth and death rates respectively, as well as the effects of hosts moving from one group to another. 

\subsection{Mosquito population structure and dynamics}\label{subsec.Assumption}

\noindent
The population of disease vectors (adult female anopheles mosquitoes) is characterized by several classes, where each mosquito's class membership is determined by its own history of past and present activity. Newly-emerged adult mosquitoes initially enter the Susceptible class: the rate of entry (that is, the recruitment rate) is $\Gamma$. 
Adult mosquitoes alternate between two activities: {\it questing} (that is, seeking a host to bite for a blood meal) and {\it resting} (to lay down eggs, or to digest a blood meal). In~\cite{jckam201411} it was assumed that all susceptible  mosquitoes are in the questing state---the current model improves on this by introducing an additional  compartment for susceptible mosquitoes in the resting state  that have  successfully obtained blood meal(s) and are not yet infected. 
   
At any given instant $t$, questing mosquitoes are equally likely to attempt to feed on any human, regardless of his/her protection method.
 Thus for any attempted blood meal, the time-dependent probability that the human host belongs to the $i^{th}$ group is 
$b_i(t)\equiv{H_i(t)}/{H(t)}$.  During a blood meal attempt involving a human in the $i^{th}$ group, the  mosquito is killed with probability $k_i$, and successfully feeds and enters the resting state  with probability $f_i$. 
Letting $a$ denote the average number of bites per mosquito per unit time (the {\em entomological  inoculation rate}), it follows that  at any given instant $t$, the incidence rate of successful blood meals is $\varpi(t) \equiv \dsum_{i=0}^nab_i(t)f_i$, 
while the  additive death rate caused by the questing activity of mosquitoes is $d(t)\equiv\dsum_{i=0}^nab_i(t)k_i$. 
If we let $I_i$ and $c_i$ denote respectively the number of infectious humans in group $i$ and the probability that a bite of a mosquito on an infectious human in group $i$ will infect the mosquito, then the incidence rate for mosquitoes becoming infected  is $\varphi(t) \equiv {\dsum_{i=0}^n} ac_if_i{I_i(t)}/{H(t)}$. 

Susceptible questing mosquitoes that become infected enter the first exposed resting class ($\mathrm{E}_r^{(\text 1)}$), while those which have not experienced successful infection stay uninfected and enter the susceptible resting class ($\mathrm{S}_r$). Following initial infection, the mosquito must remain alive for a certain period before becoming infectious (this period is called the {\em extrinsic incubation period} in the biological and medical literature ~\cite{010047862}).
During this period, the mosquito experiences a positive number of resting/question cycles. In our model, we suppose that a mosquito becomes infectious after a fixed number $\ell$ of resting/questing cycles following initial infection.  These successive resting/questing cycles are modeled as a sequence of $2\ell $ Exposed states, and are denoted by $ \mathrm{E}^{(\text 1)}_q,\mathrm{E}^{(\text 2)}_r, \cdots, \mathrm{E}^{(\ell)}_q,\mathrm{E}^{(\ell+ 1)}_r$. If a mosquito survives through all of these states, it then enters the Infectious class, which is further divided into questing and resting subclasses ($\mathrm{I}_q$ and $\mathrm{I}_r$, respectively). Once a mosquito enters the Infectious class, it remains there for the rest of its life, alternating between questing and resting states.

\begin{figure}[hpbt]
\begin{minipage}[c]{6cm}
\centerline{\hbox{\includegraphics[scale=0.48]{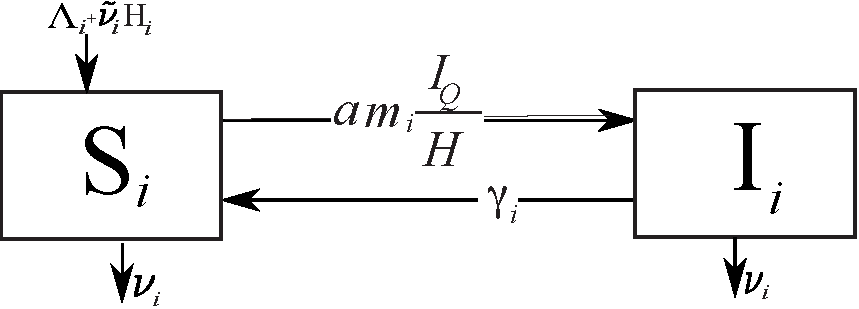}}} 
\caption{\footnotesize Dynamics of the $i^{th}$ human
subgroup\label{fig:figMulticomAppli1}}
\end{minipage}
\begin{minipage}{11cm}
\centerline{\hbox{\includegraphics[scale=0.40]{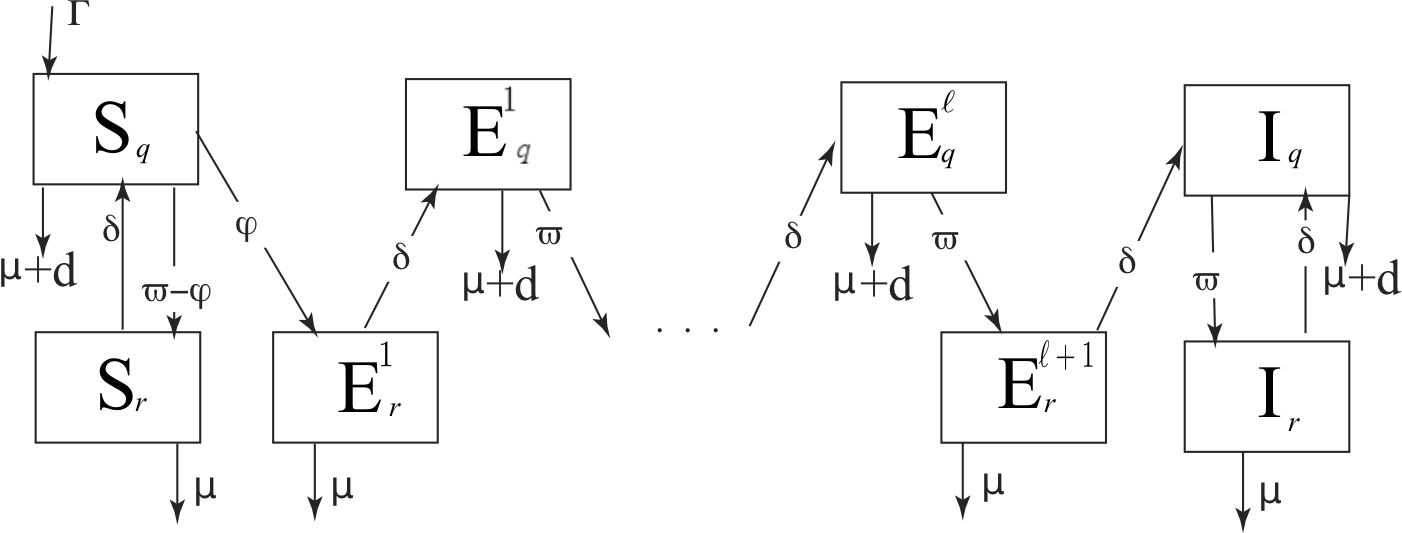}}} 
\caption{\footnotesize Dynamics of the mosquito  population\label{fig:figMulticomAppli2}}
\end{minipage}
\end{figure}

The overall dynamics of the mosquito population is depicted in the multicompartment diagram in  Figure~\ref{fig:figMulticomAppli2}: The fundamental model parameters are summarized in Table~\ref{tab.tabvd}, while derived parameters are  in Table~\ref{tab.tabvd2}. Some of the entries in Table~\ref{tab.tabvd2} are time-dependent, and are used to simplify the notation in our model and the derived system of differential equations.

\begin{table}[htbp]

\captionsetup{skip=0pt}\caption{Fundamental model parameters}\label{tab.tabvd}

\begin{tabular}{p{1.2cm}p{15.25cm}}
\hline
Param. &  Description\hfill\,  \\
\hline
&\hfill Rate parameters that characterize the mosquito population\hfill\,  \\
\hline
$a$ & Rate of bite attempts  for questing vectors  
\\$\delta$ & Rate at which resting vectors move to the questing state
\\$\Gamma$ & Recruitment rate of vectors 
\\$\mu$ & Natural death rate of vectors
\\\hline
&\hfill Parameters  that characterize the mosquito population's interaction with the $i^{\textrm{th}}$ host group\hfill\,  \\
\hline
$c_i$ & Probability that a vector that  bites an infected host of the $i^{th}$ host group and survives is infected
\\$f_i$ & Probability that a vector attempting to bite the $i^{th}$ host group survives and obtains a blood meal
\\$k_i$ & Probability that a vector attempting to bite $i^{th}$ host group is killed
\\\hline
&\hfill Parameters  that characterize the $i^{th}$ host group\hfill\,  \\
\hline
$m_i$ & Probability that a host in group $i$ is infected due to a bite attempt of an infectious vector
\\$\Lambda_i$ & Migration  rate (hosts $/$ time)  
\\$\gamma_i$ & Transition rate from Infectious to Susceptible  state
\\$\nu_i$ & outgoing rate in $i^{th}$ host group
\\$\tilde\nu_i$ & incoming rate rate in $i^{th}$ host group 
\\
\hline
\end{tabular}
\end{table}

\begin{table}[htbp]
\caption{Derived model parameters and time-dependent functions}\label{tab.tabvd2}

\begin{tabular}{p{1.2cm}p{1.7cm}p{13.5cm}}
\hline
Param. &  Formula &  Description  \\
\hline
\\$b_i$ & $\frac{H_i}{H}$ &Proportion of hosts in group $i$ at a given time
\\$d$ & $\dsum_{i=0}^nab_ik_i$ &Death rate of vectors due to questing activity
\\$\bar{c}_i$ & $1 - c_i$ &Probability that a  vector which successfully bites infectious host   of the $i^{th}$  group fails to get infected
\\$f_q$ & $\dfrac{\varpi}{\hat\mu+\varpi}$ & Questing frequency of mosquitoes (i.e. questing mosquito survival proportion) 
\\$f_r$ & $\dfrac{\delta}{\mu+\delta}$ & Resting frequency of mosquitoes  (i.e. resting mosquito survival proportion)
\\$r_i$ & $1 - f_i$ &Repelling effectiveness of measures used for $i^{th}$ host group
\\$\hat{\mu}$ & $\mu + d$ & Death rate of questing vectors
\\ $\varphi$ & ${\dsum_{i=0}^n} ac_if_i\frac{I_i}{H}$ &  Incidence rate of infection for questing susceptible vectors
\\ $\bar{\varphi}$ & ${\dsum_{i=0}^n} ab_ic_if_i$ &  Maximum incidence rate of infection for questing susceptible vectors
\\ $\varpi$ & $\dsum_{i=0}^nab_if_i$ & Incidence rate of successful blood meal for questing vectors
\\
\hline
\end{tabular}
\end{table}

\subsection{Model equations}\label{subsec.Assumption3}
\noindent  The  system of ordinary differential equations that characterize the model are given as follows:
\begin{equation}
\left\{
\begin{aligned}
\dot S_i &~~=~~\Lambda_i+\tilde{\nu_i}H_i - \left(\nu_i+a\,m_i \frac{I_q}{H}\right)S_i +\gamma_i I_i & i = 0,\;1,\;\cdots,\; n \\
\dot S_q &~~=~~\Gamma- (\hat\mu+\varpi) S_q + \delta S_r & \,\\
\dot S_r &~~=~~ (\varpi-\varphi) S_q-(\mu+\delta)S_r\\
\dot E^{(1)}_r &~~=~~\varphi  S_q-(\mu+\delta) E^{(1)}_r & \,\\
\dot E^{(j)}_q &~~=~~\delta E^{(j)}_r - (\hat\mu+\varpi)E^{(j)}_q & j = 1,\;2,\;\cdots,\; \ell \\
\dot  E^{(j+1)}_r &~~=~~\varpi  E^{(j)}_q-(\mu+\delta) E^{(j+1)}_r &j = 1,\;2,\;\cdots,\; \ell \\
\dot I_i &~~=~~a\,m_i \frac{I_q}{H}S_i -\left(\gamma_i+\nu_i\right) I_i & i = 0,\;1,\;\cdots,\; n \\
\dot I_q &~~=~~\delta E^{(\ell+ 1)}_r - (\hat\mu+\varpi) I_q +\delta I_r& \, \\
\dot I_r &~~=~~\varpi I_q - (\mu+\delta)I_r & \,
    \end{aligned}
\right.
     \label{eq:eqbednet_}
\end{equation} 

\noindent The system \eqref{eq:eqbednet_} together with initial conditions completely specifies the evolution of the multicompartment system shown in Figures~\ref{fig:figMulticomAppli1} and~\ref{fig:figMulticomAppli2}. 

\section{Well-posedness, dissipativity and equilibria of the system}\label{dfestabana}

\noindent 
In this section we demonstrate well-posedness of the model by demonstrating invariance of the set of nonnegative states, as well as boundedness properties of the solution. We also calculate the equilibria of the system, whose stability properties will be examined in the following section.

\subsection{Positive invariance of the nonnegative cone in state space}\label{subsec.pinn}
 
The
system~\eqref{eq:eqbednet_} can be rewritten  in matrix form as 
\begin{equation} \label{eq:eqmodel}\dot{\mathbf x}=\mathbf A(\mathbf x)\mathbf x + \mathbf  b
 \Leftrightarrow \left\{\begin{array}{ccc}\dot{\mathbf x}_S & = & \mathbf A_S(\mathbf x)\mathbf x_S  + \mathbf A_{S,\,I}(\mathbf x)\mathbf x_I  +   \mathbf b_S\\
   \dot{\mathbf x}_I & = & \;\;\;\;\;\;\mathbf A_I(\mathbf x)\mathbf x_I
\end{array}\right. 
\Leftrightarrow \left\{\begin{array}{ccr}\dot{\mathbf x}_S & = & \mathbf A_S(\mathbf x)\, \left(\mathbf x_S - \mathbf x^*_S\right)  +  \hat{\mathbf A}_{S,\,I}(\mathbf x)\,\mathbf x_I \\
   \dot{\mathbf x}_I & = & \;\;\;\;\;\;\mathbf A_I(\mathbf x)\,\mathbf x_I
\end{array}\right. ,
\end{equation}
where 
\begin{equation}\label{eq:eqmodel1}
\mathbf A(\mathbf x)=\begin{pmatrix}
\mathbf A_S(\mathbf x) &\mathbf A_{S,\,I}(\mathbf x)\cr \mathbf 0&\mathbf A_I(\mathbf x)
\end{pmatrix};~~~~ \mathbf b = \left(\mathbf b_S;~\mathbf 0\right)~\text{where}~\mathbf b_{S}=\left(\Lambda_0;~\cdots;~\Lambda_n;~\Gamma;~0\right);~~~~
\mathbf x^*_S \equiv  A_S(\mathbf x^*)^{-1}  \mathbf b_S.
\end{equation}
$\mathbf x^*_S$ is a vector whose components are components of vector $\mathbf x_S$ in eq.~\eqref{eq:naive_and_non} at the disease free equilibrium; its computation is carried out in Proposition~\ref{prop:stdst0}.

 Equation~\eqref{eq:eqmodel} is defined for values of the state variable  $\mathbf x=(\mathbf x_S;\;\mathbf x_I)$ lying in the nonnegative cone of $\R^u$ ($
u=2n+ 2\ell +7$), which we denote as $\R^u_+$.  Here $\mathbf x_S$  and  $\mathbf x_I$ represent respectively the naive and non-naive components of the system state: explicitly, 
\begin{equation}\label{eq:naive_and_non}
\mathbf x_S\equiv \left((S_i)_{0\leq i\leq
n};~~S_q;\;S_r\right) ; \qquad
\mathbf x_I\equiv \left(\;(E_r^{(j)};~~E_q^{(j)})_{1\leq j\leq \ell};~~E_r^{(\ell+ 1)};~~(I_i)_{0\leq i\leq
n};~~I_q;~~I_r\right).
\end{equation}
This notation is consistent with \cite{KamSal07}, and some results from this previous reference  are used in our analysis. 
	
\noindent The matrix $\mathbf A_S(\mathbf x) =\diag\left(\mathbf A_{S_h}(\mathbf x),\;\mathbf A_{S_v}(\mathbf x)\right)$ with 
\begin{equation}\label{eq:A_S_v}
\mathbf A_{S_h}(\mathbf x) = -\diag\left(\nu_i-\tilde\nu_i+a\,m_i \frac{I_q}{H}\right)_{0\leq i\leq
n}\hbox{ and }\mathbf A_{S_v}(\mathbf x) =  \left(\begin{array}{cc} -(\hat\mu+\varpi) &   \delta \\ \varpi-\varphi &  -(\mu+\delta) \end{array}\right),
\end{equation}
$\mathbf A_{S,\,I}(\mathbf x)$ is the $(n+3)\times (n+2(\ell+ 2))$ matrix with components equal to zero except for the first $n+1$ components of the $(2\ell +2)^{th}$ column, which are given by $\nu_i+\gamma_i$, $0\leq i\leq n$. 

\noindent The    matrix $\mathbf A_I(\mathbf x)$ may be written in block form  as 
\begin{equation}
\mathbf A_I(\mathbf x) =    \left(\begin{array}{cc} \mathbf A_{I_E}(\mathbf x)
&   \mathbf A_{I_{I,\,E}}(\mathbf x) \\ \mathbf A_{I_{E,\,I}}(\mathbf x) &  \mathbf
A_{I_I}(\mathbf x) \end{array}\right),
\label{eq:mati}
\end{equation}
where the four matrix blocks may be described as follows: 
\medskip

First, the  $(2\ell +1)\times(2\ell +1)$ matrix $\mb A_{I_E}(\mathbf x)$  expresses the interaction between exposed components of the system. It is a 2-banded matrix whose diagonal and subdiagonal elements are given by  the vectors $\mathbf d_0$ and $\mathbf d_{-1}$ respectively, defined by
\begin{equation}\label{eq.eqdiagelts}\mathbf d_0 = {\left(\right.}\underbrace{-(\mu+\delta),\; -(\hat\mu+\varpi),\;\cdots,-(\mu+\delta),\;
-(\hat\mu+\varpi)}_{2\ell \;\; components},\;-(\mu+\delta) {\left.\right)};\qquad \mathbf d_{-1} = {\left(\right.} \underbrace{ \delta, \;\varpi, \;\cdots,
\;\delta,\;\varpi}_{2\ell \;\;components}{\left.\right.).}\end{equation} 

Next, the  $(2\ell +1)\times (n+3)$ matrix  
\[
 \mathbf A_{I_{I,\,E}}(\mathbf x)=a\frac{S_q}{H}\left(\begin{array}{ccccc}\,c_0f_0 &\cdots & c_nf_n&0&0\\0 & \cdots&0&0&0\\\vdots&\cdots&\vdots&\vdots&\vdots\\0 & \cdots&0&0&0\end{array}\right)
\]
  gives the dependence of the exposed components 
$E_r^{(j)}\; (j=1,\cdots,\; \ell+ 1)$, $E_q^{(j)},\;\; (j=1,\cdots,\; \ell)$ on the infectious components $I_i (i=0,\;\cdots,\; n)$, $I_r$ and $I_q$. 

Next, the $(n+3)\times (2\ell +1)$ matrix  $ \mathbf A_{I_{E,\,I}}(\mathbf x)$  gives the dependence of infectious components on exposed components. All entries are zero except the $(n+2,\; 2\ell +1)$ entry, which is equal to $\delta$ reflecting the transition rate of vectors from state $E^{(\ell+ 1)}_r$ to state 
$I_q$.
 
Lastly, the $(n+3)\times (n+3)$  matrix
$ \mathbf A_{I_I}(\mathbf x)$ may be written in block  form as  
$\mathbf A_{I_I}(\mathbf x)=\left(\begin{array}{cc}
\mathbf A_{I_{I_h}} &   \mathbf A_{I_{I_{v,\,h}}} \\     \mathbf 0 &  \mathbf A_{I_{I_v}}
\end{array}\right)$,  with 
\begin{align*}
\mathbf A_{I_{I_h}} = -\mathrm{diag} \left(
 \nu_i+\gamma_i\right)_{0\leq i\leq n};~~~~
\mathbf A_{I_{I_v}} = \left(\begin{array}{cc}-\left(\hat\mu+\varpi\right) &
\delta\\\varpi & - \left( \mu+\delta\right)\end{array}\right);~~~~
\mathbf A_{I_{I_v,\,h}}&=\frac{a}{H}\left(\begin{array}{cc}
 S_0\,m_0 & 0 \\
\vdots & \vdots \\
 S_{n}\,m_{n} & 0 
\end{array}
\right).
\end{align*}
For a given $\mathbf
x\in\mathbb
R^u_+$, the matrices  $\mathbf A_S(\mathbf x)$, $\mathbf A_I(\mathbf x)$ and $\mathbf A(\mathbf x)$ are Metzler matrices (see Appendix~\ref{appx:defs}), and the vector  $\mathbf b  \in \mathbb R_+^u$.

\noindent The following proposition establishes that  system~\eqref{eq:eqmodel} is epidemiologically well-posed. 
\begin{prop}\label{prop:invrnce} The nonnegative cone $\mathbb R^u_+$  is positively invariant for
the system~\eqref{eq:eqmodel}.\end{prop}

\noindent \emph{Proof:}~~The proof is similar to the standard proof that systems determined by  Metzler matrices preserve invariance of the nonnegative cone. It can be shown directly that if $\mathbf x$ is on the boundary of $\mathbb R_+^u$, then $\dot{\mathbf x}$ is in $\mathbb R_+^u$, hence the trajectories never leave $\mathbb R_+^u$.
\edem

\subsection{Disease-free equilibrium (DFE) of the system} 
The system~\eqref{eq:eqmodel} admits two steady states. The trivial steady state that is the DFE is established in  Proposition~\ref{prop:stdst0} below, while the nontrivial steady state will be established in Proposition~\ref{prop:stdst1} after some necessary preliminaries. 

Before characterizing the DFE, we first introduce some useful notation. The {\em questing frequency} $f_q$ and {\em resting frequency} $f_r$ are defined respectively as:
\[
f_q\equiv \frac{\varpi}{\hat{\mu}+\varpi}; \qquad f_r\equiv \frac{\delta}{\mu+\delta}.
\]
$f_q$ may be interpreted as the proportion of questing mosquitoes that pass on to the resting state; while $f_r$ is conversely the proportion of resting mosquitoes that pass on to the questing state. In \cite{jckam201411} these parameters are constants of the model,  but in the current model $f_q$ depends on the system state. The value of $f_q$ at the DFE is denoted by $f^*_q$.   In the following we shall frequently make use of the following replacements:
\begin{equation}\label{eq:frfq}
\hat\mu+\varpi=\frac{\varpi}{f_q}; \qquad \mu+\delta=
\frac{\delta}{f_r}.
\end{equation}
This new notation enables us to give a simple expression for the DFE and to shorten expressions in many other computations throughout this paper.
\begin{prop}\label{prop:stdst0}
The system~\eqref{eq:eqmodel} admits a trivial equilibrium (the disease-free equilibrium (DFE))  given by $\mathbf x^*=\left(\mathbf x_S^*;\;\mathbf x_I^*\right)\in\mathbb R^u_+$,  
 with $\mathbf x_I^*=\mathbf 0\in\mathbb R^{u-n-3}$;~ $\mathbf x_S^* =\left(\mathbf x_{S_h}^*;\;\mathbf x_{S_v}^*\right)$, where 
\begin{equation}\label{eq:eqexpdfe}\mathbf x_{S_h}^* = \left(\frac{\Lambda_0}{\tilde \nu_0-\nu_0};\,\frac{\Lambda_1}{\tilde \nu_1-\nu_1};\,\cdots;\,\frac{\Lambda_n}{\tilde \nu_n-\nu_n}\right)\;\;\hbox{ and }\;\; \mathbf x_{S_v}^* = (S_q^*,S_r^*) =\left( \frac{f^*_q\Gamma}{\varpi^*(1-f^*_qf_r)};\;\frac{f_rf^*_q\Gamma}{\delta(1-f^*_qf_r)}\right).\end{equation} 
\end{prop}

\noindent{\em Proof }:~~
The DFE corresponds to  a state $\mathbf x^*$ in which all components representing non-naive classes are equal to zero: that is, $\mathbf x^* = \left(\mathbf x^*_S;\;\mathbf x^*_I\right)$ with $\mathbf x^*_I\equiv 0$. The steady-state equation for the system~\eqref{eq:eqmodel} with the constraint $\mathbf x_I\equiv0$ is  
\begin{equation}\label{eq:Asys}
\mathbf
A_S(\mathbf x_S\,;\mathbf 0)\,.\,\mathbf x_S  +  \mathbf b_S=\mathbf 0\Leftrightarrow \left\{\begin{array}{l}
\mathbf A_{S_h}(\mathbf x_S\,;\mathbf 0)\,.\,\mathbf x_{S_h}  +  \mathbf b_{S_h}=\mathbf 0\cr \mathbf A_{S_v}(\mathbf x_S\,;\mathbf 0)\,.\,\mathbf x_{S_v}  +  \mathbf b_{S_v}=\mathbf 0
\end{array}\right. .
\end{equation} 
This system may be solved in two stages, since the subsystem $\mathbf A_{S_h}(\mathbf x_S\,;\mathbf 0)\,.\,\mathbf x_{S_h}  +  \mathbf b_{S_h}=\mathbf 0$ is uncoupled. The solution of this subsystem is $\mathbf x^*_{S_h} = \left(\frac{\Lambda_0}{\tilde \nu_0-\nu_0};\,\frac{\Lambda_1}{\tilde \nu_1-\nu_1};\,\cdots;\,\frac{\Lambda_n}{\tilde \nu_n-\nu_n}\right)$. Using this solution in system \eqref{eq:Asys}, we obtain the equation 
$$\mathbf A_{S_v}(\mathbf x_{S_h}\,;\mathbf x_{S_v}\,;\mathbf 0)\,.\,\mathbf x_{S_v}  +  \mathbf b_{S_v}=\mathbf 0
\implies
\mathbf x^*_{S_v}=-{\mathbf A_{S_v}(\mathbf x^*_{S_h}\,;\mathbf x_{S_v}\,;\mathbf 0)}^{-1}\,.\, \mathbf b_{S_v}.$$
Using expression~\eqref{eq:A_S_v} for $\mathbf A_{S_v}$ (with $\varphi=0$ at DFE), and recalling that $\mathbf b_{S_v} = (\Gamma ; 0)$, we find the solution
\begin{equation}\label{eq:xsv}
\mathbf x^*_{S_v} = (S_q^*,S_r^*)= \left(\dfrac{\mu + \delta}{(\hat\mu+\varpi)(\mu + \delta) - \varpi \delta};\;\dfrac{\varpi}{(\hat\mu+\varpi)(\mu + \delta) - \varpi \delta}\right) = \left(\dfrac{f^*_q\Gamma}{\varpi^*(1-f^*_qf_r)};\;\dfrac{f_rf^*_q\Gamma}{\delta(1-f^*_qf_r)}\right),
\end{equation}
where we have used the replacements \eqref{eq:frfq} to obtain the final equality in \eqref{eq:xsv}.

\noindent As a corollary we have

\begin{prop}\label{prop:stbsysred} The system \begin{equation}\label{eq.sysred}
\dot{\mathbf x} = \mathbf A_S(\mathbf x^*)\,.\,\left(\mathbf x-\mathbf x^*_S\right)
\end{equation}
 is GAS at $\mathbf x^*_S$ on $\mathbb R_+^{n+3}$.\end{prop}
The proof is straightforward, based on Proposition~\ref{prop:blockdecomposition}.

\subsection{Boundedness and dissipativity of the trajectories}
We have the following proposition.
\begin{prop}
 \label{prop:dissip}  The simplex
\begin{equation}\Omega =\left\{\mathbf x\in\mathbb R^u_+\;\left|\;\left(S_q\leq S_q^* \right)\;\wedge \;\left(S_r\leq S_r^*\right)\;\wedge\;\left(H_i= H^*_i,\;\; 1\leq i\leq
n \right) \wedge\; \left( M_I\leq\frac{\bar \varphi}{\mu}S_q^* \right) \; \right.
\right\}\label{eq.simplex}\end{equation} 
is a compact forward-invariant and absorbing set for the
system~\eqref{eq:eqbednet_},  where
\[
 M_I\equiv {\dsum_{j=1}^\ell} \left(E_q^{(j)} + E_r^{(j)}\right) + E_r^{(\ell+ 1)} + I_q + I_r~;~~~~\bar \varphi\equiv a{\dsum_{i=0}^n}b^*_ic_if_i,
\]
and where $(S_q^*, S_r^*)$ are the DFE components for naive questing and resting mosquitoes respectively (given in \eqref{eq:xsv}).
\end{prop}

Note that  $M_I$ is the overall population of non-naive mosquitoes;  $\bar\varphi$  is the maximum incidence rate of infection for questing susceptible mosquitoes; and $b_i^* = H_i^*/H^*$.

The proof is given in Appendix~\ref{ssec.supdissip}.  As a result of Proposition~\ref{prop:dissip}, we may limit our study to the simplex specified in eq.~\eqref{eq.simplex}.

\subsection{Computation of the threshold condition}\label{sec.algo}
The following propostion gives a formula for the basic reproduction number $\mathcal R_0$, and  shows that the condition $\mathcal R_0<1$ is a necessary and sufficient condition  for local stability of the DFE.



\begin{prop}\label{prop:basicrepn}
The basic reproduction number $R_0$ of the system \eqref{eq:eqmodel} is 
\begin{equation}
\label{eq:R0} \mathcal
R_0 \equiv \frac{(f^*_qf_r)^{\ell+ 1}}{(1-f^*_qf_r)^2}\dfrac{f^*_q}{{\varpi^*}^2}\dfrac{\Gamma}{H^*}a^2\sum_{i=0}^n\frac{b^*_ic_if_im_i}{\gamma_i+\nu_i},
\end{equation}  
where $f^*_q$ and $f_r$ are the questing and resting frequencies respectively of mosquitoes at the DFE, $H^*\equiv {\dsum_{i=0}^n}\Lambda_i/({\nu_i-\tilde \nu_i})\equiv {\dsum_{i=0}^n}H^*_i$ is the total host population at the DFE, and $b^*_i\equiv {H^*_i}/{H^*}$ is the proportion of hosts in group $i$ at the DFE. Then $\mathcal R_0<1$ is a necessary and sufficient condition  for local stability of the DFE. 

\end{prop}

\noindent The proof of Proposition~\ref{prop:basicrepn} is given in Appendix~\ref{ssec.supprrprepn}.

\subsection{ Endemic equilibrium (EE) of the System} 
For our system, there is a unique endemic equilibrium that is  specified by the following proposition.
\begin{prop}\label{prop:stdst1}
System~\eqref{eq:eqmodel} admits a unique endemic equilibrium (EE) $\mathbf x^\star\in{\mathbb R_{>0}}^+$ with components given by

\begin{equation}\label{eq:eqexpMinfctstateb}
\begin{array}{l}
S_q^\star=\frac{f^*_q}{\varpi^*(1-f^*_qf_r)}\Gamma - \frac{I_q^\star}{(f^*_qf_r)^l};\qquad S_r^\star=\frac{\delta f^*_q}{f_r(1-f^*_qf_r)}\Gamma - \frac{\varpi^*}{\delta}\frac{(1-(f^*_qf_r)^{\ell+ 1}+f^*_qf_r)I_q^\star}{ f^*_q}; \\ 
E_q^{(j)\star}=\dfrac{1-f^*_qf_r }{(f^*_qf_r)^{\ell+ 1-j}}I_q^\star; 
\qquad E_r^{(j)\star}=\frac{\varpi^* }{\delta }\dfrac{1-f^*_qf_r}{ f^*_q(f^*_qf_r)^{\ell+ 1-j}} I_q^\star\qquad(1\leq j\leq \ell);
\\
I_r^\star=\frac{\varpi^* }{\delta }f_rI_q^\star; \qquad E_r^{(\ell+ 1)\star}=\frac{\varpi^* }{\delta }\frac{1-f^*_qf_r}{f^*_q} I_q^\star;\\
S_i^\star=\dfrac{(\nu_i+\gamma_i) H^* }{ (\nu_i + \gamma_i) H^* +a\,m_i I_q^\star} H^*_i; \qquad I_i^\star = \dfrac{a\,m_iI_q^\star }{(\nu_i+\gamma_i) H^* +a\,m_i I_q^\star} H^*_i ~~ (0\leq i\leq n),
\end{array}
\end{equation}

where $I_q^\star\in~\big]0,\;\bar{I}_q^\star\big[$ is the unique finite root of the equation \begin{equation}\label{eq:eqeqIQ}\dsum_{i=0}^n \frac{a^2\,b^*_im_ic_if_i}{(\nu_i+\gamma_i)H^*+am_ix}
=\dfrac{{\varpi^*}^2(1-f^*_qf_r)^2 }{f^*_q(f^*_qf_r)^{\ell+ 1} \Gamma-\varpi^* f^*_qf_r (1-f^*_qf_r)x}.
\end{equation}
and
\begin{equation}\label{eq:Iq-bar}
\bar{I}_q^\star \equiv \frac{\Gamma}{\varpi^*}\frac{f^*_q(f^*_qf_r)^{\ell+ 1}}{1-f^*_qf_r}
\end{equation}
\end{prop}

\noindent The proof of Proposition~\ref{prop:stdst1} is given in Appendix~\ref{subsec.sptopropee}.

\begin{rmq}\em \label{rmk.rplctm} According to \eqref{eq:eqeqIQ},  the dynamics of the mosquito population (expressed in the parameters $f_r, f_q$, and $\ell$) as well as the protection means used by the population (expressed in the parameter $\varpi$) 
strongly influence the location of the EE. This justifies our initial supposition that mosquito dynamics and host protection means are important practical factors in determining the prevalence of infection.
\end{rmq}
%

\section{Stability of system equilibria}\label{sec.analysis}
In this section we analyze the stability of the system equilibria  given in Propositions~\ref{prop:stdst0} and~~\ref{prop:stdst1}.

\subsection{Stability analysis of the disease free equilibrium (DFE)}\label{subsec.dfestabanan}
\noindent We have the following result about the global stability of the disease free equilibrium:
\begin{thm} \label{thm.defstab} When $\mathcal R_0 \leq 1$, then the DFE is GAS  in $\mathbb R_+^u$.
\end{thm}

\noindent{\em Proof }:~~
Our proof relies on Theorem 4.3 of \cite{KamSal07}, which establishes global asymptotic stability (GAS) for epidemiological systems that can be expressed in the matrix form~\eqref{eq:eqmodel}. This theorem is restated as Theorem~\ref{thm:kamsal} in the Appendix: for the proof, the reader may consult \cite{KamSal07}. To complete the proof, we need only to establish for the system~\eqref{eq:eqmodel} that the five conditions ({\bf h}1--{\bf h}5) required in Theorem~\ref{thm:kamsal} are satisfied when $\mathcal R_0\leq 1$:

\begin{enumerate}[({\bf h}1)]
\item 
This condition  is satisfied for the system~\eqref{eq:eqmodel} as a result of Proposition~\ref{prop:dissip}. 
\item
We note first that $n_S = n+3$, and the canonical projection of $\Omega$ on $\mathbb R_+^{n+3}$ is $\mathbb I = \{\mathbf x_{S_h}\}\times\left [0,\;\;S^*_q\right]\times \left [0,\;\;S^*_r\right]$. The system~\eqref{eq:eqmodel} reduced to this subvariety is given in \eqref{eq.sysred}, and this system is GAS at its equilibrium ($\mathbf x^*_S$) as a result of  Proposition~\ref{prop:stbsysred}. 

\item
 We consider first the case $\ell=1$ and $n=1$. In this case, the matrix  $\mathbf A_I(\mathbf x)$ in the system \eqref{eq:eqmodel} is $$\mathbf A_I(\mathbf x) = \left(\begin{array}{ccccccc}-(\mu+\delta)&0&0&ac_0f_0\frac{S_q}{H} &ac_1f_1\frac{S_q}{H}&0&0\\\delta&-(\hat\mu+\varpi)&0&0&0&0&0 \\0&\varpi&-(\mu+\delta)&0&0&0&0\\ 0&0&0&-(\nu_0+\gamma_0)&0&am_0\frac{S_0}{H}&0 \\ 0&0&0&0&-(\nu_1+\gamma_1)&am_1\frac{S_1}{H}&0 \\ 0&0&\delta&0&0&-(\hat\mu+\varpi) &\delta \\ 0&0&0&0&0&\varpi&-(\mu+\delta)\end{array}\right).$$ 
In this case, the two properties required for condition ($\mathbf h3$) follow immediately: the off-diagonal terms of the matrix   $\mathbf A_I(\mathbf x)$ are nonnegative, and  the matrix is irreducible as can be seen from the associated directed graph $G(\mathbf A_I(\mathbf x))$ in Figure~\ref{graph}. 
For general $l$ and $n$, the proof of ({\bf h}3) is similar.

\begin{figure}[H]
\centering
\includegraphics[scale=0.50]{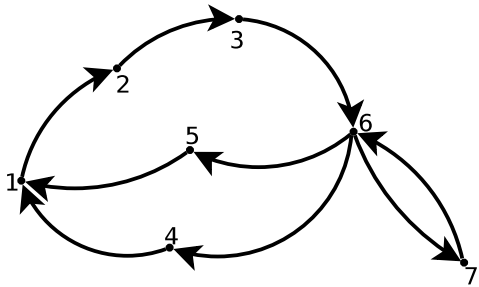}
\caption{Graph associated with the matrix $\mathbf A_I(\mathbf x)$}\label{graph}
\end{figure}

\item
Defining $\overline{\mathbf A}_I \equiv \mathbf A_I(\mathbf x^*)$, we have $\mathbf A_I(\mathbf x)\leq\overline{\mathbf A}_I$  $\forall \,\mathbf x\in\Omega$,   and $\mathbf x^*\in\left(\mathbb R_+^{n+3}\times\{\mathbf 0\}\right)\cap\Omega$. Thus the upper bound of $\mathfrak M$ is attained at the DFE which is a point on the boundary of $\Omega$, and condition ($\mathbf h 4$) is satisfied.

\item
We first observe that $\overline{\mathbf A}_I$ is the block matrix of the Jacobian matrix of the system~\eqref{eq:eqbednet_} corresponding to the Infected submanifold, taken at the DFE. As  noted in~\cite{KamSal07}, the condition that all eigenvalues of $\overline{\mathbf A}_I$ have negative real parts, which is equivalent to  the condition that $\overline{\mathbf A}_I$ is a stable Metzler matrix, is also equivalent to the condition $\mathcal R_0\leq1$. This fact is developed in the proof of Proposition~\ref{prop:basicrepn} (see Appendix) where we compute the value of $\mathcal R_0$ by establishing necessary and sufficient conditions for the stability of the Metzler matrix $\overline{\mathbf A}_I$.
\end{enumerate}
Since the five conditions for Theorem 4.3 of \cite{KamSal07} are satisfied, the theorem follows. \edem

\subsection{Stability analysis of the endemic equilibrium (EE)} \label{sec:eeqstana}
\noindent In this section we analyze the behavior of the system under the condition $\mathcal R_0>1$. From Proposition~\ref{prop:basicrepn} it follows  that the DFE is not stable in this case. As stated in Proposition~\ref{prop:stdst1}, the system~\eqref{eq:eqbednet_} also admits  a unique nontrivial biologically feasible  equilibrium  (or endemic equilibrium (EE)). It remains to address the stability of the EE, which determines the behavior of the system when the disease persists. Our main result in this regard is the following theorem.

\begin{thm}\label{thm:thmstabee}
When $\mathcal R_0>1$,  the EE $\mathbf x^\star$ of the system~\eqref{eq:eqbednet_} defined in \eqref{eq:eqexpMinfctstateb} is GAS on  $\left(\mathbb R_{>0}\right)^u$.
\end{thm}

\begin{rmq}\label{rmq:rmqglstb}\em The above theorem implies that the GAS of the EE is in the nonnegative  cone  $\mathbb R^u_+$, since  the positive cone $\left(\mathbb R_{>0}\right)^u$ is absorbing for the system~\eqref{eq:eqbednet_}.\end{rmq}

\noindent {\em Proof }:~~Considering the system~\eqref{eq:eqbednet_} when $\mathcal R_0 > 1$, there is a unique endemic equilibrium $\mathbf x^\star$ with respective components given as in eq.~\eqref{eq:eqexpMinfctstateb}.
Let  the function $V_{ee}$ be defined on $\left(\mathbb R_{>0}\right)^u$ as follows:

\begin{equation}\label{eq:eqliapvee}
\begin{array}{rcl}V_{ee}(\mathbf x) & = & (S_q-S^\star_q\ln S_q)+(S_r-S^\star_r\ln S_r)+\, \sigma_r^{(1)}(E^{(1)}_r-E^{(1)\star}_r\ln E^{(1)}_r)\\&& +\,\dsum_{j=1}^\ell\left(\sigma_q^{(j)}(E^{(j)}_q-E^{(j)\star}_q\ln E^{(j)}_q)+\sigma_r^{(j+1)}(E^{(i+1)}_r-E^{(j+1)\star}_r\ln E^{(j+1)}_r)\right)+\,\tau_q(I_q-I_q^{\star}\ln \,I_q)\\&& +\tau_r(I_r-I_r^{\star}\ln \,I_r)+\,\dsum_{i=0}^{n}\upsilon_i\left((S_i-S_i^{\star}\ln \,S_i)+(I_i-I_i^{\star}\ln \,I_i)\right),
\end{array}
\end{equation}
where  $\sigma^{(j)}_r = (f^*_qf_r)^{1-j}$, for $j=1,\;2,\;\cdots,\;\ell+ 1$, $\sigma^{(j)}_q = (f^*_qf_r)^{1-j}/f_r$, for $j=1,\;2,\;\cdots,\;\ell$, $\tau_q = (f^*_qf_r)^{-\ell} /f_r$,   $\tau_r =(f^*_qf_r)^{-\ell} $,  
$\upsilon_i=a\frac{S_q^\star}{H^*}\frac{f_i}{\nu_i - \tilde \nu_i}$ for $i=0,\;1,\;\cdots,\;n$
(the motivation for these coefficients, and the derivation of expression~\eqref{eq:eqliapeeder2} below for the derivative, are both provided in Appendix~\ref{sec.supplyap}). $V_{ee}(\mathbf x)$ is a  $\mathcal C^{\infty}$ positive definite function defined on $\left(\mathbb R_{>0}\right)^u$, whose derivative along the trajectories of the system~\eqref{eq:eqbednet_} is given by:

\begin{equation}\label{eq:eqliapeeder2}
\begin{array}{rcl}
{\frac{dV_{ee}}{d\,t}}(\mathbf x(t)) & = & \hat\mu S^\star_q\left(2-\frac{S_q}{S^\star_q}-\frac{S^\star_q}{S_q}\right) + \delta S^\star_r  \left( \frac{S^\star_q}{S_q}+\frac{S_r}{S_r^\star}-\frac{S^\star_q}{S_q}\frac{S_r}{S_r^\star}-1\right)  +\, I_q^\star\frac{\varpi \sigma_r^{(1)}}{(f_qf_r)^{l}} \left(2 - \frac{I_r}{I_r^\star}\frac{I_q^\star}{I_q}- \frac{I_q}{I_q^\star}\frac{I^\star_r}{I_r}\right)\\
&&+\,{\dsum_{i=0}^n}\upsilon_i\Bigg[ \hat\nu_i\Big[S_i^\star \Big(4-\frac{S^\star_i}{S_i}-\frac{S^\star_q}{S_q}-\frac{S_i}{S^\star_i}\frac{S_q}{S^\star_q}\frac{S^\star_r}{S_r}-\frac{S_r}{S^\star_r}\Big)\\&& +\,c_i I^\star_i\Big(2\ell +5 -\frac{S^\star_i}{S_i}-\frac{S_i}{S^\star_i}\frac{I^\star_i}{I_i}\frac{I_q}{I^\star_q} -\frac{S^\star_q}{S_q} - \frac{I_i}{I^\star_i}\frac{S_q}{S^\star_q}\frac{E^{(1)\star}_r}{E^{(1)}_r}- {\dsum_{j=1}^{\ell}}\frac{E^{(j)}_q}{E^{(j)\star}_q}\frac{E^{(j+1)\star}_r}{E^{(j+1)}_r}-\, {\dsum_{j=1}^{\ell}}\frac{ E^{(j)}_r}{ E^{(j)\star}_r}\frac{E^{(j)\star}_q}{E^{(j)}_q} -\frac{E^{(\ell+ 1)}_r}{E^{(\ell+ 1)\star}_r}\frac{I^\star_q}{I_q}\Big) 
\\&&+\,\bar c_iI^\star_i\Big(4+\frac{I_q}{I^\star_q}-\frac{S^\star_i}{S_i}-\frac{S_i}{S^\star_i}\frac{I^\star_i}{I_i}\frac{I_q}{I^\star_q}-\frac{S^\star_q}{S_q}-\frac{I_i}{I^\star_i}\frac{S_q}{S^\star_q}\frac{S^\star_r}{S_r}-\,\frac{S_r}{S^\star_r}\Big)\Big] + \hat\gamma_iI^\star_i\left(1 + \frac{I_q}{I^\star_q} - \frac{I_i}{I^\star_i} \frac{S^\star_i}{S_i} - \frac{I_q}{I^\star_q} \frac{S_i}{S^\star_i}\frac{I^\star_i}{I_i}\right)  \Bigg]\\
&=&   f_1(\mathbf x)+{\dsum_{i=0}^n}\upsilon_i G_{1_i}(\mathbf x),
\end{array}
\end{equation}
where 
for each $i$, $\bar c_i$ is the complementary probability of $c_i$ (i.e. $\bar c_i+c_i=1$), $ \hat \nu_i = \nu_i - \tilde \nu_i$, $\hat{\gamma_i}\equiv\tilde\nu_i+\gamma_i$;
\begin{equation*}
f_1(\mathbf x) = \hat\mu S^\star_q\left(2-\frac{S_q}{S^\star_q}-\frac{S^\star_q}{S_q}\right) + \delta S^\star_r  \left( \frac{S^\star_q}{S_q}+\frac{S_r}{S_r^\star}-\frac{S^\star_q}{S_q}\frac{S_r}{S_r^\star}-1\right)  +\, I_q^\star\frac{\varpi \sigma_r^{(1)}}{(f_qf_r)^{l}} \left(2 - \frac{I_r}{I_r^\star}\frac{I_q^\star}{I_q}- \frac{I_q}{I_q^\star}\frac{I^\star_r}{I_r}\right);
\end{equation*}
and
\begin{equation}\label{eq:G1exp1}
G_{1_i}(\mathbf x) = g_{1_i}(\mathbf x)+h_{1_i}(\mathbf x)+p_{1_i}(\mathbf x),
\end{equation}
where
\begin{align*}
{g_1}_i(\mathbf x)= &c_i\hat\nu_i I^\star_i\left(2\ell +5 -\frac{S^\star_i}{S_i}-\frac{S_i}{S^\star_i}\frac{I^\star_i}{I_i}\frac{I_q}{I^\star_q} -\frac{S^\star_q}{S_q} -\; \frac{I_i}{I^\star_i}\frac{S_q}{S^\star_q}\frac{E^{(1)\star}_r}{E^{(1)}_r}- {\dsum_{j=1}^{\ell}} \frac{E^{(j)}_q}{E^{(j)\star}_q} \frac{E^{(j+1)\star}_r}{E^{(j+1)}_r}-\, {\dsum_{j=1}^{\ell}}\frac{ E^{(j)}_r}{ E^{(j)\star}_r}\frac{E^{(j)\star}_q}{E^{(j)}_q} -\frac{E^{(\ell+ 1)}_r}{E^{(\ell+ 1)\star}_r}\frac{I^\star_q}{I_q}\right)\\
&+\hat\nu_i\,S_i^\star \left(4-\frac{S^\star_i}{S_i}-\frac{S^\star_q}{S_q}-\frac{S_i}{S^\star_i}\frac{S_q}{S^\star_q}\frac{S^\star_r}{S_r}-\frac{S_r}{S^\star_r}\right),\\
{h_1}_i(\mathbf x)=& \bar c_i\hat\nu_iI^\star_i \left(4+\frac{I_q}{I^\star_q} -\frac{S^\star_i}{S_i} -\frac{S_i}{S^\star_i} \frac{I^\star_i}{I_i}\frac{I_q}{I^\star_q}-\frac{S^\star_q}{S_q}-\frac{I_i}{I^\star_i}\frac{S_q}{S^\star_q}\frac{S^\star_r}{S_r}-\,\frac{S_r}{S^\star_r}\right),\\
{p_1}_i(\mathbf x)=&\hat\gamma_iI^\star_i\left(1 + \frac{I_q}{I^\star_q} - \frac{I_i}{I^\star_i} \frac{S^\star_i}{S_i} - \frac{I_q}{I^\star_q} \frac{S_i}{S^\star_i}\frac{I^\star_i}{I_i}\right).
\end{align*}

\noindent For each $i$,  by adding and subtracting $\nu_iI^\star_i \left( 1-\frac{I^\star_i}{I_i} \frac{S_i}{S^\star_i}\right)$
the function ${G_1}_i(\mathbf x)$ may be rewritten as 
\begin{equation}\label{eq:eqliapeeder20}
\begin{array}{rcl}
{G_1}_i(\mathbf x) & = &  \hat\nu_i\Big[S_i^\star \Big(4-\frac{S^\star_i}{S_i}-\frac{S^\star_q}{S_q}-\frac{S_i}{S^\star_i}\frac{S_q}{S^\star_q}\frac{S^\star_r}{S_r}-\frac{S_r}{S^\star_r}\Big)\\&& +\,c_i I^\star_i\Big(2\ell +6 -\frac{S^\star_i}{S_i} -\frac{S_i}{S^\star_i} \frac{I^\star_i}{I_i} - \frac{I_q}{I^\star_q} -\frac{S^\star_q}{S_q} - \frac{I_i}{I^\star_i}\frac{S_q}{S^\star_q}\frac{E^{(1)\star}_r}{E^{(1)}_r}- {\dsum_{j=1}^{\ell}}\frac{E^{(j)}_q}{E^{(j)\star}_q}\frac{E^{(j+1)\star}_r}{E^{(j+1)}_r}-\, {\dsum_{j=1}^{\ell}}\frac{ E^{(j)}_r}{ E^{(j)\star}_r}\frac{E^{(j)\star}_q}{E^{(j)}_q} -\frac{E^{(\ell+ 1)}_r}{E^{(\ell+ 1)\star}_r}\frac{I^\star_q}{I_q}\Big) 
\\&&+\,\bar c_iI^\star_i\Big(5-\frac{S^\star_i}{S_i}-\frac{S_i}{S^\star_i}\frac{I^\star_i}{I_i}-\frac{S^\star_q}{S_q}-\frac{I_i}{I^\star_i}\frac{S_q}{S^\star_q}\frac{S^\star_r}{S_r}-\,\frac{S_r}{S^\star_r}\Big)\Big]  +\, \bar\gamma_iI^\star_i\Big(\frac{I_q}{I^\star_q} + \frac{S_i}{S^\star_i}\frac{I^\star_i}{I_i} - 1 -\frac{S_i}{S^\star_i}\frac{I^\star_i}{I_i}\frac{I_q}{I^\star_q}\Big) \\&& + \hat\gamma_iI^\star_i\left(2  - \frac{I_i}{I^\star_i} \frac{S^\star_i}{S_i} -  \frac{S_i}{S^\star_i}\frac{I^\star_i}{I_i}\right)  \\
&=&   \tilde{{g_1}}_i(\mathbf x)+\tilde{{p_1}}_i(\mathbf x),
\end{array}
\end{equation}
with $\bar{\gamma_i}\equiv\nu_i+\gamma_i$ ,
\begin{align*}
\tilde{{g_1}}_i(\mathbf x)= &I^\star_i\Bigg[\hat\nu_i \Bigg[c_i\left(2\ell +6 -\frac{S^\star_i}{S_i} - \frac{I_q}{I^\star_q} -\frac{S_i}{S^\star_i}\frac{I^\star_i}{I_i} -\frac{S^\star_q}{S_q} -\; \frac{I_i}{I^\star_i}\frac{S_q}{S^\star_q}\frac{E^{(1)\star}_r}{E^{(1)}_r}- {\dsum_{j=1}^{\ell}} \frac{E^{(j)}_q}{E^{(j)\star}_q} \frac{E^{(j+1)\star}_r}{E^{(j+1)}_r}-\, {\dsum_{j=1}^{\ell}}\frac{ E^{(j)}_r}{ E^{(j)\star}_r}\frac{E^{(j)\star}_q}{E^{(j)}_q} -\frac{E^{(\ell+ 1)}_r}{E^{(\ell+ 1)\star}_r}\frac{I^\star_q}{I_q}\right)\\
& + \,\bar c_i\left(5+\frac{I_q}{I^\star_q}-\frac{S^\star_i}{S_i}-\frac{S_i}{S^\star_i}\frac{I^\star_i}{I_i}-\frac{S^\star_q}{S_q}-\frac{I_i}{I^\star_i}\frac{S_q}{S^\star_q}\frac{S^\star_r}{S_r}-\,\frac{S_r}{S^\star_r}\right)\Bigg] + \hat\gamma_i\left(2  - \frac{I_i}{I^\star_i} \frac{S^\star_i}{S_i} -  \frac{S_i}{S^\star_i} \frac{I^\star_i}{I_i}\right) \Bigg] \\& +\,\hat\nu_i\,S_i^\star \left( 4-\frac{S^\star_i}{S_i} - \frac{S^\star_q}{S_q}-\frac{S_i}{S^\star_i}\frac{S_q}{S^\star_q}\frac{S^\star_r}{S_r}-\frac{S_r}{S^\star_r}\right),\\
\tilde{{p_1}}_i(\mathbf x)=& \bar\gamma_iI^\star_i\left(\frac{I_q}{I^\star_q} + \frac{I_i}{I^\star_i} \frac{S^\star_i}{S_i} -1 - \frac{I_q}{I^\star_q} \frac{S_i}{S^\star_i}\frac{I^\star_i}{I_i}\right).
\end{align*}


Alternatively, using the identity $\delta S_r^\star = (\varpi^\star-\varphi^\star)S_q^\star-\mu S_r^\star$ we may write
\begin{equation}
\begin{array}{rcl}
{\dfrac{dV_{ee}}{d\,t}}(\mathbf
x(t)) & = & \hat\mu S^\star_q\Big(2-\frac{S_q}{S^\star_q}-\frac{S^\star_q}{S_q}\Big) + \mu S^\star_r  \Big(1 + \frac{S^\star_q}{S_q}\frac{S_r}{S_r^\star} - \frac{S^\star_q}{S_q}-\frac{S_r}{S_r^\star}\Big)  +\, I_q^\star\frac{\varpi \sigma_r^{(1)}}{(f_qf_r)^{l}} \Big(2 - \frac{I_r}{I_r^\star}\frac{I_q^\star}{I_q}- \frac{I_q}{I_q^\star}\frac{I^\star_r}{I_r}\Big)\\
&&+\,{\dsum_{i=0}^n}\upsilon_i\Bigg[\hat\nu_i\Big[S_i^\star \Big(3-\frac{S^\star_i}{S_i}-\frac{S_i}{S^\star_i}\frac{S_q}{S^\star_q}\frac{S^\star_r}{S_r}-\frac{S^\star_q}{S_q}\frac{S_r}{S^\star_r}\Big)\\
&& +\,c_i I^\star_i\Big(2\ell +5 -\frac{S^\star_i}{S_i}-\frac{S_i}{S^\star_i}\frac{I^\star_i}{I_i}\frac{I_q}{I^\star_q} -\frac{S^\star_q}{S_q} -\; \frac{I_i}{I^\star_i}\frac{S_q}{S^\star_q}\frac{E^{(1)\star}_r}{E^{(1)}_r}- {\dsum_{j=1}^{l}}\frac{E^{(j)}_q}{E^{(j)\star}_q}\frac{E^{(j+1)\star}_r}{E^{(j+1)}_r}-\, {\dsum_{j=1}^{\ell}}\frac{ E^{(j)}_r}{ E^{(j)\star}_r}\frac{E^{(j)\star}_q}{E^{(j)}_q} -\frac{E^{(\ell+ 1)}_r}{E^{(\ell+ 1)\star}_r}\frac{I^\star_q}{I_q}\Big) 
\\&&+\,\bar c_iI^\star_i\Big(3+\frac{I_q}{I^\star_q}-\frac{S^\star_i}{S_i}-\frac{S_i}{S^\star_i}\frac{I^\star_i}{I_i}\frac{I_q}{I^\star_q}-\frac{I_i}{I^\star_i}\frac{S_q}{S^\star_q}\frac{S^\star_r}{S_r}-\,\frac{S^\star_q}{S_q}\frac{S_r}{S^\star_r}\Big)\Big] + \hat\gamma_iI^\star_i\left(1 + \frac{I_q}{I^\star_q} - \frac{I_i}{I^\star_i} \frac{S^\star_i}{S_i} - \frac{I_q}{I^\star_q} \frac{S_i}{S^\star_i}\frac{I^\star_i}{I_i}\right)\Bigg]\\
&=&   f_2(\mathbf x)+{\dsum_{i=0}^n}\upsilon_i G_{2_i},
\end{array}
\label{eq:eqliapeeder21}
\end{equation}
with
\begin{align*}
f_2(\mathbf x) = &\hat\mu S^\star_q\left(2-\frac{S_q}{S^\star_q}-\frac{S^\star_q}{S_q}\right) + \mu S^\star_r  \left(1 +\frac{S^\star_q}{S_q}\frac{S_r}{S_r^\star}- \frac{S^\star_q}{S_q}-\frac{S_r}{S_r^\star}\right)  +\, I_q^\star\frac{\varpi \sigma_r^{(1)}}{(f_qf_r)^{l}} \left(2 - \frac{I_r}{I_r^\star}\frac{I_q^\star}{I_q}- \frac{I_q}{I_q^\star}\frac{I^\star_r}{I_r}\right); 
\end{align*}
and
\begin{equation}\label{eq:G2exp2}
{G_2}_i = {g_2}_i(\mathbf x)+{h_2}_i(\mathbf x)+{p_1}_i(\mathbf x),
\end{equation}
where
\begin{align*}
{g_2}_i(\mathbf x) =&c_i\hat\nu_i I^\star_i\Big(2\ell +5 -\frac{S^\star_i}{S_i}-\frac{S_i}{S^\star_i}\frac{I^\star_i}{I_i}\frac{I_q}{I^\star_q} -\frac{S^\star_q}{S_q} -\; \frac{I_i}{I^\star_i}\frac{S_q}{S^\star_q}\frac{E^{(1)\star}_r}{E^{(1)}_r}- {\dsum_{j=1}^{\ell}}\frac{E^{(j)}_q}{E^{(j)\star}_q}\frac{E^{(j+1)\star}_r}{E^{(j+1)}_r}-\, {\dsum_{j=1}^{\ell}}\frac{ E^{(j)}_r}{ E^{(j)\star}_r}\frac{E^{(j)\star}_q}{E^{(j)}_q} -\frac{E^{(\ell+ 1)}_r}{E^{(\ell+ 1)\star}_r}\frac{I^\star_q}{I_q}\Big)\\
&+\hat\nu_i\,S_i^\star \left(3-\frac{S^\star_i}{S_i}-\frac{S_i}{S^\star_i}\frac{S_q}{S^\star_q}\frac{S^\star_r}{S_r}-\frac{S_r}{S^\star_r}\frac{S^\star_q}{S_q}\right); \\
{h_2}_i(\mathbf x)=&\bar c_i\hat\nu_iI^\star_i\left(3+\frac{I_q}{I^\star_q}-\frac{S^\star_i}{S_i}-\frac{S_i}{S^\star_i}\frac{I^\star_i}{I_i}\frac{I_q}{I^\star_q}-\frac{I_i}{I^\star_i}\frac{S_q}{S^\star_q}\frac{S^\star_r}{S_r}-\,\frac{S_r}{S^\star_r}\frac{S^\star_q}{S_q}\right).
\end{align*}

\noindent For each $i$, by adding and subtracting $\nu_iI^\star_i \left( 1-\frac{I^\star_i}{I_i} \frac{S_i}{S^\star_i}\right)$ the function ${G_2}_i$  may be rewritten as
\begin{equation}\label{eq:eqliapeeder23}
\begin{array}{rcl}
{G_2}_i(\mathbf x) & = &  \hat\nu_i\Big[S_i^\star \Big(3-\frac{S^\star_i}{S_i}-\frac{S_i}{S^\star_i}\frac{S_q}{S^\star_q}\frac{S^\star_r}{S_r} - \frac{S^\star_q}{S_q} \frac{S_r}{S^\star_r}\Big)\\&& +\,c_i I^\star_i\Big(2\ell +6 -\frac{S^\star_i}{S_i} -\frac{S_i}{S^\star_i} \frac{I^\star_i}{I_i} - \frac{I_q}{I^\star_q} -\frac{S^\star_q}{S_q} - \frac{I_i}{I^\star_i}\frac{S_q}{S^\star_q}\frac{E^{(1)\star}_r}{E^{(1)}_r}- {\dsum_{j=1}^{\ell}}\frac{E^{(j)}_q}{E^{(j)\star}_q}\frac{E^{(j+1)\star}_r}{E^{(j+1)}_r}-\, {\dsum_{j=1}^{\ell}}\frac{ E^{(j)}_r}{ E^{(j)\star}_r}\frac{E^{(j)\star}_q}{E^{(j)}_q} -\frac{E^{(\ell+ 1)}_r}{E^{(\ell+ 1)\star}_r}\frac{I^\star_q}{I_q}\Big) 
\\&&+\,\bar c_iI^\star_i\Big(4-\frac{S^\star_i}{S_i}-\frac{S_i}{S^\star_i}\frac{I^\star_i}{I_i}-\frac{I_i}{I^\star_i}\frac{S_q}{S^\star_q}\frac{S^\star_r}{S_r}-\,\frac{S^\star_q}{S_q}\frac{S_r}{S^\star_r}\Big)\Big]  +\, \bar\gamma_iI^\star_i\Big(\frac{I_q}{I^\star_q} + \frac{S_i}{S^\star_i}\frac{I^\star_i}{I_i} - 1 -\frac{S_i}{S^\star_i}\frac{I^\star_i}{I_i}\frac{I_q}{I^\star_q}\Big) \\&& + \hat\gamma_iI^\star_i\left(2  - \frac{I_i}{I^\star_i} \frac{S^\star_i}{S_i} -  \frac{S_i}{S^\star_i}\frac{I^\star_i}{I_i}\right)  \\
&=&   \tilde{{g_2}}_i(\mathbf x)+\tilde{{p_1}}_i(\mathbf x),
\end{array}
\end{equation}
with $\bar{\gamma_i}\equiv\nu_i+\gamma_i$ , and
\begin{align*}
\tilde{{g_2}}_i(\mathbf x)= &I^\star_i\Bigg[\hat\nu_i \Bigg[c_i\left(2\ell +6 -\frac{S^\star_i}{S_i} - \frac{I_q}{I^\star_q} -\frac{S_i}{S^\star_i}\frac{I^\star_i}{I_i} -\frac{S^\star_q}{S_q} -\; \frac{I_i}{I^\star_i}\frac{S_q}{S^\star_q}\frac{E^{(1)\star}_r}{E^{(1)}_r}- {\dsum_{j=1}^{\ell}} \frac{E^{(j)}_q}{E^{(j)\star}_q} \frac{E^{(j+1)\star}_r}{E^{(j+1)}_r}-\, {\dsum_{j=1}^{\ell}}\frac{ E^{(j)}_r}{ E^{(j)\star}_r}\frac{E^{(j)\star}_q}{E^{(j)}_q} -\frac{E^{(\ell+ 1)}_r}{E^{(\ell+ 1)\star}_r}\frac{I^\star_q}{I_q}\right)\\
& + \,\bar c_i\left(4+\frac{I_q}{I^\star_q}-\frac{S^\star_i}{S_i}-\frac{S_i}{S^\star_i}\frac{I^\star_i}{I_i}-\frac{I_i}{I^\star_i}\frac{S_q}{S^\star_q}\frac{S^\star_r}{S_r}-\,\frac{S^\star_q}{S_q}\frac{S_r}{S^\star_r}\right)\Bigg] + \hat\gamma_i\left(2  - \frac{I_i}{I^\star_i} \frac{S^\star_i}{S_i} -  \frac{S_i}{S^\star_i} \frac{I^\star_i}{I_i}\right) \Bigg] \\& +\,\hat\nu_i\,S_i^\star \left( 3-\frac{S^\star_i}{S_i} -\frac{S_i}{S^\star_i} \frac{S_q}{S^\star_q} \frac{S^\star_r}{S_r}- \frac{S^\star_q}{S_q} \frac{S_r}{S^\star_r} \right).
\end{align*}
We split $\left(\mathbb R_{>0}\right)^u$ into two overlapping subsets, 
$$\Omega_1=\left\{\mathbf x\in\left(\mathbb R_{>0}\right)^u\;|\;\frac{S^\star_q}{S_q}+\frac{S_r}{S_r^\star}-\frac{S^\star_q}{S_q}\frac{S_r}{S_r^\star}-1\leq 0\right\}\;\hbox{ and }\;\Omega_2 = \left\{\mathbf x\in\left(\mathbb R_{>0}\right)^u\;|\;\frac{S^\star_q}{S_q}+\frac{S_r}{S_r^\star}-\frac{S^\star_q}{S_q}\frac{S_r}{S_r^\star}-1\geq 0\right\},$$
and we shall  consider ${\frac{dV_{ee}}{d\,t}}(\mathbf x(t))$ on $\Omega_1$ and  $\Omega_2$ as given in ~\eqref{eq:eqliapeeder2} and ~\eqref{eq:eqliapeeder21}, respectively. 

On the entire set $\left(\mathbb R_{>0}\right)^u$ we have for each $i$, ${g_k}_i(\mathbf x)\leq0$ and ${\tilde {g_k}}_i(\mathbf x)\leq0$  by Corollary~\ref{cor.agmi} of the arithmetic--geometric means inequality (Lemma~\ref{lem.agmi}) for $k\in\{1,~2\}$. 
The same corollary implies $f_k(\mathbf x)\leq0$  on $\Omega_k$  for $k\in\{1,~2\}$. 
On the entire set $\left(\mathbb R_{>0}\right)^u$ and for each $i$, given that $(I_i-I^\star_i)(I_q-I^\star_q)\geq0$ we may show that $\tilde{{p_1}}_i(\mathbf x)\leq0$ by  applying Corollary~\ref{cor.agmi0} to the ratios $\frac{I^\star_q}{I_q}$ and $\frac{I_q}{I^\star_q}$ with respective associated weights $\frac{I_q}{I^\star_q}$ and $\frac{S_i}{S^\star_i}\frac{I^\star_i}{I_i}$. 
On the other hand, when  $(I_i-I^\star_i)(I_q-I^\star_q)\leq0$ we may show $h_1(\mathbf x)\leq0$, $h_2(\mathbf x)\leq0$ and ${p_1}_i(\mathbf x)\leq 0$ by applying the same corollary to the following data (given in pairs as (number, weight))    $\left(\frac{S_i^\star}{S_i},~1\right)$, $\left(\frac{I_i^\star}{I_i}\frac{S_i}{S^\star_i},~ \frac{I_q}{I^\star_q}\right)$, $\left(\frac{S^\star_q}{S_q},~ 1\right)$, $\left(\frac{I_i}{I^\star_i}\frac{S_q}{S^\star_q} \frac{S^\star_r}{S_r},~ 1 \right)$, $\left(\frac{S_r}{S^\star_r},~1\right)$ for $h_1$;  $\left(\frac{S_i^\star}{S_i},~1\right)$, $\left(\frac{I_i^\star}{I_i}\frac{S_i}{S^\star_i},~ \frac{I_q}{I^\star_q}\right)$, $\left(\frac{I_i}{I^\star_i}\frac{S_q}{S^\star_q} \frac{S^\star_r}{S_r},~1\right)$, $\left(\frac{S^\star_q}{S_q}\frac{S_r}{S^\star_r},~ 1\right)$ for $h_2$; and $\left(\frac{S_i}{S^\star_i} \frac{I^\star_i}{I_i},~\frac{I_q}{I^\star_q}\right)$, $\left(\frac{S^\star_i}{S_i}\frac{I_i}{I^\star_i},~1\right)$  for $p_1$. 

In view of expressions \eqref{eq:G1exp1}, \eqref{eq:eqliapeeder20} for $G_{1_i}$ and expressions \eqref{eq:G2exp2}, \eqref{eq:eqliapeeder23} for $G_{2_i}$, the results of the previous paragraph show that  $G_{k_i} \le 0$ on the entire set $(\mathbb R_{>0})^u$ for $k \in \{1,2\}$. Since we have also shown that 
$f_k(\mathbf x)\leq0$  on $\Omega_k$  for $k\in\{1,~2\}$, in view of expressions \eqref{eq:eqliapeeder2} and \eqref{eq:eqliapeeder21} for ${\frac{dV_{ee}}{d\,t}}(\mathbf x(t))$ we may conclude that ${\frac{dV_{ee}}{d\,t}}(\mathbf x(t)) \le 0$ for all $\mathbf x(t) \in  (\mathbb R_{>0})^u$. 
 
%
%

\noindent In order to determine the subset of $(\mathbb R_{>0})^u$ where $\frac{dV_{ee}}{d\,t}(\mathbf x(t))= 0$, we make use of ~\eqref{eq:eqliapeeder2} to conclude that
$$\frac{dV_{ee}}{d\,t}(\mathbf x(t))= 0\;\Leftrightarrow(f_1(\mathbf x)=0)\wedge ({g_1}_i(\mathbf x),\;i=0,\;\cdots,\;n)\wedge ({h_1}_i(\mathbf x),\;i=0,\;\cdots,\;n)\wedge ({p_1}_i(\mathbf x),\;i=0,\;\cdots,\;n).$$ 
By Lemma~\ref{lem.agmi},  $f_1(\mathbf x)=0$ if and only if $S_q=S^\star_q\wedge  I_qI^\star_r=I^\star_qI_r$. 
Given that $f_1(\mathbf x)=0$, Lemma~\ref{lem.agmi} also implies that ${h_1}_i(\mathbf x)=0$ if and only if $(S_r=S^\star_r)\wedge  (S_i=S^\star_i) \wedge (I_i=I^\star_i) \wedge  (I_q=I^\star_q)$, and thus $I_r=I^\star_r$.
Finally, assuming $f_1(\mathbf x)=0$ and ${h_1}_i(\mathbf x)=0$, Lemma~\ref{lem.agmi} also gives  
$${g_1}_i(\mathbf x)=0\;\;\hbox{ if and only if }\;\;1=\frac{E^{(1)\star}_r}{E^{(1)}_r} = \frac{ E^{(1)}_r}{ E^{(1)\star}_r}\frac{E^{(1)\star}_q}{E^{(1)}_q}= \frac{E^{(1)}_q}{E^{(1)\star}_q}\frac{E^{(2)\star}_r}{E^{(2)}_r}=\cdots =  \frac{ E^{(l)}_r}{ E^{(l)\star}_r}\frac{E^{(l)\star}_q}{E^{(l)}_q}=  \frac{ E^{(\ell+ 1)\star}_r}{ E^{(\ell+ 1)}_r}\frac{E^{(l)}_q}{E^{(l)\star}_q} =\frac{E^{(\ell+ 1)}_r}{E^{(\ell+ 1)\star}_r},$$ 
which implies $E^{(j)}_r=E^{(j)\star}_r, \;j=1,\;\cdots,\;\ell+ 1$ and $E^{(j)}_q=E^{(j)\star}_q, \;j=1,\;\cdots,\;\ell$. Thus, $\frac{dV_{ee}}{d\,t}(\mathbf x(t))= 0$ if and only if $\mathbf x = \mathbf x^\star$. 

From the above discussion we may conclude that $V_{ee}$ is a strict Lyapunov function for the system~\eqref{eq:eqbednet_} on $(\mathbb R_{>0})^u$. The LaSalle invariance principle implies the global asymptotic stability of the EE $\mathbf x^\star$ of the system~\eqref{eq:eqbednet_} on the set $(\mathbb R_{>0})^u$~\cite{Bhatia70, Las68, MR0481301, MR0594977}.\edem

\section{Sensitivity analysis of $\mathcal R_0$ and endemic equilibrium infection levels}\label{sec:secsam}
The parameters of greatest interest in evaluating strategies for malaria control are the reproduction number  $\mathcal R_0$ and the EE host infection levels $I_i^\star$. Ideally, one would hope to achieve a reproduction number $\mathcal R_0 < 1$, which would imply the eventual elimination of malaria. In the short range, a more immediate objective should be to reduce infection levels as much as possible. 

Various practical control strategies will impact different model parameters. Control strategies are listed in Table~\ref{tab.tabvd3}, along with affected model parameters. 

\begin{table}[htbp]
\caption{Derived model parameters and time-dependent functions}\label{tab.tabvd3}

\begin{tabular}{ | l | c | c | c | c | c | c | }

\cline{2-7}
\multicolumn{1}{c|}{} & \multicolumn{6}{c|}{\emph{Affected model parameters}}\\
\hline
\emph{Control method} & $\Gamma$ & $\mu$ & $f_i$ & $k_i$ & $m_i$ & $\gamma_i$  \\
\hline
Outdoor spraying with larvicides & \checkmark & & & & &  \\
Breeding habitat reduction (e.g. draining standing water) & \checkmark & & & & & \\
Outdoor vector control & & \checkmark & & & &\\
Indoor residual spraying & & \checkmark & & & & \\
Insecticide-treated bed nets (ITN), Long-lasting insecticidal nets (LLIN) & & & \checkmark & \checkmark & \checkmark & \\
Untreated bed nets & & & \checkmark & & \checkmark & \\
Repellents (topical repellents, mosquito coils) & & & \checkmark & & \checkmark & \\
Preventative drugs & & & & & \checkmark & \\
Rapid diagnosis and treatment (RDT) & & & & & & \checkmark \\

\hline
\end{tabular}
\end{table}
 In order  to quantify the effects of the parameters in Table~\ref{tab.tabvd3} on the two critical performance measures $\mathcal R_0$ and $I_i^\star$ we will make use of sensitivity indices, which measure the percentage change in a dependent variable in response to an incremental percentage change in a system parameter.  When the variable is a differentiable function of the parameter, the sensitivity index may be formally defined as follows:
\begin{dfn} The normalized forward sensitivity index of a variable $u$ that depends differentiably on a parameter $p$ is defined as:
	\begin{equation}
	\label{eq:seind}
	\varUpsilon_p^u := \frac{\partial u}{\partial p}\times\frac{p}{u}.\end{equation}
\end{dfn}

Our analysis is facilitated by the following observations, which are easily proved using basic calculus:
\begin{itemize}
\item
Sensitivity of additive terms: $\varUpsilon_p^{\sum_i u_i} =\dfrac{\sum_i \varUpsilon_p^{u_i} u_i}{\sum_i u_i}$; 
\item
Additive sensitivity of multiplicative terms:  $\varUpsilon_p^{uv} = \varUpsilon_p^{u} + \varUpsilon_p^{v}$;
\item
Negative sensitivity of reciprocal terms:  $\varUpsilon_p^{u^{-1}} = -\varUpsilon_p^{u}$;
\item
Multiplicative sensitivity of compositions: i.e. if $u = u(x)$ and $x = x(p)$, then $\varUpsilon_p^{u} = \varUpsilon_x^{u}\varUpsilon_p^{x}$.
\end{itemize}

\subsection{Sensitivities of $\mathcal R_0$ with respect to controllable parameters}\label{subsec:secsamnpkc}
To facilitate the calculation of sensitivities, we introduce some intermediate parameters:
\begin{align}
\rho^\star &\equiv \dfrac{\hat{\mu}^\star}{\varpi^\star};\\
 \widehat{\mathcal R}_0 &\equiv  \frac{(f^*_qf_r)^{\ell+ 1}}{(1-f^*_qf_r)^2}\dfrac{f^*_q}{{\varpi^*}^2} = f_r^{\ell+ 1}(1 + \rho^\star)^{-\ell}  \left( 1-f_r + \rho^\star \right)^{-2} (\varpi^\star)^{-2};\\
u_i &\equiv \frac{c_i f_i m_i}{\gamma_i+\nu_i};\\
U &\equiv \sum_{i=1}^n b_i^* u_i.
\end{align}
The parameter $\rho^\star$ is interpretable as the failure/success ratio of questing mosquitos: that is, the fraction of mosquitos at each questing stage that fail to feed and survive divided by the fraction of mosquitos that succeed and pass on to the next resting stage. Note also that $U$ is the weighted average of $u_i$ values, where the weight is the EE population proportion.  
In terms of these parameters, we may rewrite $\mathcal R_0$ from \eqref{eq:R0}  as:
\begin{align}\label{eq:R0_rho}
\mathcal R_0  &= \frac{a^2\Gamma}{H^*} \widehat{\mathcal R}_0  U.
\end{align}
We begin with the sensitivities of $\mathcal R_0$ on $k_i$ and $f_i$. First we calculate the sensitivities with respect to the intermediate parameters $\hat{\mu}^*$ and $\varpi^*$: note that $\hat{\mu}^*$ which depend on  $k_i$ and $f_i$.  Using the additive property of sensitivities mentioned above, it follows directly from \eqref{eq:R0_rho}  that
\begin{equation}
\varUpsilon_{\rho^\star}^{\mathcal R_0} =\varUpsilon_{\rho^\star}^{\widehat{\mathcal R}_0} =  \frac{ -\ell  \rho^\star}{1+\rho^\star} - 2\frac{ \rho^\star}{1 - f_r+\rho^\star},
\end{equation}   
which may be rewritten as
\begin{equation}
\varUpsilon_{\rho^\star}^{\mathcal R_0}=\varUpsilon_{\rho^\star}^{\widehat{\mathcal R}_0} = \frac{-\ell  \rho^\star}{1+\rho^\star} - 2 + \frac{2t}{1+\rho^\star},
\end{equation}   
where 
\begin{equation*}
t \equiv \frac{1-f_r + \rho^* - f_r \rho^*}{1-f_r + \rho^*} = \frac{1-f_r}{1-f_q^*f_r}.
\end{equation*}
Note that  $0\le t \le 1$:  $t \approx 1$ when $f_r \approx 0$ (low resting survival rate) or when $f_q^* \approx 1$ (high feeding success); and $t \approx 0$ when $f_r \approx 1$ (high resting survival rate).

Using $\varUpsilon_{\hat{\mu}^\star}^{\rho^\star} = 1$, $\varUpsilon_{\varpi^\star}^{\rho^\star} = -1$, $f_q^\star = (1 + \rho^\star)^{-1}$ and the multiplicative property of sensitivities,  we have
\begin{align*}
\varUpsilon_{\hat{\mu}^\star}^{\mathcal R_0} &= \varUpsilon_{\hat{\mu}^\star}^{\widehat{\mathcal R}_0} = \varUpsilon_{\rho^\star}^{\widehat{\mathcal R}_0}\varUpsilon_{\hat{\mu}^\star}^{\rho^\star} = - \ell (1-f_q^\star) - 2 + 2tf_q^\star = - \ell -2 + f_q^\star(\ell+ 2t),\\
\varUpsilon_{\varpi^\star}^{\mathcal R_0} &=\varUpsilon_{\varpi^\star}^{\widehat{\mathcal R}_0}  = -\varUpsilon_{\hat{\mu}^\star}^{\widehat{\mathcal R}_0} - 2 =  \ell (1-f_q^\star) - 2tf_q^\star.
\end{align*}

The sensitivity $\varUpsilon_{\hat{\mu}^\star}^{\mathcal R_0}$ is negative, as expected: if the kill rate of questing mosquitos is increased, we would expect $\mathcal R_0$ to decrease. Surprisingly, in the case where  $f_q^\star \approx 1$ (high questing success rate) so that $t \approx 1$, then it is possible to achieve a negative sensitivity of $\mathcal R_0$ on $\varpi^*$: in other words, decreasing the success rate of questing mosquitos can actually increase the reproduction number!  However, in the usual case the equation for $\varUpsilon_{\varpi^\star}^{\mathcal R_0}$ indicates a positive value, as expected.

We are now ready to obtain sensitivities based on $k_i$ and $f_i$. According to the definitions in Table \ref{tab.tabvd2}, the parameter $\hat{\mu}^*$ depends only on $k_i$, while the parameter $\varpi^\star$ depends only on $f_i$.  We have
\begin{equation*}
\varUpsilon_{k_i}^{\hat{\mu}^\star} = \frac{a b_i^*k_i}{\hat{\mu}^*}  ~~ \text{and}~~ \varUpsilon_{f_i}^{\varpi^\star} = \frac{a b_i^*f_i}{\varpi^*} = \frac{ b_i^*f_i}{\sum_i b_i^*f_i}  
\end{equation*}
This leads finally to the following expressions for $\varUpsilon_{k_i}^{\mathcal R_0}$ and for $\varUpsilon_{f_i}^{\mathcal R_0}$:

\begin{align}
\varUpsilon_{k_i}^{\mathcal R_0} = \varUpsilon_{\hat{\mu}^\star}^{\mathcal R_0}\varUpsilon_{k_i}^{\hat{\mu}^\star} &=
b_i^*\left(\frac{ a k_i }{\hat{\mu}^*}\right)\left( - (\ell+ 2)(1-f_q^\star)- 2(1-t)f_q^\star\right),\label{eq:k_iSens}\\
\varUpsilon_{f_i}^{\mathcal R_0} = \varUpsilon_{\varpi^\star}^{\mathcal R_0}\varUpsilon_{f_i}^{\varpi^\star}
+ \frac{ b_i^\star u_i}{U}  &=  b_i^* \left(\left(\frac{ f_i}{\sum_i b_i^* f_i}\right)\left(\ell (1-f_q^\star) - 2tf_q^\star\right) + \frac{  u_i}{U} \right) \quad( 0 \le t \le 1)\label{eq:f_iSens}
\end{align}
The sensitivites of $\mathcal R_0$ on $k_i$ and $ f_i$  given in \eqref{eq:k_iSens}-\eqref{eq:f_iSens} are proportional to $b_i^*$.  This reflects the fact that the impacts of these parameters are directly proportional to the size of the host groups that they impact. But although sensitivities are larger for larger host groups, presumably the effort required to change $k_i$ and $f_i$ will also be larger since a larger population is involved.

The presence of the multiplicative factor $\ell$ (number of vector stages)  in the sensitivites $\varUpsilon_{k_i}^{\mathcal R_0}$ and $\varUpsilon_{f_i}^{\mathcal R_0}$ is noteworthy.  The parameters $k_i$ and  $f_i$ affect the vector population during each  stage, which leads to an $\ell$-fold reinforcement of the parameters' impact on $\mathcal R_0$.  Since $0 \le t \le 1$, terms in \eqref{eq:k_iSens}-\eqref{eq:f_iSens} that involve $t$ or $1-t$ will have a mitigated effect. The factor $1 - f_q^*$, which appears in both sensitivities, indicates that  $k_i$ and $f_i$'s effects on $\mathcal R_0$ are suppressed if a large proportion of questing mosquitos survive to the next resting stage.    

The sensitivities $\varUpsilon_{k_i}^{\mathcal R_0}$ and $\varUpsilon_{f_i}^{\mathcal R_0}$ are proportional to $k_i$ and $f_i$ respectively (note that $u_i$ is proportional to $f_i$).  In the case of $\varUpsilon_{f_i}^{\mathcal R_0}$, this means that we should expect diminishing returns from a control strategy that targets $f_i$. On the other hand, $\varUpsilon_{k_i}^{\mathcal R_0}$ increases as $k_i$ increases towards 1  (although in practice further increases in kill rate are likely to become more difficult to achieve as the kill rate approaches 1, due to diminishing returns).   

Next we compute the sensitivity of $\mathcal R_0$ with respect to $\mu$. Both $f_r$ and $\hat{\mu}^*$ depend on $\mu$, so we will need sensitivities with respect to $f_r$. We obtain
\begin{align}
\varUpsilon_{\mu}^{\mathcal R_0} &= -\left(\ell+1 +\frac{2f_r}{1-f_r +\rho^\star}\right) \frac{\mu}{\mu+\delta}
- (\ell +2 - f_q^\star(\ell+ 2t))\frac{\mu}{\mu+d} \nonumber\\
& = -(1 - f_r)\left(\ell+1 +\frac{2f_q^*f_r}{1-f_q^*f_r}\right) 
- \frac{\mu}{\hat{\mu}^\star} \left((\ell + 2)(1 - f_q^\star) + 2f_q^\star(1-t)\right), \label{eq:muSens}
\end{align}
which  is always negative, as expected. Once again we see a multiplicative factor $\ell$, which reflects the influence of $\mu$ throughout all questing stages. The terms that are proportional to $1-f_r$ and $1-f_q^*$ are reduced when resting (resp. questing) mosquito survival rates are high. Note the factor $\mu / \hat{\mu}^*$ is the ratio of resting/questing death rates, and is always less than 1.

The remaining controllable parameters $\Gamma,m_i, \gamma_i$ are much easier to deal with, because only a single term in the sum in \eqref{eq:R0_rho} depends on each of these parameters. The calculation of the corresponding sensitivity indices is straightforward, using the additive sensitivity rule:  
\begin{equation}\label{eq:otherSens}
\varUpsilon_{\Gamma}^{\mathcal R_0} = 1;\qquad
\varUpsilon_{m_i}^{\mathcal R_0} =\frac{b_i^* u_i}{U} ;\qquad
\varUpsilon_{\gamma_i}^{\mathcal R_0} = -\frac{b_i^* u_i}{U} \left(\frac{\gamma_i}{\gamma_i+\nu_i}\right).
\end{equation}
Eq.  \eqref{eq:otherSens} implies that  $\sum_i \varUpsilon_{m_i}^{\mathcal R_0} = 1$ and $-1 <  \sum_i \varUpsilon_{\gamma_i}^{\mathcal R_0} < 0$. In practical situations, the death rate $\nu_i$ is much less than the cure rate $\gamma_i$, which implies that sensitivities of $\mathcal R_0$ on $m_i$ and $\gamma_i$ are roughly equal but opposite in sign. Both are proportional to  $\frac{b_i^* u_i}{U}$, which indicates that measures that target $m_i$ or $\gamma_i$ applied to groups for which $u_i$ is relatively large (compared to the population average $U$)  will have relatively greater effect on $\mathcal R_0$. Since $u_i$ is proportional to $m_i$ and $(\gamma_i + \nu_i)^{-1}$, it follows that control measures that target $f_i$ and $m_i$ will produce diminishing returns.

\subsection{Sensitivities of $I_q^\star$ and $I_j^\star$ with respect to controllable parameters} \label{subsec:secsamnpkc}

In cases where it is not feasible to reduce  $\mathcal{R}_0$ to a value less than 1, the target of a malaria control strategy should be to reduce $I_j^*$ as much as possible for as many host groups $j$ as possible---particularly those host groups for which infection is more dangerous, like pregnant mothers and infants. According to \eqref{eq:eqexpMinfctstateb}, the infection levels $I_j^*$ may be conveniently expressed in terms of $I_q^\star$. Thus we first calculate the sensitivities of $I_q^*$ with respect to controllable parameters.
 
Making use of \eqref{eq:eqeqIQ} and after some calculations, we find the sensitivity of $I_q^*$ with respect to $\Gamma$:
\begin{equation}
 \varUpsilon_\Gamma^{I_q^\star} = \left( \frac{a I_q^*}{Z} \sum_{i=0}^n \frac{b_i^\star z_i^2}{c_if_i } + 
(1-t) \tilde{t}  \left(1 - \frac{a I_q^*}{Z} \sum_{i=0}^n \frac{b_i^\star z_i^2}{c_if_i } \right) 
\right)^{-1}, 
\end{equation}
where

\begin{equation}
z_i \equiv \frac{m_ic_if_i}{(\nu_i+\gamma_i)H^*+am_iI_q^*}; \qquad  Z \equiv \sum_{i=1}^n b_i^*z_i ; \qquad \tilde{t} \equiv \frac{I_q^\star}{ \Gamma \widehat{\mathcal R}_0 \hat{\mu}^\star}=\frac{I_q^\star}{\bar{I}_q^\star}\frac{f_q^\star(1-f_q^\star f_r)}{1-f_q^\star},
\end{equation}
where $0 \le \tilde{t} \le 2$ (note $1 \le \frac{1-f_q^\star f_r}{1-f_q^\star}\le 2$ since $f_r \ge f_q^*$).
In terms of  $ \varUpsilon_\Gamma^{I_q^\star}$, the sensitivities with respect to $ p_i \in \{f_i,k_i,m_i,\gamma_i\}$ may be found from \eqref{eq:eqeqIQ} (after extended calculations)  as:
\begin{equation}\label{eq:sensIq_wrt_pj}
\varUpsilon_{p_i}^{I_q^\star} =
\varUpsilon_{\Gamma}^{I_q^\star} \left(\varUpsilon_{p_i}^{\widehat{\mathcal R}_0}
+ t\varUpsilon_{p_i}^{t}\tilde{t}  + \varUpsilon_{p_i}^{\hat{\mu}^\star}(1-t)\tilde{t}   
+  \frac{b_i^*z_i}{Z} \varUpsilon_{p_i}^{z_i} 
\left(1 - (1-t)\tilde{t} \right)
\right),
\end{equation}
Equation \eqref{eq:sensIq_wrt_pj} gives specifically for $k_i,f_i,m_i, \gamma_i$: 
\begin{align}
\varUpsilon_{k_i}^{I_q^\star} &=
b_i^*  \varUpsilon_{\Gamma}^{I_q^\star} \left(\frac{a k_i }{\hat{\mu}^*}\right) 
\left( -(\ell+ 2)(1-f_q^\star)- 2(1-t)f_q^\star
- t \tilde{t} \left(\frac{f_rf_q^*(1-f_q^*)}{1-f_rf_q^*}\right)  +  (1-t)\tilde{t}   
\right);\label{eq:sensIqstar_on_ki}\\
\varUpsilon_{f_i}^{I_q^\star} &=
b_i^* \varUpsilon_{\Gamma}^{I_q^\star} \left(\,\, \frac{af_i}{\varpi^*}\left(\ell (1-f_q^\star) - 2tf_q^\star
+ t\left(\frac{f_rf_q^*(1-f_q^*)}{1-f_q^*f_r}\right) \right)\tilde{t}   
+  \frac{z_i}{Z}  
\left(1 - (1-t)\tilde{t} \right)
\right);\label{eq:sensIqstar_on_fi}\\
\varUpsilon_{m_i}^{I_q^*} &= b_i^*\varUpsilon_{\Gamma}^{I_q^*} \left(\frac{z_i}{Z}\right)   \left(  1 + \frac{a f_i I_q^*}{(\nu_i+\gamma_i)H^*+am_iI_q^*} \right) (1 - (1-t)\tilde{t} ); \label{eq:sensIqstar_on_mi}\\
\varUpsilon_{\gamma_i}^{I_q^*} &= -b_i \varUpsilon_{\Gamma}^{I_q^*}\left(\frac{z_i}{Z}\right) \left( \frac{  H^*\gamma_i }{(\nu_i+\gamma_i)H^*+am_iI_q^*}\right) (1 - (1-t)\tilde{t} ).\label{eq:sensIqstar_on_gammai}
\end{align}
Similar calculations give the sensitivity of  $I_q^*$  as a function of $\mu$:
\begin{align}
\varUpsilon_\mu^{I_q^\star} &=   \varUpsilon_\Gamma^{I_q^*} \left( \varUpsilon_\mu^{\mathcal R_0}+(1-t)\tilde{t}\frac{\mu }{\hat{\mu}^\star} \right)
\end{align}

The expression in \eqref{eq:eqexpMinfctstateb} for $I_j^*$ in terms of $I_q^*$  can then be applied to find the sensitivites of $I_j^*$ for host group $j$:
\begin{align}
 \varUpsilon_{k_i}^{I_j^\star}  &= \varUpsilon_{k_i}^{I_q^*}\left(1 - \frac{I_j^*}{H^*}\right) \\
 \varUpsilon_{f_i}^{I_j^\star}  &= \varUpsilon_{f_i}^{I_q^*}\left(1 - \frac{I_j^*}{H^*}\right) \\
 \varUpsilon_{\gamma_i}^{I_j^\star}  &= \varUpsilon_{\gamma_i}^{I_q^*}\left(1 - \frac{I_j^*}{H^*}\right) - \delta_{ij} \frac{\gamma_i I_i^*}{am_i I_q^*} \\
 \varUpsilon_{m_i}^{I_j^\star}  &= (\delta_{ij}+\varUpsilon_{m_i}^{I_q^*})\left(1 - \frac{I_j^*}{H^*}\right) \\
 \varUpsilon_{\mu}^{I_i^\star}  &= \varUpsilon_{\mu}^{I_q^*}\left(1 - \frac{I_j^*}{H^*}\right) 
\end{align}

In general, we would expect that the infected population of any particular group is a relatively small proportion of the total population, so that $\left(1 - \frac{I_j^*}{H^*}\right) \approx 1$. This in turn implies that the sensitivities of $I_j^*$ from variables  $k_i, f_i$ and $\mu$ are almost identical to the corresponding sensitivities of $I_q^*$ on these variables. This reflects the fact that  $k_i, f_i$ and $\mu$ only have an indirect effect on  $I_j^*$ via their influence on the variable $I_q^*$. On the other hand,  treatement rate $\gamma_j$ and infection rate $m_j$ have direct effects on group $j$, as expected. 

From (\eqref{eq:sensIqstar_on_fi})-(\eqref{eq:sensIqstar_on_gammai})  we may see that the variable $z_i$ plays a similar role in the sensitivities of $I_q^*$ and $I_j$ as the variable $u_i$ in the sensitivities of $\mathcal R_0$.  Groups $i$  with larger $z_i$ values will exhibit larger sensitivities with respect to $f_i, m_i,$ and $\gamma_i$.

\section{Discussion}\label{sec.discuss}

We have developed and rigorously analyzed a model of the dynamics of malaria transmission within a system comprising populations of vectors and human hosts, where hosts are subdivided into several groups depending on the way they usually protect themselves against mosquito bites. The model includes essential parameters ($k_i, f_i, m_i, \mu,$ and $\gamma_i$) that are targeted by various realistic control strategies. 
These values determine the model's prediction of the basic reproduction number $\mathcal R_0$, and we have  established that the DFE of the model is GAS providing that $\mathcal R_0\leq 1$. We also have shown theat there is a unique EE when $\mathcal R_0 > 1$, as well as the level of the endemicity when $\mathcal R_0>1$.  The level of the endemicity among human population groups is largely determined by the 
size of the infected questing vector population at the endemic equilibrium, denoted as $I_q^\star$. We do not have an explicit expression for $I_q^\star$, but we have computed an upper bound (see~\eqref{eq:eqexpMinfctstateb}). This upper bound is a decreasing function of the extrinsic incubation period, which is proportional to the number of questing-resting cycles  $\ell$.

The sensitivities derived in previous sections have relatively complicated expressions. However, they reduce considerably in certain limiting cases: for example, if  $1-f_r \ll 1-f_q^*$ then $t \approx 0$  so many terms in the sensitivities disappear. 
In the general case, notwithstanding the complication of the expressions we may still extract some useful information. Most sensitivities are proportional to the controllable parameter they depend on, which implies that diminishing returns will be obtained as the parameter value decreases as control measures are applied. This argues for a comprehensive strategy that targets several parameters, rather than focusing on just one.  Table~\ref{tab.tabvd3} shows that some control measures target multiple parameters: in particular, ITNs influence three different control parameters.  The model suggests that ameliorative  effects from multiple parameters will be multiplied.  This would seem to point to ITNs as the most effective option. However, the model doesn't take into account the fact that a significant fraction of ITN owners do not use their bed nets, for various reasons \cite{atieli2011insecticide,russell2015determinants}. Naturally this compliance issue must be addressed to achieve the full potential benefit of ITNs.  

We have also seen that as far as different groups are concerned, those with large $u_i$ values (where $u_i = \frac{c_i f_i m_i}{\gamma_i + \nu_i}$) should be targeted to produce the greatest impact on $\mathcal R_0$, while those with large $z_i$ values (where $z_i =  \frac{c_i f_i m_i}{(\gamma_i + \nu_i)H^* + am_iI_q^*}$) should be targeted to produce the greatest overall impact on $I_q^*$ (and hence the overall $I_j$ levels for all groups. Finally, we note the multiplier effect of the number of malaria stages on several sensitivies, including vector kill rate ($k_i$), biting success frequency ($f_i$), and overall death rate ($\mu$).  This argues favorably for strategies that target these parameters (such as bed nets (treated or untreated), IRS, and  outdoor vector control) over strategies that reduce infection probability or recovery rate of bitten humans (like IPT and RDT). Although untreated bed nets are not as effective as ITN's (roughly half as effective, since they only affect $f_i$ and $m_i$ but not $k_i$) they still exhibit this multiplier effect---and the fact that they have no associated environmental and health risks may make them an attractive option.

\section{Conclusion and perspective}\label{sec.conclusion}
The model presented in this paper represents an improvement over previous models that do not divide the human population into groups, or take into account multiple vector stages. The model does not take geographical effects into account, and in particular does not consider migration between geographically separated populations which may have different levels of prevalence. The effects of population displacements are very significant in practice, and research is ongoing to take them into account.  

\appendix{
\section{ Useful definitions and results}\label{appx:defs}
In order to make this paper self-contained, this appendix gives definitions and summarizes prior results from the literature that were used in the above discussion. Proofs of all results in this section may be found  in~\cite{McCluskey07} or \cite{KamSal07}, as indicated.

\subsection{Useful definitions}

\begin{dfn}[Metzler matrix, Metzler stable matrix~\cite{MR1298430, MR94c:34067, 0458.93001}] \label{dfn:dfnMzlr} A given $n\times n$ real matrix is said to be a Metzler matrix if all its off-diagonal terms are nonnegative.  The matrix is called Metzler stable if in addition all of the eigenvalues have negative real parts.   \end{dfn}

\noindent Note that a square matrix $\mathbf A$  is a Metzler matrix if and only if $-\mathbf A$ is a $Z$-matrix, and $\mathbf A$  is  Metzler stable  if and only if $-\mathbf A$ is a $M$-matrix. This paper makes use of the fact (which is not difficult to prove) that the positive cone is invariant for every dynamical system described by a system of ordinary differential equations whose Jacobian matrix is a Metzler matrix.

\begin{dfn}[Irreducible matrix] \label{dfn:dfnIrmtx} A given $n\times n$ matrix $\mathbf A$ is said to be a reducible matrix if there exists a  permutation matrix $\mathbf P$ such that $\mathbf P^{\mathbf T}\,\mathbf A\,\mathbf P$ has block matrix form:  $\mathbf P^{\mathbf T}\,\mathbf A\,\mathbf P=\left(\begin{array}{cc}\mathbf A_1 & \mathbf A_{1\,2}\\\mathbf 0 & \mathbf A_{2}
\end{array}\right)$, where $\mathbf A_1$ and $\mathbf A_2$ are square matrices. A matrix $\mathbf A$ that is not reducible is said to be irreducible.\end{dfn}

\noindent Irreducibility of $\mathbf A$ can be checked using the directed graph associated with $\mathbf A = (a_{k\,j} )$. This graph (denoted as $G(\mathbf A)$) has vertices ${1,\; 2,\; \cdots,\; n}$, and the directed arc $(k,\; j)$ from $k$ to $j$ is in $G(\mathbf A)$ iff $a_{k\,j}\neq 0$. $G(\mathbf A)$ is said to be \defn{strongly connected} if any two distinct vertices of $G(\mathbf A)$ are joined by an oriented path. The matrix $\mathbf A$ is irreducible if and only if $G(\mathbf A)$ is strongly connected~\cite{Guo_li_PAMS08}.

\subsection{Arithmetic-geometric means inequality, and consequences}In demonstrating that the Lyapunov derivative is nonpositive (see Section~\eqref{sec:eeqstana}), a key tool is the arithmetic-geometric means inequality, stated as follows: 

\begin{lem}[Weighted AM-GM]\label{lem.agmi}
	Let the positive numbers $z_1,\;\cdots,\;z_d$ and the positive weights $w_1,\;\cdots,\;w_d$ be given. Set $w = w_1+\;\cdots+\;w_d$. If $w>0$, then  
	$$\sqrt[w]{z_1^{w_1}\cdots z_d^{w_d}} \leq \frac{w_1z_1+\cdots+w_dz_d}{w}$$
	
	Furthermore, exact equality occurs iff $z_1=\cdots=z_d$
\end{lem}

\noindent The classical weighted AM--GM is more general than this lemma. 

\noindent An immediate consequence of weighted AM-GM is:

\begin{cor}[\cite{McCluskey07}]\label{cor.agmi} Let $z_1,\;\cdots,\;z_d$ be positive real numbers such that their product is 1.Then
	$$d - z_1 -\cdots - z_d\leq 0.$$  Furthermore, exact equality  occurs iff $z_1=\cdots=z_d$.
\end{cor}

\noindent This is obtained from Lemma~\ref{lem.agmi} by choosing $w_i=1$ for all  $1$.

\noindent Another useful consequence  is 

\begin{cor}\label{cor.agmi0} Let the positive numbers $z_1,\;\cdots,\;z_d$ and the positive weights $w_1,\;\cdots,\;w_d$ be given. Assume for a given $i$, $1\leq i \leq d$ there is a $v\in\mathbb R$ such that ${z_i}^v=z_1^{w_1}\cdots z_{i-1}^{w_{i-1}}z_{i+1}^{w_{i+1}}\cdots z_d^{w_d}$.  Set $w = w_1+\;\cdots+\;w_d$. If $w>0$ and $(z_i-1)(w+v)\leq 0$, then
	
	$$w - w_1z_1 -\cdots - w_dz_d\leq 0.$$  Furthermore, exact equality  occurs iff $z_1=\cdots=z_d$.
\end{cor}
\noindent {\em Proof }:~~ For numbers and associated weights as stated in Corollary, applying the Lemma~\ref{lem.agmi} gives

$$z_i^{\frac{w_i+v}{w}}\leq\frac{w_1z_1+\cdots+w_dz_d}{w};$$ since for given $x$ and $\beta$ positive real numbers and $\alpha$ a real number with $|\alpha|<\beta$ we have the relation $1\leq x^{\frac{\alpha}{\beta}}$ iff $x\leq 1\wedge\alpha\geq0$ or $x\geq 1\wedge\alpha\leq0$, which holds iff $(x-1)\alpha\leq0$. Thus assuming  $(z_i-1)(w_i+v)\leq 1$,  since this is equivalent to $1\leq z_i^{\frac{w_i+v}{w}}$, the  result follows straightforwardly. \edem

\subsection{Results on the global asymptotic stability (GAS) of the DFE of epidemiological model} Theorem~\ref{thm:kamsal} and Proposition~\ref{prop:blockdecomposition} given in this section  are key tools in demonstrating Theorem~\ref{thm.defstab} in Section
`\ref{subsec.dfestabanan} and Proposition~\ref{prop:basicrepn} in Section~\ref{sec.algo},respectively. 

Theorem~\ref{thm.defstab} is conventionally proven by constructing an adequate Lyapunov function for the model at the equilibrium concerned. Theorem~\ref{thm:kamsal} establishes the existence of such a Lyapunov function in cases similar to the situation in this paper.

\begin{thm}[\cite{KamSal07}]\label{thm:kamsal}Consider the system  
 \begin{equation} \label{eq:eqmodelc} \left\{\begin{array}{ccr}\dot{\mathbf x}_S & = & \mathbf
A_S(\mathbf x)\,.\, \left(\mathbf x_S - \mathbf x^*_S\right)  +  \mathbf A_{S,\,I}(\mathbf x)\,.\,\mathbf x_I \\
   \dot{\mathbf x}_I & = & \;\;\;\;\;\;\mathbf A_I(\mathbf x)\,.\,\mathbf x_I
\end{array}\right. \end{equation}
defined on a positively invariant set $\Omega\subset\mathbb R_+^{n_S\times n_I}$.  Given that:
\begin{enumerate}[{\bf h}1:]
\item The system is  dissipative on $\Omega$;
\item The equilibrium $\mathbf x_S^*$ of the subsystem $\dot{\mathbf x}_S = \mathbf A_S\left(\mathbf x_S,\;\mathbf 0\right)\,.\,(\mathbf x_S-\mathbf x^*_S)$ of system~\eqref{eq:eqmodelc} is globally asymptotically stable   on the canonical projection of $\Omega$ on $\mathbb R_+^{n_S}$;
\item The matrix $\mathbf A_I(\mathbf x)$ is a Metzler matrix  and irreducible for each $\mathbf x\in\Omega$;
\item There is an upper--bound matrix $\overline{\mathbf A}_I$ (in the sense of pointwise order) for the set of $n_I\times n_I$ square matrices  $\mathfrak{M} = \{\mathbf A_I(\mathbf x)\;/\;\mathbf x\in\Omega\}$ with the property that either $\overline{\mathbf A}_I\not \in\mathfrak{M}$ or if $\overline{\mathbf A}_I \in\mathfrak{M}$, then for any $\overline{\mathbf x}\in\Omega$ such that $\overline{\mathbf A}_I = \mathbf A_I(\overline{\mathbf x})$, we have $\overline{\mathbf x}\in\mathbb R_+^{n_S}\times\{\mathbf 0\}$;
\item $\alpha(\overline{\mathbf A}_I)\leq 0$.
\end{enumerate}

\noindent Then the DFE $\mathbf x^*$ is GAS for the system~\eqref{eq:eqmodelc} in $\overline{\Omega}$.
\end{thm}  

 Proposition~\ref{prop:blockdecomposition} shows that  the Metzler stability of matrix $\mathbf M$ is equivalent to the Metzler stability of two smaller matrices,which if properly chosen may be easier to compute with.

\begin{prop}[\cite{KamSal07}]
    Let  $\mathbf M$  be a Metzler matrix,  with block decomposition
    $\mathbf M=\left(\begin{array}{cc}
    \mathbf {A} & \mathbf {B}\\
    \mathbf {C} & \mathbf {D}
    \end{array}\right)$,
     where $ \mathbf {A}$ and $ \mathbf {D}$ are square matrices.
    \noindent Then      $\mathbf  M$ is Metzler stable if and only if 
    $\mathbf A$ and  $ \mathbf  D - \mathbf  C \mathbf A^{-1} \mathbf  B $ (or $\mathbf D$ and $\mathbf A-\mathbf B \mathbf D^{-1} \mathbf C$) are Metzler stable.
    \label{prop:blockdecomposition}
\end{prop}

\section{Proofs of Propositions in Section~\ref{dfestabana}}

\subsection{Proof of Proposition~\ref{prop:dissip}}\label{ssec.supdissip}

Considering equations of the  system~\eqref{eq:eqbednet_} that describe the dynamic of the transmission in the vectors population, we write the system:  
\begin{equation}
   \left\{
    \begin{array}{lcl}
        \dot S_q & = &\Gamma- (\hat\mu+\varpi) S_q+\delta S_r\\
        \dot S_r & = & (\varpi-\varphi) S_q-(\mu+\delta)S_r\\
        \dot M_{I_v} & = & \varphi  S_q - \mu M_{I_v} - dM_{I_q}\\
    \end{array}
    \right. 
     \label{eq:eqbednet_p}
\end{equation} 
constituted of the two first equations of the system~\eqref{eq:eqbednet_}, and the last made by adding equations that describe the evolution of all other components (components concerning non naive vectors) of the state of the model concerning vectors, to which we give the identification $M_{I_v}$.  $M_{I_q}$ stands  for the overall population of non naive vectors in the questing states. The evolution of other components of the state of the model are keep implicit since they are not needed in the proof.  Let $\left(S_q(t),\;S_r(t),\; M_{I_v}(t)\right)$ be the solution of the system~\eqref{eq:eqbednet_p} beginning at a given initial state $(S_q(0),\;S_r(0),M_{I_v}(0))\in\mathbb R^3_+$. Consider now the system
\begin{equation}
   \left\{
    \begin{array}{lcl}
        \dot S_q & = &\Gamma- (\hat\mu+\varpi) S_q+\delta S_r\\
        \dot S_r & = & \varpi S_q-(\mu+\delta)S_r\\
    \end{array}
    \right. 
     \label{eq:eqbednet_pu}
\end{equation} 

\noindent This is the canonical projection of the system~\eqref{eq:eqbednet_} on the disease free subvariety. As stated in Proposition~\ref{prop:stbsysred}, the unique equilibrium of system~\eqref{eq:eqbednet_pu} that is $\mathbf x_S^*$ is globally asymptotically stable in $\mathbb R^2_+$.  If $(\bar S_q(t),\;\bar S_r(t))$ is the solution of the system~\eqref{eq:eqbednet_pu}, with the initial state $(S_q(0);\;S_r(0))$,  
we have $0\leq S_r(t)\leq\bar S_r(t)$, since $(\varpi-\varphi)S_q-(\mu+\delta)S_r\leq \varpi S_q-(\mu+\delta)S_r$; we have also $0\leq S_q(t)\leq\bar S_q(t)$ since at any $t$, $\Gamma-(\hat\mu+\varpi)S_q+\delta S_r(t)\leq \Gamma-(\hat\mu+\varpi)S_q+\delta \bar S_r(t)$. For any $\varepsilon>0$, there exists a $t_\varepsilon>0$ such that for all $t>t_\varepsilon$, we have  $0\leq S_r(t)\leq\bar S_r(t)<S_r^*+\varepsilon$, and $0\leq S_q(t)\leq\bar S_q(t)<S_q^*+\varepsilon$. For the third equation of eq.~\eqref{eq:eqbednet_p}, it follows that $0\leq M_{I_v}(t)\leq\bar M_{I_v}(t)<\frac{\bar \varphi}{\mu}S^*_q+\varepsilon$ where $\bar M_{I_v}(t)$ is the solution of the equation $\dot M_{I_v}=\bar\varphi S_q^*-\mu M_{I_v}$ with initial condition $M_{I_v}^0$. This proves the attractiveness of the set $\Omega$. The invariance of $\Omega$ is straightforward. \edem   

\subsection{Proof of Proposition~\ref{prop:basicrepn}}\label{ssec.supprrprepn}

We prove first the stability condition $R_0 < 1$ for the system~\eqref{eq:eqmodel} at the DFE where $R_0$ is given by~\eqref{eq:R0}. Then we show that $R_0$ can be interpreted as the basic reproduction number. 

On the infection-free subvariety of $\left(\mathbb R_+\right)^u$, system~\eqref{eq:eqmodel} reduces to system~\eqref{eq.sysred} which  has a unique equilibrium $\mathbf x^*_S$ that is GAS according to Proposition~\ref{prop:stbsysred}. Thus to ensure local stability of~\eqref{eq:eqmodel} at the DFE, it is necessary and sufficient that  the submatrix $\mathbf A_I(\mathbf x^*)$ defined by~\eqref{eq:eqmodel1} is stable, since  $\mathbf A_I(\mathbf x^*)$ is the Jacobian matrix of the system~\eqref{eq:eqmodel} reduced to the infected subvariety. 

\noindent The stability of $\mathbf A_I(\mathbf x^*)$ is established as follows. Since  $\mathbf A_I(\mathbf x^*)$ is a Metzler  matrix, according to Proposition~\ref{prop:blockdecomposition}  we have that $\mathbf A_I(\mathbf x^*)$ is Metzler stable if and only if  $\mathbf A_{I_E}(\mathbf x^*)$ and $\mathbf N(\mathbf x^*) \equiv \mathbf A_{I_I}(\mathbf x^*) - \mathbf A_{I_{I,\,E}}(\mathbf x^*){\mathbf A_{I_E}(\mathbf x^*)}^{-1}\mathbf A_{I_{E,\,I}}(\mathbf x^*)$ and  are Metzler stable. Since $\mathbf A_{I_E}(\mathbf x^*)$ is always Metzler stable, we may focus on $\mathbf N(\mathbf x^*)$, which is a $n+3\times n+ 3$ matrix that can be written as 
$$\mathbf N(\mathbf x^*)=\left(\begin{array}{cc} \mathbf N_{1\,1}(\mathbf x^*)&\mathbf N_{1\,2}(\mathbf x^*)\\\mathbf N_{2\,1}(\mathbf x^*)&\mathbf N_{2\,2}(\mathbf x^*) \end{array}\right),$$ 
with
$\mathbf N_{1\,1}(\mathbf x^*)=\mathbf A_{I_{I_h}}$, $\mathbf N_{2\,2}(\mathbf x^*)=\mathbf A_{I_{I_v}}$,
 $\mathbf N_{1\,2}(\mathbf x^*)=\mathbf A_{I_{I_{v,\,h}}}$  and $\mathbf N_{2\,1}(\mathbf x^*)=f_r^{\ell+ 1}{f_q^*}^l\dfrac{a}{H}S_q^*\begin{pmatrix}c_0f_0&\cdots&c_nf_n\cr0&\cdots&0\end{pmatrix}$.

\noindent Once again we may apply Proposition~\ref{prop:blockdecomposition}, this time to  $\mathbf N(\mathbf x^*)$, and conclude that 
  $\mathbf N(\mathbf x^*)$ is Metzler stable if and only if $\mathbf N_{1\,1}(\mathbf x^*)$ and $\mathbf L(\mathbf x^*) \equiv  \mathbf N_{2\,2}(\mathbf x^*)-\mathbf N_{2\,1}(\mathbf x^*){\mathbf N_{1\,1}(\mathbf x^*)}^{-1}\mathbf N_{1\,2}(\mathbf x^*)$ are Metzler stable. Since $\mathbf N_{1\,1}(\mathbf x^*)$ is always Metzler stable, we only need to establish the Metzler stability of $\mathbf L(\mathbf x^*)$, which may be written more explicitly as
\begin{equation}\label{eq:Lmx}  
\mathbf L(\mathbf x^*) =\left(\begin{array}{cc}\xi(\mathbf
x^*)-(\hat\mu^*+\varpi^*)&\delta\\\varpi^*&-(\mu+\delta)\end{array}\right),
\end{equation}
where
\begin{equation}\label{eq:coeffspec}
\xi(\mathbf x^*)=f_r^{\ell+ 1}{f_q^*}^l \dfrac{S^*_q}{H}a^2\sum_{i=0}^n\frac{b_i^*c_if_im_i}{\nu_i+\gamma_i}
\end{equation}
\noindent 
We make one final application of Proposition~\ref{prop:blockdecomposition}. Since $\mathbf L(\mathbf
x^*)_{2,\,2}$ is negative and hence Metzler stable,  the Metzler stability of   $\mathbf
A_I(\mathbf x^*)$ holds if and only if
$$\mathbf
L(\mathbf x^*)_{1\,1}-\mathbf L(\mathbf x^*)_{1\,2}\cdot{\mathbf L(\mathbf x^*)_{2\,2}}^{-1}\cdot\mathbf L(\mathbf x^*)_{2\,1}<0,$$  
which in view of (\ref{eq:Lmx}) and (\ref{eq:coeffspec}) may be rewritten as
\begin{equation}\label{eq:MetzlerCondition}
\dfrac{\mu+\delta}{(\mu+\delta)(\hat\mu^*+\varpi^*)
-\varpi^*\delta}f_r^{\ell+ 1}{f_q^*}^\ell\dfrac{S^*_q}{H}a^2\sum_{i=0}^n\frac{b^*_ic_if_im_i}{\nu_i+\gamma_i}<1.
\end{equation}
Using expressions from \eqref{eq:xsv}, we may simplify \eqref{eq:MetzlerCondition} to obtain the following necessary and sufficient condition for Metzler stability of the matrix $\mathbf A_I(\mathbf x^*)$ :
 
\begin{equation}\label{eq:thrshold}
\frac{({f_q^*}f_r)^{\ell+ 1}}{(1-f_q^*f_r)^2}\dfrac{f_q^*}{{\varpi^*}^2}\dfrac{\Gamma}{H^*}a^2\sum_{i=0}^n\frac{b^*_ic_if_im_i}{\nu_i+\gamma_i}<1.
\end{equation}


We now show directly that the coefficient in the left of the condition~\eqref{eq:thrshold} is the basic reproduction number $\mathcal R_0$, which can be represented as
\begin{equation}\label{eq:R0formula}
\mathcal R_0 = \sum_{i=0}^n \mathcal R_0^{vh_i} \mathcal R_0^{h_iv},
\end{equation}
where $ \mathcal R_0^{vh_i}$ is the average number of hosts in group $i$ infected by a single infectious vector during the course of its remaining lifetime, and 
$\mathcal R_0^{h_i v}$  is the average number of vectors which become infectious due to bites of a single infectious host in group $i$. We may characterize $\mathcal R_0^{vh_i}$ as
\begin{align*}
\mathcal R_0^{vh_i} =  am_i\frac{S_i^*}{H^*}\cdot\frac{1}{\hat\mu^*+ \varpi^*}\sum_{j=0}^{\infty} (f_q^* f_r)^{j}  =   \frac{f_q^*} {1-f_q^* f_r} \cdot  \frac{a b_i^* m_i}{\varpi^*},  
\end{align*}
where $am_i\frac{S^*_i}{H^*}$ is the mean number of susceptible hosts in group $i$ per unit time that have caught the disease due to their contact with a single infectious questing vector; and  $\frac{1}{\hat\mu^*+ \varpi^*}\dsum_{j=0}^{\infty} (f_q^* f_r)^{j}$ is the mean total questing duration for an infectious questing vector, since after each successful blood meal the infectious vector will enter the infectious resting state before returning to the infectious questing state, and the mean  duration of the $j^{th}$ infectious questing period is $\frac{(f_q^* f_r)^{j-1}}{\hat\mu^*+ \varpi^*}\equiv\frac{f_q^*}{\varpi^*}(f_q^* f_r)^{j-1}$.

\noindent We may also characterize $\mathcal R_0^{h_iv}$ as
\begin{equation}\label{eq:R0hiv}
\mathcal R_0^{h_iv} = \frac{af_ic_i}{H^*}S_q^* \cdot  f_r(f_q^* f_r)^{\ell} \cdot \frac{1}{\nu_i + \gamma_i} = f_r(f_q^* f_r)^{\ell}\frac{af_ic_iS_q^*}{H^*(\nu_i + \gamma_i)}.  
\end{equation}
In this expression, $\frac{ac_if_iS_q^*}{H^*}$ is the rate at which susceptible questing vectors are infected through bites of a single infectious individual of the $i^{th}$ host group, and  $f_r(f_rf_q^*)^l$ is the frequency of survival through all $2\ell +1$ exposed states of the mosquitoes dynamics. It follows that the product of these two factors gives the rate of production of infectious mosquitoes due to the presence of a single infectious individual in the $i^{th}$ host group. The additional factor $\frac{1}{\nu_i+\gamma_i}$ in \eqref{eq:R0hiv} represents the mean duration of infectiousness of an infected host.

Plugging these formulas for $\mathcal R_0^{h_iv}$ and $\mathcal R_0^{h_iv}$ into expression \eqref{eq:R0formula}
for $\mathcal R_0$ yields the left-hand side of \eqref{eq:thrshold}, thus completing the proof.

The above expression for $\mathcal R_0$, 
$$\mathcal R_0={\dsum_{i=0}^n}\mathcal R_0^{vh_i}\mathcal R_0^{h_iv},$$
occurs commonly when dealing with vector-borne diseases, and represents the average number of secondary cases of infectious vectors (respectively hosts) that are occasioned by one infectious vector (respectively host) introduced in a population of susceptible vectors (respectively hosts). i.e. $\mathcal R_0$ for the population of vectors and also for the population of hosts. When  the next generation matrix technique of van den Driessche et {\it al.}~\cite{VddWat02} to compute this number, the square root of this number is obtained.  J. Li et {\it al.}~\cite{527610} discuss the possible failure of the next generation matrix technique. especially in cases of diseases with three actors or more, like vector borne diseases.  We have also tried with the technique in~\cite{VddWat02} with reasonable choice of $l$ and $n$, and the  result was the square root of the $\mathcal R_0$ given above.

 \edem

\subsection{Proof of Proposition~\ref{prop:stdst1}}\label{subsec.sptopropee}
The purpose of Proposition~\ref{prop:stdst1} is to specify all  possible endemic equilibria (EE) of the system~\eqref{eq:eqbednet_}.
At an endemic state, at  least one of the infected or infectious components of the solution  $\mathbf x^\star$ must nonzero.  Since the continuing presence of disease requires questing infectious vectors that successfully transmits disease to a host in one of the host subpopulation, we may assume that $I_q^\star\neq0$ where $I_q^\star$ is the component of $\mathbf x^\star$  corresponding to the Infectious questing mosquitoes.

The dynamics of the overall population of mosquitoes is given by the set of differential equations \begin{equation}\left\{\begin{array}{ccl}\dot M_q&=&\Gamma-(\hat\mu+\varpi)M_q+\delta M_r\\\dot M_r &=&\varpi M_q-(\mu+\delta)M_r\end{array}\right.,
\end{equation}
where $M_q$ and $M_r$ are the total numbers of resting and questing mosquitoes, respectively.
It follows that the steady-state values  $M_q$,  $M_r$  must satisfy $M_q^*=\frac{(\mu+\delta)\Gamma}{(\mu+\delta)(\hat\mu^*+\varpi^*)-\varpi^*\delta}=\frac{f^*_q}{\varpi^*(1-f^*_qf_r)}\Gamma$  and   $M_r^\star=\frac{\delta f^*_q}{f_r(1-f^*_qf_r)}\Gamma $. Since $\hat{\mu}$ and $\varpi$ depend only on the host dynamics and not on the presence or absence of infection, it follows that $\hat{\mu}^*$ and $\varpi^*$ for any possible EE must be the same as in the DFE.

Using system \eqref{eq:eqbednet_}, it is possible to solve for the various components of the infected mosquito population at the EE  in terms of $I_q^\star$:
 \begin{equation}\label{eq:eqexpMinfctstate_}\begin{array}{ll}
I_r^\star=\frac{\varpi^*}{(\mu+\delta)}I_q^\star=\frac{\varpi^* }{\delta }f_rI_q^\star\hbox{, } & E_r^{(\ell+ 1)\star}=\frac{\varpi^*}{\delta }\frac{1-f^*_qf_r}{f^*_q} I_q^\star,\\ E_q^{(i)\star}=\frac{1-f^*_qf_r }{(f^*_qf_r)^{\ell+ 1-i}}I_q^\star\hbox{, } & E_r^{(i)\star}=\frac{\varpi^*}{\delta }\frac{1-f^*_qf_r}{ f^*_q(f^*_qf_r)^{\ell+ 1-i}} I_q^\star\quad\hbox{, for } 1\leq i\leq l.\end{array}.
\end{equation}
From \eqref{eq:eqexpMinfctstate_} it follows that $\dsum_{i=1}^lE_q^{(i)\star}+I_q^\star=\frac{I_q^\star}{(f^*_qf_r)^l}$, and  since $M_q^*=S^\star_q+\dsum_{i=1}^lE_q^{(i)\star}+I_q^\star$ we have
\begin{equation*}
S_q^\star=\frac{f^*_q}{\varpi^*(1-f^*_qf_r)}\Gamma - \frac{I_q^\star}{(f^*_qf_r)^l}.\label{eq:eqexpMsucpqst}
\end{equation*}
Similarly, we find
\begin{equation}
S_r^\star=M_r^\star-\dsum_{i=1}^{\ell+ 1}E_r^{(i)\star}-I_r^\star=\frac{\delta f^*_q}{f_r(1-f^*_qf_r)}\Gamma - \frac{\varpi^*}{\delta}\frac{(1-(f^*_qf_r)^{\ell+ 1}+f^*_qf_r)I_q^\star}{ f^*_q}\label{eq:eqexpMsucprst}.
\end{equation}
Equations \eqref{eq:eqexpMinfctstate_} and \eqref{eq:eqexpMsucprst} uniquely specify all vector components at EE in terms of $I_q^\star$.

Concerning host populations in the model at EE, for each $i$, the components $I^\star_i$ and $S^\star_i$ satisfy $a\,m_i \frac{I_q^\star}{H^*}S^\star_i -(\nu_i+\gamma_i) I_i^\star=0$ and $I_i^\star+S^\star_i = H^*_i=\frac{\Lambda_i}{\nu_i-\tilde \nu_i}$.  This yields:
$$I_i^\star=\dfrac{a\,m_iI_q^\star }{H^*(\nu_i+\gamma_i)+am_iI_q^\star}H^*_i~~\hbox{and}~~S^\star_i = \dfrac{H^*(\nu_i+\gamma_i) }{H^*(\nu_i+\gamma_i)+am_iI_q^\star}H^*_i.$$

\noindent All of the components at EE given above require the existence of  a feasible nonzero $I_q^\star$. To determine this component, we use  two  expressions of $\varphi$ that hold at the EE.
The first comes  from the equality $\varphi^\star S^\star_q-(\mu+\delta)E_r^{(1)\star}=0$ (third equation of~\eqref{eq:eqbednet_}), so that: \begin{equation}\label{eq:eqphi}\varphi^\star = \frac{\delta}{f_r}\frac{E_r^{(1)\star}}{S^\star_q}=\frac{{\varpi^*}^2(1-f^*_qf_r)^2 I_q^\star}{f^*_q(f^*_qf_r)^{\ell+ 1}\Gamma-\varpi^* f^*_qf_r (1-f^*_qf_r)I_q^\star}.\end{equation}
The second comes from  the expression for $\varphi$ in Table~\ref{tab.tabvd2}:  $\varphi^\star= \frac{a}{H^*}\dsum_{i=0}^n\,c_if_iI_i^\star$.   Rewriting this in terms of  $I^\star_q$ gives:\begin{equation}\label{eq:eqphi1}\varphi^\star=I_q^\star\dsum_{i=0}^n \frac{a^2\,b^*_ic_if_im_i}{H^*(\nu_i+\gamma_i)+am_iI_q^\star}.\end{equation}

Setting~\eqref{eq:eqphi} equal to \eqref{eq:eqphi1} gives  the following equation with $I^\star_q$ as unknown: \begin{equation}\label{eq:eqeqIQ1}\begin{array}{ccl}\dsum_{i=0}^n \frac{a^2\,b^*_ic_if_im_i}{H^*(\nu_i+\gamma_i)+am_iI_q^\star}
&=&\dfrac{{\varpi^*}^2(1-f^*_qf_r)^2 }{f^*_q(f^*_qf_r)^{\ell+ 1}\Gamma-\varpi^* f^*_qf_r (1-f^*_qf_r)I_q^\star}\end{array},
\end{equation}
which is the determining equation for $I_q^*$ as specified in \eqref{eq:eqeqIQ}.

We now show that \eqref{eq:eqeqIQ1} has a unique biologically-feasible solution. For simplicity we express the solution in terms of the new variables 
\begin{equation}\label{eq:xdef}
x \equiv a\frac{I^\star_q}{H^*}; \qquad
\alpha_i \equiv b^*_ic_if_i; \qquad s \equiv \frac{\varpi^*}{a}\frac{1-f^*_qf_r}{f^*_q(f^*_qf_r)^l}\frac{H^*}{\Gamma},
\end{equation}
where $x$ uniquely determines $I^\star_q$.  
From expression \eqref{eq:R0} for $\mathcal R_0$, it follows
 \begin{equation}\label{eq:eqrelpo}s\dfrac{1-f^*_qf_r}{f^*_qf_r}\dfrac{\varpi^*}{a}=\dfrac{{\varpi^*}^2(1-f^*_qf_r)^2}{f^*_q(f^*_qf_r)^{\ell+ 1}}\dfrac{H^*}{a^2\,\Gamma}=\frac{1}{\mathcal R_0}\dsum_{i=0}^{n}\frac{\alpha_i m_i}{\nu_i+\gamma_i}.\end{equation}
The expression~\eqref{eq:eqeqIQ1} then becomes:
 \begin{equation}\label{eq:eqeqIQnew}\dsum_{i=0}^{n}\frac{\alpha_i m_i}{\nu_i+\gamma_i}\left(\dfrac{1}{\mathcal R_0 (1-sx)}-\frac{\nu_i+\gamma_i}{\nu_i+\gamma_i+m_ix} \right)=0\end{equation}

\noindent Using the same strategy as in~\cite{jckam201411}, we set:

\begin{eqnarray*}\label{eq:eqpolIQnew}
 T(x)& \equiv&{\dsum_{i=0}^{n}}\frac{\alpha_im_i}{\nu_i+\gamma_i}\left(\dfrac{1}{\mathcal R_0 (1-sx)}-\frac{\nu_i+\gamma_i}{\nu_i+\gamma_i+m_ix} \right)\\&=&\dfrac{1}{\mathcal R_0(1-sx)}{\dsum_{i=0}^{n}}\frac{\alpha_im_i}{\nu_i+\gamma_i}\left(\frac{(\nu_i+\gamma_i)(1-\mathcal R_0)+(m_i+(\nu_i+\gamma_i)\mathcal R_0s)x}{\nu_i+\gamma_i+m_ix}\right).
\end{eqnarray*} 
Since $T(x)$ is a rational function, it is of class $\mathcal C^\infty$ on $\R_+\setminus\left\{\frac{1}{s}\right\}$. Furthermore, we  have  
$T(0)=\dfrac{1-\mathcal R_0}{\mathcal R_0}{\dsum_{i=0}^{n}}\frac{\alpha_im_i}{\nu_i+\gamma_i},$ 
which implies that $T(0)<0$ whenever $\mathcal R_0>1$. We have also that $\dlim_{x\rightarrow\frac{1}{s}^-}T(x)=+\infty$, $\dlim_{x\rightarrow\frac{1}{s}^+}T(x)=-\infty$, and $\dlim_{x\rightarrow+\infty}T(x)=0$.
The derivative of the function $T$ is 
$$\frac{d\,T}{dx}(x)={\dsum_{i=0}^{n}}\frac{\alpha_im_i^2}{\nu_i+\gamma_i} \frac{\mathcal R_0(\nu_i+\gamma_i)(1-sx)^2+s(m_ix+\nu_i+\gamma_i)x}{\mathcal R_0(1-sx)^2(\nu_i+\gamma_i+m_ix)^2},$$ 
which is positive on $\R_+$ so that $T$ is an increasing function on $\mathbb R_+\setminus\left\{\frac{1}{s}\right\}$. The intermediate value theorem implies that there are two solutions for equation~\eqref{eq:eqeqIQnew}: a biologically feasible solution in the interval $\left]0,\;\;\frac{a}{\varpi^*}\frac{f^*_q(f^*_qf_r)^l}{1-f^*_qf_r}\frac{\Gamma}{H^*}\right[$, and a second solution at  infinity, which is not biologically feasible. In view of the definition of $x$ in \eqref{eq:xdef}.

\noindent Moreover, from the formulas of components of the endemic  equilibrium of the system \eqref{eq:eqbednet_} (see eq.~\eqref{eq:eqexpMinfctstateb}),  it follows that 
\[S_q^\star \varphi^\star = \Gamma \frac{f_q^*\varphi^\star}{ \varpi^*(1 - f_r f_q^*) +  f_q^*f_r \varphi^*}. \]
Since $x / (a + bx)$ is a monotone increasing function of $x$ and $\varphi^\star < \bar{\varphi}$, we have
\[S_q^\star \varphi^\star < \Gamma \frac{f_q^*\bar{\varphi}}{ \varpi^*(1 - f_r f_q^*) +  f_q^*f_r \bar{\varphi}}. \]
Using the equality $ \delta E_r^{(\ell+ 1)\star} = f_r (f_r f_q^*)^l S_q^\star \varphi^\star = \frac{\varpi^*}{f^*_q }(1-f^*_qf_r) I_q^\star$,   we have: 
\[
I_q^\star < \Gamma \frac{f_q^*(f_r f_q^*)^{\ell+ 1}}{\varpi^*(1 - f_q^*f_r)}  \left( \frac{ \bar{\varphi}}{\varpi^*(1 - f_r f_q^*) +  f_q^*f_r \bar{\varphi}} \right),\]
that can also be written,
\[
I_q^\star < \Gamma \frac{ f_q^*(f_r f_q^*)^{\ell+ 1}}{\varpi^*(1 - f_q^*f_r)}  \left[ \frac{\bar{\varphi}/\varpi^*}{(1 -f_r f_q^*) +  f_r f_q^*\bar{\varphi}/\varpi^*  } \right].\]
Since $\bar{\varphi}/\varpi^* < 1$, we may once again use the monotonicity of $x/(a+bx)$ to obtain
\begin{equation}\label{eq:Iqbar_final}
I_q^\star < \Gamma \frac{ f_q^*(f_r f_q^*)^{\ell+ 1}}{\varpi^*(1 - f_q^*f_r)}\equiv \bar I^\star_q .
\end{equation}
This ends the proof. \edem

\section{The Lyapunov function in the  proof of the theorem ~\ref{thm:thmstabee}}\label{sec.supplyap}

\noindent The Lyapunov function technique is commonly used  in the study of the stability  of endemic equilibria of epidemiological systems in the literature~\cite{Guo_li_CAMQ_06, Guo_li_PAMS08, 0999.92036, KoroMMB04, koroMain04, 1022.34044,   MaLiuLi03, McClu06, 1008.92032, 1076.37012, 1056.92052, MR2518930, 2011.10.085}. For the current model, we define a function  $V_{ee}$ on the space state of the model, $\left(\mathbb R_{>0}\right)^u$:

\begin{equation}\label{eq:eqliapee}
\begin{array}{rcl}V_{ee}(\mathbf x) & \equiv & \left(S_q-S^\star_q\ln S_q\right) +\,\tilde\sigma_r\left(S_r-S^\star_r\ln S_r\right)+\, \sigma_r^{(1)}\left(E^{(1)}_r-E^{(1)\star}_r\ln E^{(1)}_r\right)\\&& +\,\dsum_{j=1}^\ell\left(\sigma_q^{(j)}\left(E^{(j)}_q-E^{(i)\star}_q\ln E^{(j)}_q\right)+\sigma_r^{(j+1)}\left(E^{(j+1)}_r-E^{(j+1)\star}_r\ln E^{(j+1)}_r\right)\right)+\,\tau_q\left(I_q-I_q^{\star}\ln \,I_q\right)\\&& +\tau_r\left(I_r-I_r^{\star}\ln \,I_r\right)+\,\dsum_{i=0}^{n}\vartheta_i\left(S_i-S_i^{\star}\ln \,S_i\right)+\,\dsum_{i=0}^{n}\upsilon_i\left(I_i-I_i^{\star}\ln \,I_i\right),
\end{array}
\end{equation}
where the coefficients $\tilde\sigma_r$; $\sigma_r^{(j)}$  for $j=1,\;2,\;\cdots,\; \ell+ 1$;  $\sigma_q^{(j)}$  for $j=1,\;2,\;\cdots,\; \ell$;  $\tau_r$; $\tau_q$;   $\upsilon_i$, $\vartheta_i$, for $i=0,\;1,\;\cdots,\; n$   are positive constants to be determined such that the derivative of $V_{ee}$ along the trajectories of the system~\eqref{eq:eqbednet_}  is negative.
This form of the Lyapunov function as well as some of the techniques used in our solution were inspired by  Guo $et~al.$~\cite{Guo_li_CAMQ_06, Guo_li_PAMS08}, who use a graph-theoretic approach to compute the derivative of the Lyapunov function.  In the system of Guo $et~al.$, each group has the same compartmental description, as well as the same mode of influence exchange with other groups (susceptible individuals of a given group are transferred to the next class of the group via contact with all infectious individuals in the system). In our model, not all groups have the same compartmental description: host groups have an SIS structure, while the vector group is more complex. Furthermore,  the mode of exchanging influence between groups is also different: there is no exchange of influence between hosts groups; the vector group influences hosts groups  through contact of susceptible hosts with infectious questing vectors; and all host groups influence questing vectors.


The function $V_{ee}$ defined in \eqref{eq:eqliapee} is $\mathcal C^\infty$ and is positive definite on $\left(\mathbb R_{>0}\right)^u$ as long as all coefficients are positive.  Its derivative  along the trajectories of the system~\eqref{eq:eqbednet_} is:
\begin{small}\begin{equation*}\begin{array}{rcl}\frac{dV_{ee}}{d\,t}(\mathbf
x(t)) & = & \left(1-\frac{S^\star_q}{S_q}\right)\left(\Gamma-(\hat\mu+\varpi) S_q+\delta S_r\right) +\, \tilde\sigma_r\left(1-\frac{S^{\star}_r}{S_r}\right)\left(\left(\varpi-\varphi\right)S_q-\frac{\delta}{f_r} S_r \right)+\, \sigma_r^{(1)}\left(1-\frac{E^{(1)\star}_r}{E^{(1)}_r}\right)\left(\varphi S_q-\frac{\delta}{f_r} E^{(1)}_r \right)\\&& +\,{\dsum_{j=1}^\ell}\sigma_q^{(j)}  \left(1-\frac{E^{(j)\star}_q}{E^{(j)}_q}\right) \left( \delta E^{(j)}_r - \frac{\varpi}{f_q}E^{(j)}_q \right) +\,{\dsum_{j=1}^\ell}\sigma_r^{(j+1)} \left(1-\frac{E^{(j+1)\star}_r}{E^{(j+1)}_r}\right) \left(   \varpi  E^{(j)}_q-\frac{\delta}{f_r}E^{(j+1)}_r  \right)\\&&+\,\tau_r\left(1-\frac{I^\star_r}{I_r}\right)\left(  \varpi I_q - \frac{\delta}{f_r}I_r \right)  +\,\tau_q\left(1-\frac{I^\star_q}{I_q}\right)\left( \delta E^{(\ell+ 1)}_r - \frac{\varpi}{f_q} I_q +\delta I_r \right)\\&&+\,{\dsum_{i=0}^{n}}\vartheta_i\left(1-\frac{S^\star_i}{S_i}\right)\left(\Lambda_i+\tilde{\nu_i}H_i - \left(\nu_i+a\,m_i \frac{I_q}{H}\right)S_i +\gamma_i I_i\right) +\,{\dsum_{i=0}^{n}}\upsilon_i\left(1-\frac{I^\star_i}{I_i}\right) \left( \frac{a}{H}\,m_iS_i I_q -  (\nu_i+\gamma_i)I_i  \right).
\end{array}
\end{equation*}\end{small}

Since according to Proposition~\ref{prop:dissip} we may restrict ourselves to $\mathbf x \in \Omega$, we have  $H_i=S_i+I_i=H^*_i$ for each i. Furthermore, from Table~\ref{tab.tabvd2} we find that  $\varpi$, $d$ and $f_q$, and $b_i$ are all time-independent: hence  we may write $\varpi^*=\varpi$, $d^*=d$, $f^*_q=f_q$. After substituting $\Gamma=(\hat\mu+\varpi^*)S_q^\star-\delta S_r^\star$, $\Lambda_i = am_i\frac{I^\star_q}{H^*}S^\star_i + (\nu_i-\tilde{\nu}_i)S^\star_i - (\gamma_i + \tilde \nu_i)I^\star_i$  and rearranging we obtain

\begin{equation*}
\begin{array}{rcl}\dfrac{dV_{ee}}{d\,t}(\mathbf x(t))   & = & \hat\mu S_q^\star\left(2-\frac{S_q^\star}{S_q}-\frac{S_q}{S_q^\star}\right) +\varpi^\star S_q^\star-\varpi^\star S_q^\star\frac{S^\star_q}{S_q}-\varpi S_q+\varpi S_q^\star + \delta S^\star_r  \left( \frac{S^\star_q}{S_q}-\frac{S^\star_q}{S_q}\frac{S_r}{S_r^\star}\right) +\,\tilde \sigma_r\left(\varpi-\varphi\right)S_q \left( 1 - \frac{S^\star_r}{S_r} \right)\\&& +\,\left(\frac{\tilde\sigma_r}{f_r}-1\right)\delta \left(S_r^\star-S_r\right) +\, \sigma_r^{(1)}\varphi S_q \left(1-\frac{E^{(1)\star}_r}{E^{(1)}_r}\right) +\, \sigma_r^{(1)}\left(\frac{\delta}{f_r}E^{(1)\star}_r- \frac{\delta}{f_r}E^{(1)}_r\right)  +\,{\dsum_{i=1}^l} \sigma_q^{(i)} \delta E^{(i)}_r\left(1-\frac{E^{(i)\star}_q}{E^{(i)}_q}\right) \\&& +\,{\dsum_{j=1}^{\ell}} \sigma_q^{(j)}\left( \frac{\varpi}{f_q}E^{(j)\star}_q- \frac{\varpi}{f_q}E^{(j)}_q \right) +\,{\dsum_{j=1}^{\ell}}\sigma_r^{(j+1)}E^{(j)}_q\varpi  \left(1-\frac{E^{(j+1)\star}_r}{E^{(j+1)}_r}\right) +\,{\dsum_{j=1}^{\ell}}\sigma_r^{(j+1)}\left(\frac{\delta}{f_r}E^{(j+1)\star}_r-\frac{\delta}{f_r}E^{(j+1)}_r\right)  \\&&   +\,\tau_q\left(E^{(\ell+ 1)}_r + I_r\right) \delta\left(1-\frac{I^\star_q}{I_q}\right) +\,\tau_q\left( \frac{\varpi}{f_q}I_q^\star - \frac{\varpi}{f_q}I_q \right) +\,\tau_r \varpi I_q \left(1- \frac{I^\star_r}{I_r} \right) +\,\tau_r \left(\frac{\delta}{f_r}I_r^\star- \frac{\delta}{f_r}I_r\right) 
\\&&+\,{\dsum_{i=0}^{n}}\vartheta_i\left(\nu_i - \tilde{\nu}_i\right)S^\star_i \left(2-\frac{S^\star_i}{S_i} \right) - \,{\dsum_{i=0}^{n}} \vartheta_i\left(\nu_i - \tilde{\nu}_i\right)H_i  +\,{\dsum_{i=0}^{n}}(\upsilon_i(\nu_i+\gamma_i) - \vartheta_i\left(\gamma_i+\tilde \nu_i\right) )  I^\star_i  -\,{\dsum_{i=0}^{n}} \vartheta_i\left(\gamma_i+\tilde \nu_i\right) I_i\frac{S^\star_i}{S_i} \\&& + \,{\dsum_{i=0}^{n}} \vartheta_i\left(\gamma_i+\tilde \nu_i\right) I^\star_i \frac{S^\star_i}{S_i} +\,{\dsum_{i=0}^{n}} \vartheta_i a\,m_i \frac{I^\star_q}{H^*} S^\star_i   +\,{\dsum_{i=0}^{n}} \vartheta_i  a\,m_i \frac{I_q}{H} S^\star_i-\,{\dsum_{i=0}^{n}} \vartheta_i   a\,m_i \frac{I^\star_q}{H^*} S^\star_i\frac{S^\star_i}{S_i}   \\&& +\,{\dsum_{i=0}^{n}} (\upsilon_i - \vartheta_i) \left( a\,m_i \frac{I_q}{H} S_i - (\gamma_i + \nu_i)I_i\right) -\,{\dsum_{i=0}^{n}} \upsilon_ia m_i\frac{I_q}{H}S_i\frac{I^\star_i}{I_i}
\end{array}
\end{equation*}
We may write 
\begin{equation}\label{eq:Fexp}
\dfrac{dV_{ee}}{d\,t}(\mathbf x(t))    =  \hat\mu S_q^\star\left(2-\frac{S_q^\star}{S_q}-\frac{S_q}{S_q^\star}\right) + F(\mathbf x),
\end{equation}
where 
\begin{equation*}
\begin{array}{rcl}F(\mathbf x)   &=&   \varpi^\star S_q^\star\left(1-\frac{S^\star_q}{S_q}\right)-\varpi S_q+\varpi S_q^\star + \delta S^\star_r  \left( \frac{S^\star_q}{S_q}+\frac{S_r}{S_r^\star}-\frac{S^\star_q}{S_q}\frac{S_r}{S_r^\star}-1\right) +\tilde \sigma_r\left(\varpi-\varphi\right)S_q \left( 1 - \frac{S^\star_r}{S_r} \right) +\,\tilde\sigma_r\frac{\delta}{f_r} S^\star_r\left(1-\frac{S_r}{S_r^\star}\right) \\&&+ \, \sigma_r^{(1)}\varphi S_q \left(1-\frac{E^{(1)\star}_r}{E^{(1)}_r}\right) +\,\delta{\dsum_{j=1}^\ell}\left(  E^{(j)\star}_r \left(\frac{\sigma_r^{(1)}}{f_r}- \sigma_q^{(j)}\frac{E^{(j)}_r}{E^{(j)\star}_r}\frac{E^{(j) \star}_q}{E^{(j)}_q}\right)  +\, \left(\sigma_q^{(j)} -\,\frac{\sigma_r^{(j)}}{f_r} \right)E^{(j)}_r \right) \\&&+\,\delta \left(\tau_q -\,\frac{\sigma_r^{(\ell+ 1)}}{f_r} \right)E^{(\ell+ 1)}_r  +\,\delta  E^{(\ell+ 1)\star}_r \left(\frac{\sigma_r^{(\ell+ 1)}}{f_r}- \tau_q\frac{E^{(\ell+ 1)}_r}{E^{(\ell+ 1)\star}_r}\frac{I^{ \star}_q}{I_q}\right)  +\, \delta\left(\tau_q -\frac{\tau_r}{f_r} \right)I_r +\, \delta I^\star_r\left(\frac{\tau_r}{f_r}-\tau_q  \frac{I_r}{I^\star_r}\frac{I^\star_q}{I_q}\right) \\&&+\,\varpi{\dsum_{j=1}^{\ell}}\left( E^{(j)\star}_q \left(\frac{\sigma_q^{(j)}}{f_q}-\sigma_r^{(j+1)}\frac{E^{(j)}_q}{E^{(j)\star}_q}\frac{E^{(j+1)\star}_r}{E^{(j+1)}_r} \right) +\,  \left(\sigma_r^{(j+1)}- \frac{\sigma_q^{(j)}}{f_q}\right)E^{(j)}_q\right) +\, \varpi I^\star_q \left(\frac{\tau_q}{f_q}- \tau_r\frac{I_q}{I^\star_q}\frac{I^\star_r}{I_r} \right)  + \, \varpi I_q\left( \tau_r- \frac{\tau_q}{f_q}\right) \\&&  +\,{\dsum_{i=0}^{n}}\vartheta_i\left(\nu_i - \tilde{\nu}_i\right)S^\star_i \left(2-\frac{S^\star_i}{S_i} \right) - \,{\dsum_{i=0}^{n}} \vartheta_i\left(\nu_i - \tilde{\nu}_i\right)H_i  +\,{\dsum_{i=0}^{n}}(\upsilon_i\nu_i - \vartheta_i\tilde \nu_i )  I^\star_i  -\,{\dsum_{i=0}^{n}} \vartheta_i\left(\gamma_i+\tilde \nu_i\right) I_i\frac{S^\star_i}{S_i} + \,{\dsum_{i=0}^{n}} \vartheta_i\left(\gamma_i+\tilde \nu_i\right) I^\star_i \frac{S^\star_i}{S_i}  \\&&  +\,{\dsum_{i=0}^{n}} \vartheta_i a\,m_i \frac{I^\star_q}{H^*} S^\star_i   +\,{\dsum_{i=0}^{n}} \vartheta_i  a\,m_i \frac{I_q}{H} S^\star_i-\,{\dsum_{i=0}^{n}} \vartheta_i   a\,m_i \frac{I^\star_q}{H^*} S^\star_i\frac{S^\star_i}{S_i}   \\&& +\,{\dsum_{i=0}^{n}} (\upsilon_i - \vartheta_i) \left( a\,m_i \frac{I_q}{H} S_i - (\gamma_i + \nu_i)I_i + \gamma_iI^\star_i\right) -\,{\dsum_{i=0}^{n}} \upsilon_ia m_i\frac{I_q}{H}S_i\frac{I^\star_i}{I_i}.
\end{array}
\end{equation*}
Note that  $\frac{dV_{ee}}{d\,t}(\mathbf x(t)) \leq  F(\mathbf x)$, since $2-\frac{S_q^\star}{S_q}-\frac{S_q}{S_q^\star}\leq 0$ for all $\mathbf x\in\left(\mathbb R_{>0}\right)^u$.

The expression for $F(\mathbf x)$ may be simplified by choosing
\begin{equation}
 \tau_q =\dfrac{\tau_r}{f_r}; ~\sigma^{(\ell+ 1)}_r = f_r\tau_q ;~  \sigma^{(j)}_q=f_q\sigma^{(j+1)}_r,  \sigma^{(j)}_r = f_r\sigma^{(j)}_q~ \textrm{~for~}j=\ell,~\ell-1,~\cdots 1,\\ ~\textrm{and~}\\  \vartheta_i = \upsilon_i ~ \textrm{~for~} i = 0,~1,~\cdots,~n
\end{equation}
from which follows
 \begin{equation}\label{eq.coefvee} \begin{array}{l}\sigma^{(j)}_r = \sigma^{(1)}_r(f_qf_r)^{1-j},\;\hbox{ for} \;j=1,\;2,\;\cdots,\;\ell+ 1; \quad\sigma^{(j)}_q = \sigma^{(1)}_r(f_qf_r)^{1-j}f_r^{-1},\; \hbox{for}\; j=1,\;2,\;\cdots,\;\ell;\\   \tau_q = \sigma^{(1)}_rf_r^{-1}(f_qf_r)^{-\ell} ; \qquad \tau_r =\sigma^{(1)}_r(f_qf_r)^{-\ell} .\end{array}
\end{equation} 
Using these substitutions and the fact that the size of the fraction of population in the $i^{th}$ hosts group $H_i=S_i+I_i$ is constant for $i=0,\;\cdots,\;n$, we get:

\begin{small}\begin{equation*}
\begin{array}{rcl}
F(\mathbf x)  &=&  \varpi^\star S_q^\star\left(1-\frac{S^\star_q}{S_q}\right)+\left((\sigma_r^{(1)}-\tilde\sigma_r)\varphi+(\tilde\sigma_r -1)\varpi\right) S_q+\varpi S_q^\star + \delta S^\star_r  \left( \frac{S^\star_q}{S_q}+\frac{S_r}{S_r^\star}-\frac{S^\star_q}{S_q}\frac{S_r}{S_r^\star} - 1\right)+\,\tilde\sigma_r\frac{\delta}{f_r} S_r^\star\left(1-\frac{S_r}{S_r^\star}\right) \\&& -\,\tilde \sigma_r\left(\varpi-\varphi\right)S_q \frac{S^\star_r}{S_r}    - \, \sigma_r^{(1)}\varphi S_q \frac{E^{(1)\star}_r}{E^{(1)}_r} +\, \varpi \sigma_r^{(1)}{\dsum_{j=1}^{\ell}}\frac{E^{(i)\star}_q}{(f_qf_r)^{j}} \left(1- \frac{E^{(j)}_q}{E^{(j)\star}_q} \frac{E^{(j+1)\star}_r}{E^{(j+1)}_r}\right)  +\,\delta \sigma_r^{(1)}{\dsum_{j=1}^{\ell} }\frac{ E^{(j)\star}_r}{f_r(f_qf_r)^{j-1}}\left(1-\, \frac{ E^{(j)}_r}{ E^{(j)\star}_r}\frac{E^{(j)\star}_q}{E^{(j)}_q}\right)\\&& +\delta  \sigma_r^{(1)} \frac{ E^{(\ell+ 1)\star}_r}{f_r(f_qf_r)^{l}} \left(1 - \frac{E^{(\ell+ 1)}_r}{ E^{(\ell+ 1)\star}_r} \frac{I^\star_q}{I_q}\right)  +\, \delta  \sigma_r^{(1)}\frac{I_r^\star}{f_r(f_qf_r)^{l}} \left(1 - \frac{I_r}{I_r^\star}\frac{I^\star_q}{I_q}\right) +\,\varpi \sigma_r^{(1)} \frac{I_q^\star}{(f_qf_r)^{l}}   \left(\frac{1}{f_rf_q}- \frac{I_q}{I_q^\star}\frac{I^\star_r}{I_r} \right)  + \,\varpi  \sigma_r^{(1)} \frac{f_rf_q-1}{(f_qf_r)^{\ell+ 1}}I_q\\&& +\,{\dsum_{i=0}^{n}}\upsilon_i\left(\nu_i - \tilde{\nu}_i\right)S^\star_i \left(2-\frac{S^\star_i}{S_i} \right) - \,{\dsum_{i=0}^{n}} \upsilon_i\left(\nu_i - \tilde{\nu}_i\right)H_i  +\,{\dsum_{i=0}^{n}}\upsilon_i(\nu_i - \tilde \nu_i )  I^\star_i  -\,{\dsum_{i=0}^{n}} \upsilon_i\left(\gamma_i+\tilde \nu_i\right) I_i\frac{S^\star_i}{S_i} + \,{\dsum_{i=0}^{n}} \upsilon_i\left(\gamma_i+\tilde \nu_i\right) I^\star_i \frac{S^\star_i}{S_i}  \\&&  +\,{\dsum_{i=0}^{n}} \upsilon_i a\,m_i \frac{I^\star_q}{H^*} S^\star_i   +\,{\dsum_{i=0}^{n}} \upsilon_i  a\,m_i \frac{I_q}{H} S^\star_i-\,{\dsum_{i=0}^{n}} \upsilon_i   a\,m_i \frac{I^\star_q}{H^*} S^\star_i\frac{S^\star_i}{S_i}    -\,{\dsum_{i=0}^{n}} \upsilon_ia m_i\frac{I_q}{H}S_i\frac{I^\star_i}{I_i} 
\end{array}
\end{equation*}\end{small}
Using the fact that all time derivatives in system~\eqref{eq:eqbednet_} are zero at the EE, we find:  $\varpi^\star I_q^\star = \frac{\delta}{f_r}I_r^\star$, $\delta S_r^\star=f_r\left(\varpi^\star-\varphi^\star\right)S_q^\star$, $\varpi E_q^{(j)\star}=\frac{\delta}{f_r}E_r^{(j+1)\star}$  for $j = 1,\;2,\;\cdots,\; \ell$,  and $am_i\frac{I_q^\star}{H^*} S^\star_i = (\nu_i+\gamma_i)I^\star_i$ for $i=0,\;1,\;\cdots,\; n$. This leads to

\begin{equation*}
\begin{array}{rcl}
F(\mathbf x)  &=&  \varpi^\star S_q^\star\left(1-\frac{S^\star_q}{S_q}\right) +\left((\sigma_r^{(1)}-\tilde\sigma_r)\varphi+(\tilde\sigma_r -1)\varpi\right) S_q +\varpi S_q^\star + \delta S^\star_r  \left( \frac{S^\star_q}{S_q}+\frac{S_r}{S_r^\star}-\frac{S^\star_q}{S_q}\frac{S_r}{S_r^\star}-1\right) +\,\tilde\sigma_r\frac{\delta}{f_r} S_r^\star\left(1-\frac{S_r}{S_r^\star}\right)\\&&-\;\tilde \sigma_r\left(\varpi-\varphi\right)S_q \frac{S^\star_r}{S_r}  - \, \sigma_r^{(1)}\varphi S_q \frac{E^{(1)\star}_r}{E^{(1)}_r} +\,\sigma_r^{(1)} \frac{\delta}{f_r} E^{(1)\star}_r\left(1-\, \frac{ E^{(1)}_r}{ E^{(1)\star}_r}\frac{E^{(1)\star}_q}{E^{(1)}_q}\right)  +\,\varpi \sigma_r^{(1)}  \frac{1-f_rf_q}{(f_qf_r)^{\ell+ 1}} \left(I^\star_q - I_q\right)    \\&&  +\, \varpi \sigma_r^{(1)}{\dsum_{j=1}^{\ell-1}}\frac{E^{(i)\star}_q}{(f_qf_r)^{i}}\left(2- \frac{E^{(j)}_q}{E^{(j)\star}_q}\frac{E^{(j+1)\star}_r}{E^{(j+1)}_r}-\, \frac{ E^{(j+1)}_r}{ E^{(j+1)\star}_r}\frac{E^{(j+1)\star}_q}{E^{(j+1)}_q}\right) +\,  \varpi \sigma_r^{(1)}\frac{E^{(l)\star}_q}{(f_qf_r)^{l}}\left(2- \frac{E^{(l)}_q}{E^{(l)\star}_q}\frac{E^{(\ell+ 1)\star}_r}{E^{(\ell+ 1)}_r}- \frac{E^{(\ell+ 1)}_r}{E^{(\ell+ 1)\star}_r}\frac{I^\star_q}{I_q}\right) \\&& +\, \varpi \sigma_r^{(1)}\frac{I_q^\star}{(f_qf_r)^{l}} \left(2 - \frac{I_r}{I_r^\star}\frac{I_q^\star}{I_q}- \frac{I_q}{I_q^\star}\frac{I^\star_r}{I_r}\right) +\,{\dsum_{i=0}^{n}}\upsilon_i\left(\nu_i - \tilde{\nu}_i\right)S^\star_i \left(2-\frac{S^\star_i}{S_i} \right) - \,{\dsum_{i=0}^{n}} \upsilon_i\left(\nu_i - \tilde{\nu}_i\right)H_i  +\,{\dsum_{i=0}^{n}}\upsilon_i(\nu_i - \tilde \nu_i )  I^\star_i   \\&& -\,{\dsum_{i=0}^{n}} \upsilon_i\left(\gamma_i+\tilde \nu_i\right) I_i\frac{S^\star_i}{S_i} - \,{\dsum_{i=0}^{n}} \upsilon_i\left(\nu_i - \tilde \nu_i\right) I^\star_i \frac{S^\star_i}{S_i}  +\,{\dsum_{i=0}^{n}} \upsilon_i a\,m_i \frac{I^\star_q}{H^*} S^\star_i   +\,{\dsum_{i=0}^{n}} \upsilon_i  a\,m_i \frac{I_q}{H} S^\star_i    -\,{\dsum_{i=0}^{n}} \upsilon_ia m_i\frac{I_q}{H}S_i\frac{I^\star_i}{I_i} 
\end{array}
\end{equation*}
 Using  relations between components of $\mathbf x^\star$  given in~\eqref{eq:eqexpMinfctstateb} we may derive $$\varphi^\star S^\star_q = \frac{\delta E^{(1)\star}_r}{f_r} = \frac{\delta E^{(2)\star}_r}{f_r(f_qf_r)} = \cdots = \frac{\delta E^{(\ell+ 1)\star}_r}{f_r(f_qf_r)^l} = \frac{\varpi(1-f_qf_r) I^{\star}_q}{(f_qf_r)^{\ell+ 1}} \hbox{ and } \frac{\delta E^{(j+1)\star}_r}{f_r(f_qf_r)^j} = \frac{\varpi E^{(j)\star}_q}{(f_qf_r)^j},\;\;\; j=1,\;\cdots,\;l,$$ 
and after few algebraic rearrangements, the above becomes:

\begin{equation*}
\begin{array}{rcl}
F(\mathbf x)  &=&  \left((\sigma_r^{(1)}-\tilde\sigma_r)\varphi+(\tilde\sigma_r -1)\varpi\right) S_q +\varpi^\star S_q^\star\left(1-\frac{S^\star_q}{S_q}\right) +\varpi S_q^\star+ \delta S^\star_r  \left( \frac{S^\star_q}{S_q} + \frac{S_r}{S_r^\star}-\frac{S^\star_q}{S_q}\frac{S_r}{S_r^\star}-1\right)+\,\tilde\sigma_r\frac{\delta}{f_r} S^\star_r  \left(1-\frac{S_r}{S_r^\star}\right) \\&&-\;\tilde \sigma_r\left(\varpi-\varphi\right)S_q \frac{S^\star_r}{S_r}  - \, \sigma_r^{(1)}\varphi S_q \frac{E^{(1)\star}_r}{E^{(1)}_r}+\, \sigma_r^{(1)}\varphi^\star S^\star_q\left(2\ell +2- {\dsum_{j=1}^{\ell}}\frac{E^{(j)}_q}{E^{(j)\star}_q}\frac{E^{(j+1)\star}_r}{E^{(j+1)}_r}-\, {\dsum_{j=1}^{\ell}}\frac{ E^{(j)}_r}{ E^{(j)\star}_r}\frac{E^{(j)\star}_q}{E^{(j)}_q}- \frac{E^{(\ell+ 1)}_r}{E^{(\ell+ 1)\star}_r}\frac{I^\star_q}{I_q} - \frac{I_q}{I^\star_q}\right)      \\&&+\, \frac{\varpi \sigma_r^{(1)}}{(f_qf_r)^{l}}I_q^\star \left(2 - \frac{I_r}{I_r^\star}\frac{I_q^\star}{I_q}- \frac{I_q}{I_q^\star}\frac{I^\star_r}{I_r}\right) +\,{\dsum_{i=0}^{n}}\upsilon_i\left(\nu_i - \tilde{\nu}_i\right)S^\star_i \left(2-\frac{S^\star_i}{S_i} \right) - \,{\dsum_{i=0}^{n}} \upsilon_i\left(\nu_i - \tilde{\nu}_i\right)H_i  +\,{\dsum_{i=0}^{n}}\upsilon_i(\nu_i - \tilde \nu_i )  I^\star_i   \\&& -\,{\dsum_{i=0}^{n}} \upsilon_i\left(\gamma_i+\tilde \nu_i\right) I_i\frac{S^\star_i}{S_i} - \,{\dsum_{i=0}^{n}} \upsilon_i\left(\nu_i - \tilde \nu_i\right) I^\star_i \frac{S^\star_i}{S_i}  +\,{\dsum_{i=0}^{n}} \upsilon_i a\,m_i \frac{I^\star_q}{H^*} S^\star_i   +\,{\dsum_{i=0}^{n}} \upsilon_i  a\,m_i \frac{I_q}{H} S^\star_i    -\,{\dsum_{i=0}^{n}} \upsilon_ia m_i\frac{I_q}{H}S_i \frac{I^\star_i}{I_i}.
\end{array}
\end{equation*}
We choose $\tilde\sigma_r=\sigma^{(1)}_r=1$ to make the initial terms vanish.
We also  choose $\upsilon_i = \frac{a}{H}\frac{f_i}{\nu_i - \tilde \nu_i}S^\star_q$, so that 
 $\varpi S^\star_q = {\dsum_{i=0}^n}\upsilon_i(\nu_i - \tilde\nu_i)H_i$. Using the fact that $\varphi = \frac{a}{H}{\dsum_{i=0}^n}f_ic_iI_i$ and $\varpi = \frac{a}{H}{\dsum_{i=0}^n}f_i\left(S_i+I_i\right)$, the above expression becomes: 

\begin{equation*}
\begin{array}{rcl}
F(\mathbf x) 
&=& \delta S^\star_r  \left( \frac{S^\star_q}{S_q}+\frac{S_r}{S_r^\star}-\frac{S^\star_q}{S_q}\frac{S_r}{S_r^\star}-1\right)  +\, \frac{\varpi}{(f_qf_r)^{l}} I_q^\star\left(2 - \frac{I_r}{I_r^\star}\frac{I_q^\star}{I_q}- \frac{I_q}{I_q^\star}\frac{I^\star_r}{I_r}\right) +\,\frac{ a}{H}{\dsum_{i=0}^n}f_iS^\star_iS^\star_q \left(4 - \frac{S^\star_q}{S_q} -\frac{S^\star_i}{S_i} -\frac{S_i}{S^\star_i} \frac{S_q}{S^\star_q}\frac{S^\star_r}{S_r}-\frac{S_r}{S^\star_r}\right)\\&& +\,\frac{ a}{H}{\dsum_{i=0}^n} c_if_iI^\star_iS^\star_q\left(2\ell +5 - \frac{S^\star_i}{S_i} -\frac{S^\star_q}{S_q} -\; \frac{I_i}{I^\star_i}\frac{S_q}{S^\star_q}\frac{E^{(1)\star}_r}{E^{(1)}_r}- {\dsum_{j=1}^{\ell}}\frac{E^{(j)}_q}{E^{(j)\star}_q}\frac{E^{(j+1)\star}_r}{E^{(j+1)}_r}-\, {\dsum_{j=1}^{\ell}}\frac{ E^{(j)}_r}{ E^{(j)\star}_r}\frac{E^{(j)\star}_q}{E^{(j)}_q} -\frac{E^{(\ell+ 1)}_r}{E^{(\ell+ 1)\star}_r}\frac{I^\star_q}{I_q}- \frac{I_q}{I^\star_q}\frac{S_i}{S^\star_i} \frac{I^\star_i}{I_i}\right) \\&& +\,\frac{ a}{H}{\dsum_{i=0}^n}\bar c_if_iI^\star_iS^\star_q\left(4 + \frac{I_q}{I^\star_q} -\frac{S^\star_q}{S_q} - \frac{S^\star_i}{S_i} -\frac{I_i}{I^\star_i}\frac{S_q}{S^\star_q}\frac{S^\star_r}{S_r}-\,\frac{S_r}{S^\star_r} - \frac{I_q}{I^\star_q}\frac{S_i}{S^\star_i} \frac{I^\star_i}{I_i}\right)  -\,{\dsum_{i=0}^{n}} \upsilon_i\left(\gamma_i+\tilde \nu_i\right) I_i\frac{S^\star_i}{S_i}   +\,{\dsum_{i=0}^{n}} \upsilon_i (\gamma_i+\tilde \nu_i)I^\star_i   \\&& +\,\frac{I_q}{I^\star_q}{\dsum_{i=0}^{n}} \upsilon_i  (\gamma_i+\tilde \nu_i)I^\star_i    -\,\frac{I_q}{I^\star_q}{\dsum_{i=0}^{n}} \upsilon_i (\gamma_i+\tilde \nu_i)I^\star_i\frac{S_i}{S^\star_i} \frac{I^\star_i}{I_i}, 
\end{array}
\end{equation*}
where $\bar c_i=1-c_i$, $i=0,\;\cdots,\;n$. Setting $\hat \nu_i = \nu_i - \tilde \nu_i$ and replacing $H_i$ by $S^\star_i+I^\star_i$ for $i=0,\;\cdots,\;n$ gives finally
\begin{equation*}
\begin{array}{rcl}
F(\mathbf x) 
&=& \delta S^\star_r  \left( \frac{S^\star_q}{S_q}+\frac{S_r}{S_r^\star}-\frac{S^\star_q}{S_q}\frac{S_r}{S_r^\star}-1\right)  +\, I_q^\star\frac{\varpi}{(f_qf_r)^{l}} \left(2 - \frac{I_r}{I_r^\star}\frac{I_q^\star}{I_q}- \frac{I_q}{I_q^\star}\frac{I^\star_r}{I_r}\right)+\,{\dsum_{i=0}^n}\hat \nu_i\upsilon_iS^\star_i \left(4-\frac{S^\star_i}{S_i}-\frac{S^\star_q}{S_q}-\frac{S_i}{S^\star_i}\frac{S_q}{S^\star_q}\frac{S^\star_r}{S_r}-\frac{S_r}{S^\star_r}\right)\\&& +\,{\dsum_{i=0}^n}\hat \nu_i \upsilon_ic_i I^\star_i\left(2\ell +5 -\frac{S^\star_i}{S_i}-\frac{S^\star_q}{S_q} -\; \frac{I_i}{I^\star_i}\frac{S_q}{S^\star_q}\frac{E^{(1)\star}_r}{E^{(1)}_r}- {\dsum_{j=1}^{\ell}} \frac{E^{(j)}_q}{E^{(j)\star}_q} \frac{E^{(j+1)\star}_r}{E^{(j+1)}_r}-\, {\dsum_{j=1}^{\ell}}\frac{ E^{(j)}_r}{ E^{(j)\star}_r}\frac{E^{(j)\star}_q}{E^{(j)}_q} -\frac{E^{(\ell+ 1)}_r}{E^{(\ell+ 1)\star}_r}\frac{I^\star_q}{I_q}-\frac{I_q}{I^\star_q}\frac{S_i}{S^\star_i} \frac{I^\star_i}{I_i}\right) 
\\&& +\,{\dsum_{i=0}^n}\hat \nu_i\upsilon_i\bar c_iI^\star_i\left(4 + \frac{I_q}{I^\star_q}-\frac{S^\star_i}{S_i}-\frac{S^\star_q}{S_q}-\frac{I_i}{I^\star_i}\frac{S_q}{S^\star_q}\frac{S^\star_r}{S_r}-\,\frac{S_r}{S^\star_r}-\frac{I_q}{I^\star_q}\frac{S_i}{S^\star_i} \frac{I^\star_i}{I_i}\right)    +\,{\dsum_{i=0}^{n}} \upsilon_i (\gamma_i+\tilde \nu_i)I^\star_i\left(1 - \frac{I_i}{I_i^\star} \frac{S_i^\star}{S_i}\right)    \\&& +\,\frac{I_q}{I^\star_q}{\dsum_{i=0}^{n}} \upsilon_i  (\gamma_i+\tilde \nu_i)I^\star_i\left(1 - \frac{I^\star_i}{I_i} \frac{S_i}{S^\star_i}\right). 
\end{array}
\end{equation*}
which together with \eqref{eq:Fexp} yields expression \eqref{eq:eqliapeeder2}.

\section{Nonstandard finite difference scheme}\label{subsec.nsfd}
The Nonstandard finite difference scheme uses for the simulations is:

\begin{equation}
   \left\{
    \begin{aligned}
    	\frac{S^{p+1}_i - S^{p}_i}{\phi(\delta t)} &~~=~~\Lambda_i+\tilde{\nu_i}H^{p}_i - \nu_iS^{p}_i- a\,m_i \frac{I^{p}_q}{H^{p}}S^{p+1}_i +\gamma_i I^{p}_i & i = 0,\;1,\;\cdots,\; n \\
        \frac{S^{p+1}_q - S^{p}_q}{\phi(\delta t)}&~~=~~\Gamma -\mu S^{p}_q- (d^{p}+\varpi^{p}) S^{p+1}_q  + \delta S^{p}_r  & \,\\
        \frac{S^{p+1}_r - S^{p}_r}{\phi(\delta t)}&~~=~~(\varpi^{p}-\varphi^p) S^{p+1}_q    -  (\mu+\delta) S^{p}_r & \,\\
        \frac{E^{(1)p+1}_r - E^{(1)p}_r}{\phi(\delta t)}&~~=~~\varphi^p  S^{p+1}_q-(\mu+\delta) E^{(1)p}_r& \,\\
       \frac{E^{(j)p+1}_q - E^{(j)p}_q}{\phi(\delta t)}&~~=~~\delta E^{(j)p}_r - \mu E^{(j)p}_q - (d^{p}+\varpi^{p})E^{(j)p+1}_q &
j = 1,\;2,\;\cdots,\; \ell  \,\\
       \frac{E^{(j+1)p+1}_r - E^{(j+1)p}_r}{\phi(\delta t)}&~~=~~\varpi^{p}  E^{(j)p+1}_q-(\mu+\delta)
E^{(i+1)p}_r& j = 1,\;2,\;\cdots,\; \ell  \,\\
       \frac{I^{p+1}_i - I^{p}_i}{\phi(\delta t)}&~~=~~a\,m_i \frac{I^{p}_q}{H^{p}}S^{p+1}_i -\left(\gamma_i+\nu_i\right) I^{p}_i & 
 i = 0,\;1,\;\cdots,\; n \,\\
       \frac{I^{p+1}_q - I^{p}_q}{\phi(\delta t)}&~~=~~\delta E^{(\ell+ 1)p}_r -\mu I^{p}_q- (d^{p}+\varpi^{p}) I^{p+1}_q +\delta
I^{p}_r& \,\\
       \frac{I^{p+1}_r - I^{p}_r}{\phi(\delta t)}&~~=~~\varpi^{p} I^{p+1}_q - (\delta+\mu)I^{p}_r
    \end{aligned}
    \right. 
     \label{eq:eqbednetnfds}
\end{equation} 
where 
$
\varphi^p = \frac{a}{H^{p}}{\dsum_{i=0}^n}f_ic_iI^p_i;\quad \varpi^p = \frac{a}{H^{p}}{\dsum_{i=0}^n}f_ic_iH^p_i; \quad d^p = \frac{a}{H^{p}}{\dsum_{i=0}^n} k_iH^p_i.
$
 
\noindent The time step function $\phi(t)$  is defined as 
\[
\phi(t) = \dfrac{1-\mathrm e^{-th}}{h}, \textrm{~~with~} h = \max(\delta+\mu,\;\nu_0+\gamma_0,\;\nu_1+\gamma_1,\;\cdots,\;\nu_n+\gamma_n,\;).
\]
Solving~\eqref{eq:eqbednetnfds} for  the $p+1^{th}$ terms give the 
following semi-implicit system of  difference equations:
\begin{equation}
	\left\{
	\begin{aligned}
		S^{p+1}_i  &~=~\frac{\phi(\delta t)\Lambda_i + (1-\phi(\delta t)\left(\nu_i-\tilde\nu_i\right))S^{p}_i + \phi(\delta t)\left(\tilde \nu_i + \gamma_i \right)I^{p}_i }{1+ \phi(\delta t)a m_i  \frac{I^{p}_q}{H^{p}}} 
		& i=0,\;1,\;\cdots,\; n \\
		S^{p+1}_q&~=~\frac{\phi(\delta t) \Gamma + (1-\phi(\delta t) \mu) S^{p}_q + \phi(\delta t)\delta  S_r^p}{1+\phi(\delta t)(d^{p}+\varpi^{p})} & \,\\
		S^{p+1}_r&~=~\phi(\delta t)(\varpi^p - \varphi^p)  S^{p+1}_q + \left(1 - \phi(\delta t)(\mu+\delta)\right) S_r^p  & \,\\
		E^{(1)p+1}_r&~=~\phi(\delta t)\varphi^p  S^{p+1}_q + \left(1 - \phi(\delta t)(\mu+\delta)\right) E^{(1)p}_r & \,\\
		E^{(j)p+1}_q&~=~\frac{\phi(\delta t) \delta E^{(j)p}_r + (1-\phi(\delta t) \mu)E^{(j)p}_q}{1+\phi(\delta t)(d^{p}+\varpi^{p})}  &
		j = 1,\;2,\;\cdots,\; \ell \,\\
		E^{(j+1)p+1}_r&~=~ \phi(\delta t)\varpi^p  E^{(j)p+1}_q + \left(1 - \phi(\delta t)(\mu+\delta)\right) E^{(j+1)p}_r  & \;j = 1,\;2,\;\cdots,\; \ell \,\\
		I^{p+1}_i&~=~am_i\frac{I^p_q}{H^p}\phi(\delta t)S^{p+1}_i + \left(1-\phi(\delta t)(\nu_i+\gamma_i)\right)I^p_i  &
		i = 0,\;1,\;\cdots,\; n \,\\
		I^{p+1}_q&~=~\frac{\phi(\delta t) \delta( E^{(\ell+ 1)p}_r + I^{p}_r) + (1 - \phi(\delta t) \mu) I^{p}_q}{1+\phi(\delta t)(d^{p}+\varpi^{p})} 
		& \,\\
		I^{p+1}_r&~=~ \phi(\delta t)\varpi^p  I^{p+1}_q + \left(1 - \phi(\delta t)(\mu+\delta)\right)I^{p}_r 
	\end{aligned}
	\right. 
	\label{eq:eqbednetnfds1}
\end{equation} 

}

\bibliographystyle{plain}

\begin{thebibliography}{10}
	
	\bibitem{akhavan1999cost}
	Dariush Akhavan, Philip Musgrove, Alexandre Abrantes, and Renato~d'A
	Gusm{\~a}o.
	\newblock Cost-effective malaria control in brazil: cost-effectiveness of a
	malaria control program in the amazon basin of brazil, 1988--1996.
	\newblock {\em Social Science \& Medicine}, 49(10):1385--1399, 1999.
	
	\bibitem{atieli2011insecticide}
	Harrysone~E Atieli, Guofa Zhou, Yaw Afrane, Ming-Chieh Lee, Isaac Mwanzo,
	Andrew~K Githeko, and Guiyun Yan.
	\newblock Insecticide-treated net (itn) ownership, usage, and malaria
	transmission in the highlands of western kenya.
	\newblock {\em Parasites \& vectors}, 4(1):113, 2011.
	
	\bibitem{awoleye2016improving}
	Olatunji~Joshua Awoleye and Chris Thron.
	\newblock Improving access to malaria rapid diagnostic test in niger state,
	nigeria: an assessment of implementation up to 2013.
	\newblock {\em Malaria research and treatment}, 2016, 2016.
	
	\bibitem{Macdonald78}
	A.~D. Barbour.
	\newblock Macdonald's model and the transmission of bilharzia.
	\newblock {\em Trans R Soc Trop Med Hyg}, 72(1):6--15, 1978.
	
	\bibitem{bayoh2010anopheles}
	M~Nabie Bayoh, Derrick~K Mathias, Maurice~R Odiere, Francis~M Mutuku, Luna
	Kamau, John~E Gimnig, John~M Vulule, William~A Hawley, Mary~J Hamel, and
	Edward~D Walker.
	\newblock Anopheles gambiae: historical population decline associated with
	regional distribution of insecticide-treated bed nets in western nyanza
	province, kenya.
	\newblock {\em Malaria journal}, 9(1):62, 2010.
	
	\bibitem{beier2008integrated}
	John~C Beier, Joseph Keating, John~I Githure, Michael~B Macdonald, Daniel~E
	Impoinvil, and Robert~J Novak.
	\newblock Integrated vector management for malaria control.
	\newblock {\em Malaria journal}, 7(1):S4, 2008.
	
	\bibitem{MR1298430}
	A~Berman and R.~J. Plemmons.
	\newblock {\em Nonnegative matrices in the mathematical sciences}, volume~9 of
	{\em Classics in Applied Mathematics}.
	\newblock Society for Industrial and Applied Mathematics (SIAM), Philadelphia,
	PA, 1994.
	\newblock Revised reprint of the 1979 original.
	
	\bibitem{besansky2004no}
	Nora~J Besansky, Catherine~A Hill, and Carlo Costantini.
	\newblock No accounting for taste: host preference in malaria vectors.
	\newblock {\em Trends in Parasitology}, 20(6):249--251, 2004.
	
	\bibitem{Bhatia70}
	N.~P. Bhatia and G.~P. Szeg{\"o}.
	\newblock {\em Stability Theory of Dynamical Systems}.
	\newblock Springer-Verlag, 1970.
	
	\bibitem{010047862}
	P~Carnevale and R~Vincent.
	\newblock {\em Les anoph{\`e}les, Biologie, transmission du Paludisme et lutte
		antivectorielle}.
	\newblock IRD, 2009.
	
	\bibitem{Chit_08}
	N.~Chitnis.
	\newblock {\em Using Mathematical models in controlling the spread of malaria}.
	\newblock PhD thesis, University of Arizona, 2005.
	
	\bibitem{chitnis2008determining}
	Nakul Chitnis, James~M Hyman, and Jim~M Cushing.
	\newblock Determining important parameters in the spread of malaria through the
	sensitivity analysis of a mathematical model.
	\newblock {\em Bulletin of mathematical biology}, 70(5):1272, 2008.
	
	\bibitem{ehiri2004mass}
	John~E Ehiri and Ebere~C Anyanwu.
	\newblock Mass use of insecticide-treated bednets in malaria endemic poor
	countries: public health concerns and remedies.
	\newblock {\em journal of public health policy}, 25(1):9--22, 2004.
	
	\bibitem{floore2006mosquito}
	Thomas~G Floore.
	\newblock Mosquito larval control practices: past and present.
	\newblock {\em Journal of the American Mosquito Control Association},
	22(3):527--533, 2006.
	
	\bibitem{9129525}
	D~Fontenille, L~Lochouarn, N~Diagne, C~Sokhna, J~J Lemasson, M~Diatta,
	L~Konate, F~Faye, C~Rogier, and J~F Trape.
	\newblock High annual and seasonal variations in malaria transmission by
	anophelines and vector species composition in {D}ielmo, a holoendemic area in
	{S}enegal.
	\newblock {\em Am J Trop Med Hyg}, 56:247--53, 1997.
	
	\bibitem{DCollCZim}
	D.~Gollin and C.~Zimmermann.
	\newblock Malaria: Disease impacts and long-run income differences.
	\newblock IZA Discussion Papers 2997, Institution for the Study of Labor (IZA),
	August 2007.
	
	\bibitem{Guo_li_CAMQ_06}
	H.~Guo, M.Y. Li, and Z.~Shuai.
	\newblock {Global stability of the endemic equilibrium of multigroup models.}
	\newblock {\em Can. Appl. Math. Q.}, 14(3):259--284, 2006.
	
	\bibitem{Guo_li_PAMS08}
	H.~Guo, M.Y. Li, and Z.~Shuai.
	\newblock A graph-theoretic approach to the method of global {L}yapunov
	functions.
	\newblock {\em Proc. Amer. Math. Soc.}, 2008.
	
	\bibitem{hawley2003community}
	William~A Hawley, Penelope~A Phillips-Howard, Feiko~O ter Kuile, Dianne~J
	Terlouw, John~M Vulule, Maurice Ombok, Bernard~L Nahlen, John~E Gimnig,
	Simon~K Kariuki, Margarette~S Kolczak, et~al.
	\newblock Community-wide effects of permethrin-treated bed nets on child
	mortality and malaria morbidity in western kenya.
	\newblock {\em The American journal of tropical medicine and hygiene},
	68(4\_suppl):121--127, 2003.
	
	\bibitem{MR94c:34067}
	J.~A. Jacquez and C.~P. Simon.
	\newblock Qualitative theory of compartmental systems.
	\newblock {\em SIAM Rev.}, 35(1):43--79, 1993.
	
	\bibitem{jckam201411}
	J.~C. Kamgang, V.~C. Kamla, and S.~Y. Tchoumi.
	\newblock Modeling the dynamics of malaria transmission with bed net protection
	perspective.
	\newblock {\em Applied Mathematics}, 5(19):3156--3205, 11 2014.
	
	\bibitem{KamSal07}
	J.~C. Kamgang and G.~Sallet.
	\newblock Computation of threshold conditions for epidemiological models and
	global stability of the disease free equilibrium.,0.
	\newblock {\em Math. Biosci.}, 213(1):1--12, 2008.
	
	\bibitem{keiser2005reducing}
	Jennifer Keiser, Burton~H Singer, and J{\"u}rg Utzinger.
	\newblock Reducing the burden of malaria in different eco-epidemiological
	settings with environmental management: a systematic review.
	\newblock {\em The Lancet infectious diseases}, 5(11):695--708, 2005.
	
	\bibitem{0999.92036}
	A.~Korobeinikov.
	\newblock {A Lyapunov function for Leslie-Gower predator-prey models.}
	\newblock {\em Appl. Math. Lett.}, 14(6):697--699, 2001.
	
	\bibitem{KoroMMB04}
	A.~Korobeinikov.
	\newblock Lyapunov functions and global properties for {SEIR} and {SEIS}
	models.
	\newblock {\em Math. Med. Biol.}, 21:75--83, 2004.
	
	\bibitem{koroMain04}
	A.~Korobeinikov and P.~K. Maini.
	\newblock A {L}yapunov function and global properties for {SIR} and {SEIR}
	epidemiological models with nonlinear incidence.
	\newblock {\em Math. Biosci. Eng.}, 1(1):57--60, 2004.
	
	\bibitem{1022.34044}
	A.~Korobeinikov and G.~C. Wake.
	\newblock {Lyapunov functions and global stability for SIR, SIRS, and SIS
		epidemiological models.}
	\newblock {\em Appl. Math. Lett.}, 15(8):955--960, 2002.
	
	\bibitem{Las68}
	J.~P. LaSalle.
	\newblock Stability theory for ordinary differential equations. stability
	theory for ordinary differential equations.
	\newblock {\em J. Differ. Equations}, 41:57--65, 1968.
	
	\bibitem{MR0481301}
	J.~P. LaSalle.
	\newblock {\em The stability of dynamical systems}.
	\newblock Society for Industrial and Applied Mathematics, Philadelphia, Pa.,
	1976.
	\newblock With an appendix: ``Limiting equations and stability of nonautonomous
	ordinary differential equations'' by Z. Artstein, Regional Conference Series
	in Applied Mathematics.
	
	\bibitem{MR0594977}
	J.~P. LaSalle.
	\newblock Stability theory and invariance principles.
	\newblock In {\em Dynamical systems (Proc. Internat. Sympos., Brown Univ.,
		Providence, R.I., 1974), Vol. I}, pages 211--222. Academic Press, New York,
	1976.
	
	\bibitem{lawrance2004mosquito}
	Clare~E Lawrance and Ashley~M Croft.
	\newblock Do mosquito coils prevent malaria? a systematic review of trials.
	\newblock {\em Journal of travel medicine}, 11(2):92--96, 2004.
	
	\bibitem{527610}
	J.~Li;, D.~Blakeley;, and R.~J. Smith.
	\newblock The failure of $r_0$.
	\newblock {\em Comp and Math Meth in Medecine}, May 2011.
	
	\bibitem{0458.93001}
	D.~G. Luenberger.
	\newblock {\em {Introduction to dynamic systems. Theory, models, and
			applications.}}
	\newblock John Wiley \& Sons Ltd., 1979.
	
	\bibitem{MaLiuLi03}
	Z.~Ma, J.~Liu, and J.~Li.
	\newblock Stability analysis for differential infectivity epidemic models.
	\newblock {\em Nonlinear Anal. : Real world applications}, 4(5):841--856, 2003.
	
	\bibitem{maia2015mosquito}
	Marta~F Maia, Merav Kliner, Martha Richardson, Christian Lengeler, and Sarah~J
	Moore.
	\newblock Mosquito repellents for malaria prevention.
	\newblock {\em Cochrane Database of Systematic Reviews}, 4(CD011595), 2015.
	
	\bibitem{McClu06}
	C.~C. McCluskey.
	\newblock Lyapunov functions for tuberculosis models with fast and slow
	progression.
	\newblock {\em Math. Biosci. Eng.}, to appear, 2006.
	
	\bibitem{McCluskey07}
	C.~C. McCluskey.
	\newblock Global stability fo a class of mass action systems allowing for
	latency in tuberculosis.
	\newblock {\em J. Math. Anal. Appl.}, 2007.
	
	\bibitem{1008.92032}
	C.C. McCluskey.
	\newblock {A model of HIV/AIDS with staged progression and amelioration.}
	\newblock {\em Math. Biosci.}, 181(1):1--16, 2003.
	
	\bibitem{1076.37012}
	C.C. McCluskey.
	\newblock {A strategy for constructing Lyapunov functions for non-autonomous
		linear differential equations.}
	\newblock {\em Linear Algebra Appl.}, 409:100--110, 2005.
	
	\bibitem{1056.92052}
	C.C. McCluskey and P.~van~den Driessche.
	\newblock {Global analysis of two tuberculosis models.}
	\newblock {\em J. Dyn. Differ. Equations}, 16(1):139--166, 2004.
	
	\bibitem{menze2016multiple}
	Benjamin~D Menze, Jacob~M Riveron, Sulaiman~S Ibrahim, Helen Irving, Christophe
	Antonio-Nkondjio, Parfait~H Awono-Ambene, and Charles~S Wondji.
	\newblock Multiple insecticide resistance in the malaria vector anopheles
	funestus from northern cameroon is mediated by metabolic resistance alongside
	potential target site insensitivity mutations.
	\newblock {\em PLoS One}, 11(10):e0163261, 2016.
	
	\bibitem{morel2005cost}
	Chantal~M Morel, Jeremy~A Lauer, and David~B Evans.
	\newblock Cost effectiveness analysis of strategies to combat malaria in
	developing countries.
	\newblock {\em Bmj}, 331(7528):1299, 2005.
	
	\bibitem{NgwaMCM00}
	A.~G. Ngwa and W.S. Shu.
	\newblock {A mathematical model for Endemic Malaria with Variable Human and
		Mosqioto Population}.
	\newblock {\em Math. Comput. Modelling}, 32:747--763, 2000.
	
	\bibitem{world2015world}
	World~Health Organization.
	\newblock {\em World malaria report 2014}.
	\newblock World Health Organization, 2015.
	
	\bibitem{pluess2010indoor}
	Bianca Pluess, Frank~C Tanser, Christian Lengeler, and Brian~L Sharp.
	\newblock Indoor residual spraying for preventing malaria.
	\newblock {\em Cochrane Database Syst Rev}, 4(4), 2010.
	
	\bibitem{ranson2011pyrethroid}
	Hilary Ranson, Raphael N’Guessan, Jonathan Lines, Nicolas Moiroux, Zinga
	Nkuni, and Vincent Corbel.
	\newblock Pyrethroid resistance in african anopheline mosquitoes: what are the
	implications for malaria control?
	\newblock {\em Trends in parasitology}, 27(2):91--98, 2011.
	
	\bibitem{10697865}
	C~Rogier, A~Tall, N~Diagne, D~Fontenille, A~Spiegel, and J~F Trape.
	\newblock { \it Plasmodium falciparum } clinical malaria: lessons from
	longitudinal studies in {S}enegal.
	\newblock {\em Parassitologia}, 41(1-3):255--9, 2000.
	
	\bibitem{Ross1911}
	R.~Ross.
	\newblock {\em The prevention of malaria}.
	\newblock John Murray, 1911.
	
	\bibitem{russell2015determinants}
	Cheryl~L Russell, Adamu Sallau, Emmanuel Emukah, Patricia~M Graves, Gregory~S
	Noland, Jeremiah~M Ngondi, Masayo Ozaki, Lawrence Nwankwo, Emmanuel Miri,
	Deborah~A McFarland, et~al.
	\newblock Determinants of bed net use in southeast nigeria following mass
	distribution of llins: implications for social behavior change interventions.
	\newblock {\em PLoS One}, 10(10):e0139447, 2015.
	
	\bibitem{russell2011increased}
	Tanya~L Russell, Nicodem~J Govella, Salum Azizi, Christopher~J Drakeley,
	S~Patrick Kachur, and Gerry~F Killeen.
	\newblock Increased proportions of outdoor feeding among residual malaria
	vector populations following increased use of insecticide-treated nets in
	rural tanzania.
	\newblock {\em Malaria journal}, 10(1):80, 2011.
	
	\bibitem{sharp2007seven}
	Brian~L Sharp, Immo Kleinschmidt, Elisabeth Streat, Rajendra Maharaj, Karen~I
	Barnes, David~N Durrheim, Frances~C Ridl, Natasha Morris, Ishen Seocharan,
	Simon Kunene, et~al.
	\newblock Seven years of regional malaria control collaboration—mozambique,
	south africa, and swaziland.
	\newblock {\em The American journal of tropical medicine and hygiene},
	76(1):42--47, 2007.
	
	\bibitem{shillcutt2008cost}
	Samuel Shillcutt, Chantal Morel, Catherine Goodman, Paul Coleman, David Bell,
	Christopher~JM Whitty, and A~Mills.
	\newblock Cost-effectiveness of malaria diagnostic methods in sub-saharan
	africa in an era of combination therapy.
	\newblock {\em Bulletin of the World Health Organization}, 86:101--110, 2008.
	
	\bibitem{MR2518930}
	J.~J. Tewa, J.~L. Dimi, and S.~Bowong.
	\newblock Lyapunov functions for a dengue disease transmission model.
	\newblock {\em Chaos Solitons Fractals}, 39(2):936--941, 2009.
	
	\bibitem{2011.10.085}
	J.J. Tewa, R.~Fokouop, Mewoli, and S.~Bowong.
	\newblock Mathematical analysis of a general class of ordinary differential
	equations coming from within-hosts models of malaria with immune effectors.
	\newblock {\em Applied Mathematics and Computation}, 218(14):7347--7361, march
	2012.
	
	\bibitem{utzinger2001efficacy}
	J{\"u}rg Utzinger, Yesim Tozan, and Burton~H Singer.
	\newblock Efficacy and cost-effectiveness of environmental management for
	malaria control.
	\newblock {\em Tropical Medicine \& International Health}, 6(9):677--687, 2001.
	
	\bibitem{VddWat02}
	P.~van~den Driessche and J.~Watmough.
	\newblock reproduction numbers and sub-threshold endemic equilibria for
	compartmental models of disease transmission.
	\newblock {\em Math. Biosci.}, 180:29--48, 2002.
	
	\bibitem{walker2007contributions}
	K~Walker and M~Lynch.
	\newblock Contributions of anopheles larval control to malaria suppression in
	tropical africa: review of achievements and potential.
	\newblock {\em Medical and veterinary entomology}, 21(1):2--21, 2007.
	
	\bibitem{white2011costs}
	Michael~T White, Lesong Conteh, Richard Cibulskis, and Azra~C Ghani.
	\newblock Costs and cost-effectiveness of malaria control interventions-a
	systematic review.
	\newblock {\em Malaria journal}, 10(1):337, 2011.
	
	\bibitem{wmr2013}
	WHO.
	\newblock World malaria report 2013.
	\newblock Technical report, WHO, Dec 2013.
	
	\bibitem{wilson2011systematic}
	Anne~L Wilson et~al.
	\newblock A systematic review and meta-analysis of the efficacy and safety of
	intermittent preventive treatment of malaria in children (iptc).
	\newblock {\em PloS one}, 6(2):e16976, 2011.
	
	\bibitem{worrall2011large}
	Eve Worrall and Ulrike Fillinger.
	\newblock Large-scale use of mosquito larval source management for malaria
	control in africa: a cost analysis.
	\newblock {\em Malaria journal}, 10(1):338, 2011.
	
	\bibitem{yohannes2005can}
	Mekonnen Yohannes, Mituku Haile, Tedros~A Ghebreyesus, Karen~H Witten, Asefaw
	Getachew, Peter Byass, and Steve~W Lindsay.
	\newblock Can source reduction of mosquito larval habitat reduce malaria
	transmission in tigray, ethiopia?
	\newblock {\em Tropical Medicine \& International Health}, 10(12):1274--1285,
	2005.
	
	\bibitem{zhu2017outdoor}
	Lin Zhu, G{\"u}nter~C M{\"u}ller, John~M Marshall, Kristopher~L Arheart,
	Whitney~A Qualls, WayWay~M Hlaing, Yosef Schlein, Sekou~F Traore, Seydou
	Doumbia, and John~C Beier.
	\newblock Is outdoor vector control needed for malaria elimination? an
	individual-based modelling study.
	\newblock {\em Malaria journal}, 16(1):266, 2017.
	
	\bibitem{Zongo09}
	P.~Zongo.
	\newblock {\em Mod{\'e}lisation math{\'e}matique de la dynamique de
		transmission du paludisme}.
	\newblock PhD thesis, Universite de Ouagadougou, 2009.
	
\end{thebibliography}

\end{document}